\documentclass[aps,prd,nofootinbib,superscriptaddress,twoside,twocolumn,floatfix,a4paper]{revtex4-2}
\usepackage{amsmath}
\usepackage{amsfonts,mathtools}
\usepackage{bm}
\usepackage{bbm, dsfont}
\usepackage{units}
\usepackage{tabularx}
\usepackage{extarrows}
\usepackage{enumitem}
\usepackage{pgf}
\usepackage{color, colortbl}
\usepackage{microtype}
\usepackage{placeins}
\usepackage{booktabs}

\usepackage{hyperref}
\hypersetup{
    colorlinks,
    linkcolor={blue},
    citecolor={green!70!black},
    urlcolor={blue!80!black},
    pdftitle={Octet baryon charges},
    pdfauthor={RQCD},
}
\newcommand{\GeV}{\text{GeV}}
\newcommand{\MeV}{\text{MeV}}	
\newcommand{\fm}{\text{fm}}
\newcommand{\latt}{\text{latt}}
\newcommand{\MS}{\overline{\text{MS}}}
\newcommand{\QED}{\text{QED}}
\newcommand{\QCD}{\text{QCD}}
\newcommand{\const}{\text{const.}}
\DeclareMathOperator{\tr}{tr}
\DeclareRobustCommand{\bmr}[1]{\bm{#1}}
\begin{document}
\title{Octet baryon isovector charges from $\bmr{N_f = 2 + 1}$ lattice QCD}
\author{Gunnar S. Bali}
\email[]{gunnar.bali@ur.de}
\affiliation{
Institut für Theoretische Physik, 
Universität Regensburg, 
93040 Regensburg, Germany.
}
\author{Sara Collins}
\email[]{sara.collins@ur.de}
\affiliation{
Institut für Theoretische Physik, 
Universität Regensburg, 
93040 Regensburg, Germany.
}
\author{Simon Heybrock}
\affiliation{
  European Spallation Source ERIC, Data Management and Software Centre (DMSC),
  Box 176, 22100 Lund, Sweden}
\author{Marius~Löffler}
\affiliation{
Institut für Theoretische Physik, 
Universität Regensburg, 
93040 Regensburg, Germany.
}
\author{Rudolf Rödl}
\affiliation{
Institut für Theoretische Physik, 
Universität Regensburg, 
93040 Regensburg, Germany.
}
\author{Wolfgang Söldner}
\email[]{wolfgang.soeldner@ur.de}
\affiliation{
Institut für Theoretische Physik, 
Universität Regensburg, 
93040 Regensburg, Germany.
}
\author{Simon Weishäupl}
\email[]{simon.weishaeupl@ur.de}
\affiliation{
Institut für Theoretische Physik, 
Universität Regensburg, 
93040 Regensburg, Germany.
}
\collaboration{RQCD Collaboration}
\date{August 31, 2023}
\begin{abstract}
  We determine the axial, scalar and tensor isovector charges of the
  nucleon, sigma and cascade baryons as well as the difference between
  the up and down quark masses, $m_u-m_d$. We employ gauge ensembles with
  $N_f=2+1$ non-perturbatively improved Wilson fermions at six values
  of the lattice spacing in the range $a\approx (0.039 - 0.098) \,
  \fm$, generated by the Coordinated Lattice Simulations (CLS)
  effort. The pion mass $M_\pi$ ranges from around $430 \, \MeV$ down
  to a near physical value of $130 \, \MeV$ and the linear spatial
  lattice extent $L$ varies from $6.5\,M_{\pi}^{-1}$ to
  $3.0\,M_{\pi}^{-1}$, where $ L M_\pi \geq 4$ for the majority of the
  ensembles. This allows us to perform a controlled
  interpolation/extrapolation of the charges to the physical mass
  point in the infinite volume and continuum limit.
  Investigating SU(3) flavour symmetry, we find moderate
  symmetry breaking effects for the axial charges at the physical
  quark mass point, while no significant effects are found for the
  other charges within current uncertainties.
\end{abstract} 

\maketitle
\section{Introduction}
A charge of a hadron parameterizes the strength of its interaction at
small momentum transfer with a particle that couples to this
particular charge.  For instance, the isovector axial charge
determines the $\beta$ decay rate of the neutron.  At the same time,
this charge corresponds to the difference between the contribution of
the spin of the up quarks minus the spin of the down quarks to the
total longitudinal spin of a nucleon in the light front frame that is
used in the collinear description of deep inelastic scattering. This
intimate connection to spin physics at large virtualities and, more
specifically, to the decomposition of the longitudinal proton spin
into contributions of the gluon total angular momentum and the spins
and angular momenta for the different quark
flavours~\cite{Ji:1996ek,Jaffe:1989jz} opens up a whole area of
intense experimental and theoretical research: the first Mellin moment
of the helicity structure functions $g_1(x)$ is related to the sum of
the individual spins of the quarks within the proton. For lattice
determinations of the individual quark contributions to its first and
third moments, see, e.g.,
Refs.~\cite{QCDSF:2011aa,Engelhardt:2012gd,Gong:2015iir,Alexandrou:2017oeh,Green:2017keo}
and Ref.~\cite{Burger:2021knd}, respectively. Due to the lack of
experimental data on $g_1(x)$, in particular at small Bjorken-$x$, and
difficulties in the flavour separation, usually additional information
is used in determinations of the helicity parton distribution
functions (PDFs) from global fits to experimental
data~\cite{deFlorian:2008mr,Blumlein:2010rn,Nocera:2014gqa,Sato:2016tuz,Ethier:2017zbq}. In
addition to the axial charge $g_A$ of the proton, this includes
information from hyperon decays, in combination with SU(3) flavour
symmetry relations whose validity need to be checked.

In this article we establish the size of the corrections to SU(3)
flavour symmetry in the axial sector and also for the scalar and the
tensor isovector charges of the octet baryons: in analogy to the
connection between axial charges and the first moments of helicity
PDFs, the tensor charges are related to first moments of transversity
PDFs. This was exploited recently in a global fit by the
JAM~Collaboration~\cite{Lin:2017stx,Gamberg:2022kdb}. Since no tensor
or scalar couplings contribute to tree-level Standard Model processes,
such interactions may hint at new physics and it is important to
constrain new interactions (once discovered) using lattice QCD input,
see, e.g., Ref.~\cite{Bhattacharya:2011qm} for a detailed
discussion. SU(3) flavour symmetry among the scalar charges is also
instrumental regarding recent tensions between different
determinations of the pion nucleon $\sigma$~term, see
Ref.~\cite{RQCD:2022xux} for a summary of latest phenomenological and
lattice QCD results and, e.g., the discussion in Sec.~10 of
Ref.~\cite{Leutwyler:2015jga} about the connection between
OZI~violation, (approximate) SU(3) flavour symmetry and the value of
the pion nucleon $\sigma$~term.  Finally, the scalar isovector charges
relate the QCD part of the mass splitting between isospin partners to
the difference of the up and down quark masses.

Assuming SU(3) flavour symmetry, the charges for the whole baryon
octet in a given channel only depend on two independent parameters.
For the proton and the axial charge, this relation reads
$g_A=F_A+D_A$, where in the massless limit $F_A$ and $D_A$ correspond
to the chiral perturbation theory (ChPT) low energy constants (LECs)
$F$ and $D$, respectively.  Already in the first lattice calculations
of the axial charge of the
proton~\cite{Fucito:1982ff,Gusken:1989ad,Woloshyn:1989ae}, that were
carried out in the quenched approximation, $F_A$ and $D_A$ have been
determined separately.  However, in spite of the long history of
nucleon structure calculations, SU(3) flavour symmetry breaking is
relatively little explored using lattice QCD: only very few
investigations of axial charges of the baryon octet exist to
date~\cite{Lin:2007ap,Erkol:2009ev,Alexandrou:2016xok,Savanur:2018jrb,Smail:2023eyk}
and only one of these includes the scalar and tensor
charges~\cite{Smail:2023eyk}.  Here we compute these charges for the
light baryon octet.  We also predict the difference between the up and
down quark masses, the QCD contributions to baryon isospin mass
splittings and isospin differences of pion baryon $\sigma$~terms.

This article is organized as follows. In Sec.~\ref{sec:hyperon} we
define the octet baryon charges and some related quantities
of interest. In Sec.~\ref{sec:lattice} the lattice set-up is described,
including the gauge ensembles employed, the computational
methods used to obtain two- and three-point correlation functions
and the excited state analysis performed to extract the ground state
matrix elements of interest. We continue with details on the
non-perturbative renormalization and order $a$ improvement, before
explaining our infinite volume, continuum limit and quark mass
extrapolation strategy. Our results for the charges in the
infinite volume, continuum limit at physical quark masses are then presented in 
Sec.~\ref{sec:results}. Subsequently, in Sec.~\ref{sec:discussion} we
discuss SU(3) symmetry breaking effects, determine the up and down 
quark mass difference from the scalar charge of the $\Sigma$ baryon,
split isospin breaking effects on the baryon masses into
QCD and QED contributions and determine isospin breaking corrections
to the pion baryon $\sigma$~terms. Throughout this section we
also compare our results to literature values, before
we give a summary and an outlook in Sec.~\ref{sec:summary}.
In the appendices further details regarding the stochastic
three-point function method are given and additional data tables and figures
are provided.

\section{Octet baryon charges\label{sec:hyperon}}

\begin{figure}
  \centering
  \includegraphics[width=0.4\textwidth]{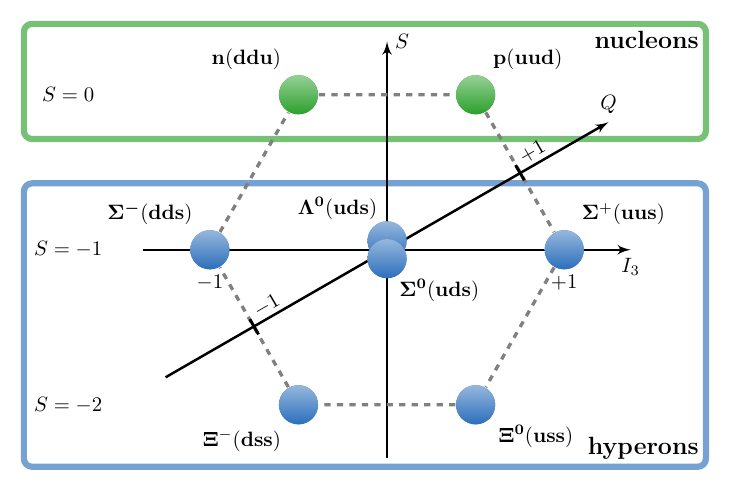}
  \caption{The spin-1/2 baryon octet where $S$, $I_3$ and $Q=(1+S)/2+I_3$ label strangeness, isospin and charge, respectively.\label{fig:b_octet}}
\end{figure}

All light baryons (i.e., baryons without charm or bottom quarks) with
strangeness $S<0$, i.e., with a net difference between the numbers of
strange ($s$) antiquarks and quarks are usually called hyperons.  The
spin-$1/2$ baryon octet, depicted in Fig.~\ref{fig:b_octet}, contains
the nucleons $N\in\{p,n\}$, besides the $S=-1$ hyperons $\Lambda^0$
and $\Sigma\in\{\Sigma^+,\Sigma^0,\Sigma^-\}$ and the $S=-2$ hyperons
$\Xi\in\{\Xi^0,\Xi^-\}$ (cascades). We assume isospin symmetry
$m_{\ell}=m_u=m_d$, where $m_{\ell}$ corresponds to the average mass
of the physical up ($u$) and down ($d$) quarks.  In this case, the
baryon masses within isomultiplets are degenerate and simple relations
exist between matrix elements that differ in terms of the isospin
$I_3$ of the baryons and of the local operator (current).

Baryon charges $g_J^{B^\prime B}$ are obtained from matrix elements of
the form
\begin{align}\label{eq:decomp}
  \langle B^\prime(p',s^\prime ) | \bar{u}\Gamma_J d | B (p, s) \rangle 
  = g_J^{B^\prime B} \bar{u}_{B^\prime} (p',s^\prime) \Gamma_J u_B (p,s)
\end{align}
at zero four-momentum transfer $q^2=(p'-p)^2=0$.  Above, $u_B (p,s)$
denotes the Dirac spinor of a baryon~$B$ with four momentum~$p$ and
spin~$s$.  We restrict ourselves to $\Delta I_3=1$ transitions within
the baryon octet.  In this case $p^\prime=p$, since in isosymmetric
QCD $m_{B^\prime}=m_B$, and it is sufficient to set
$\mathbf{p}=\boldsymbol{0}$.  Rather than using the above $I_3=1$
currents $\bar{u}\Gamma_J d$ (where the vector and axial currents
couple to the $W^-$ boson), it is convenient to define $I_3=0$
isovector currents,
\begin{align}
   \mathcal{O}_J(x) = \bar{u}(x) \Gamma_J u(x) - \bar{d}(x) \Gamma_J d(x),
\end{align}
and the corresponding charges $g_J^B$,
\begin{align}\label{eq:decomp2}
  \langle B(p,s) | \mathcal{O}_J | B (p, s) \rangle 
  = g_J^{B} \bar{u}_{B} (p,s) \Gamma_J u_B (p,s),
\end{align}
which, in the case of isospin symmetry, are trivially related to the $g_J^{B'B}$:
\begin{align}
  g_J^N&\coloneqq g_J^p=g_J^{pn},\label{eq:ccc1}\\
  g_J^\Sigma&\coloneqq g_J^{\Sigma^+}=-\sqrt{2}\,g_J^{\Sigma^+\Sigma^0},\\
  g_J^\Xi&\coloneqq g_J^{\Xi^0}=-g_J^{\Xi^0\Xi^-}.\label{eq:ccc3}
\end{align}
Note that we do not include the $\Lambda$ baryon here since in this
case the isovector combination trivially gives zero. We consider
vector~($V$), axialvector~($A$), scalar~($S$) and tensor~($T$)
operators which are defined through the Dirac matrices $\Gamma_J =
\gamma_4,\, \gamma_i\gamma_5,\, \mathds{1},\, \sigma_{ij} $ for $J \in
\{V, A, S, T\}$, with $\sigma_{\mu\nu} = \frac{1}{2} [\gamma_\mu,
  \gamma_\nu]$, where $i,j \in \{1,2,3\}$ and $i < j$.

The axial charges in the $m_s=m_{\ell}=0$ chiral limit are important
parameters in SU(3) ChPT and enter the
expansion of every baryonic quantity. These couplings can be
decomposed into two LECs $F$ and $D$ which appear in
the first order meson-baryon Lagrangian for three light quark flavours
(see, e.g., Ref.~\cite{Borasoy:1996bx}):
\begin{align}
  g_A^N=F+D,\quad g_A^{\Sigma}=2F,\quad g_A^{\Xi}=F-D.\label{eq:FD1}
\end{align}
Due to group theoretical constraints, see, e.g.,
Refs.~\cite{Cooke:2012xv,Bickerton:2019nyz}, such a decomposition also
holds for $m_s=m_{\ell}>0$, for the axial as well as for the other
charges.  We define for $m=m_s=m_{\ell}$
\begin{align}
  g_J^N(m)&=F_J(m)+D_J(m),\label{eq:symm1}\\
  g_J^{\Sigma}(m)&=2F_J(m),\\
  g_J^{\Xi}(m)&=F_J(m)-D_J(m)\label{eq:symm3},
\end{align}
where $F=F_A(0)$, $D=D_A(0)$. The vector Ward identity (conserved
vector current, CVC relation) implies that $g_V^N=g_V^{\Xi}=F_V=1$ and
$g_V^{\Sigma}=2F_V$, i.e., in this case the above relations also hold
for $m_s\neq m_{\ell}$, with $F_V(m)=1$ and $D_V(m)=0$.

In this article we determine the charges at many different positions
in the quark mass plane and investigate SU(3) flavour symmetry
breaking, i.e., the extent of violation of
Eqs.~\eqref{eq:symm1}--\eqref{eq:symm3}.  Due to this, other than for
$J\neq V$ where $D_V/F_V=0$, the functions $D_J(m)$ and $F_J(m)$ are
not uniquely determined at the physical point, where $m_s\gg
m_{\ell}$. At this quark mass point we will find the approximate
ratios $D_A/F_A\approx (1.55 \mbox{ -- } 1.95)$, $D_S/F_S\approx -(0.2 \mbox{ -- } 0.5)$ and
$D_T/F_T\approx 1.5$. The first ratio can be compared to the SU(6)
quark model expectation $D_A(m)/F_A(m)= 3/2$ (see, e.g.,
ref.~\cite{Jenkins:1991es}), which is consistent with the large-$N_c$
limit~\cite{Dashen:1993ac}.

\section{Lattice set-up\label{sec:lattice}}

In this section we present the details of our lattice set-up. First,
we describe the gauge ensembles employed and the construction of the
correlation functions. The computation of the three-point correlation
functions via a stochastic approach is briefly discussed. Following
this, we present the fitting analysis and treatment of excited state
contributions employed to extract the ground state baryon matrix
elements. The renormalization factors used to match to the continuum
$\MS$ scheme and the improvement factors utilized to ensure leading
$O(a^2)$ discretization effects are then given. Finally, the strategy
for interpolation/extrapolation to the physical point in the infinite
volume and continuum limit is outlined.

\subsection{Gauge ensembles\label{sec:ensembles}}

\begin{figure*}
  \includegraphics[width=0.98\textwidth]{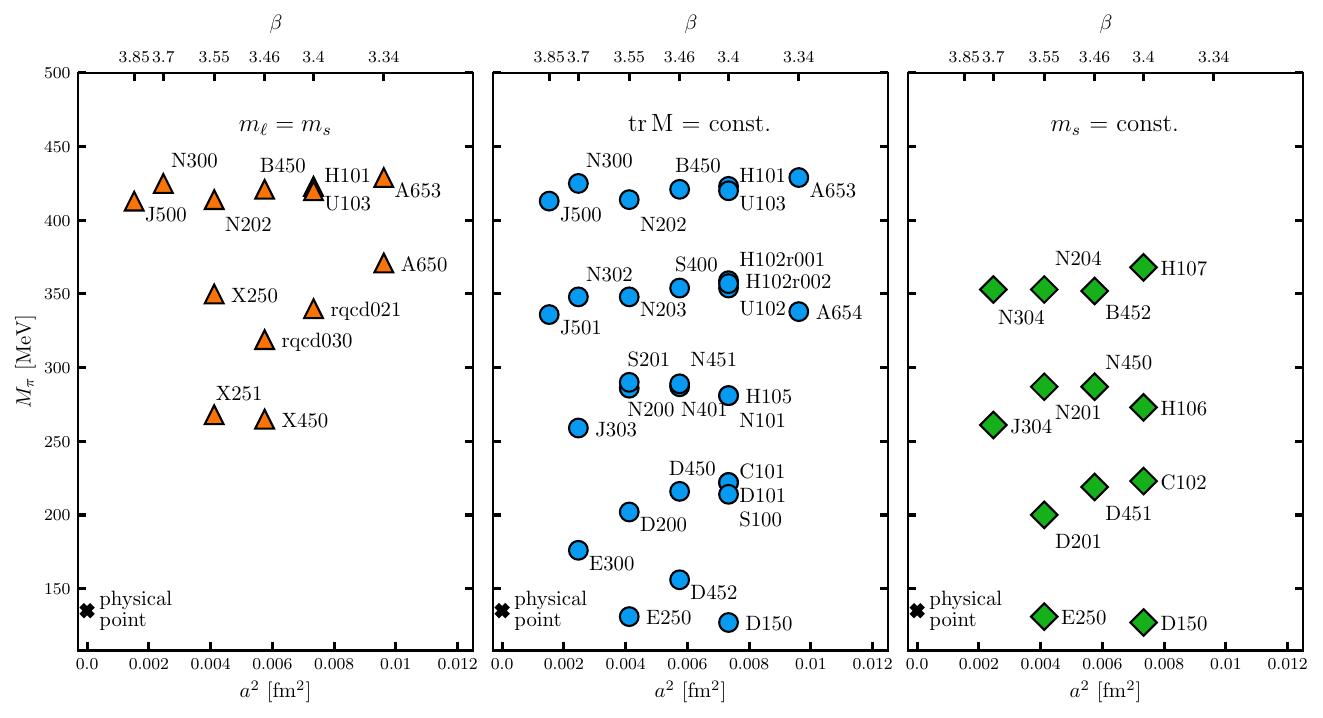}
  \caption{Parameter landscape of the ensembles listed in
    Table~\ref{tab:cls_ensembles}. The ensembles are grouped according
    to the three quark mass trajectories~(see the text and
    Fig.~\ref{fig:cls_trajectories}): (left) the symmetric
    line~($m_\ell = m_s$), (middle) the $\tr\,M = \const$ line and
    (right) the $m_s=\const$ line.\label{fig:cls_ensembles}}
\end{figure*}

We employ ensembles generated with $N_f = 2 + 1$ flavours of
non-perturbatively $\mathcal{O}(a)$ improved Wilson fermions and a
tree-level Symanzik improved gauge action, which were mostly produced
within the Coordinated Lattice Simulations (CLS)~\cite{Bruno:2014jqa}
effort.  Either periodic or open boundary conditions in
time~\cite{Luscher:2011kk} are imposed, where the latter choice is
necessary for ensembles with $a<0.06\,\fm$ in order to avoid freezing
of the topological charge and thus to ensure
ergodicity~\cite{Schaefer:2010hu}.  The hybrid Monte Carlo (HMC)
simulations are stabilized by introducing a twisted mass term for the
light quarks~\cite{Luscher:2012av}, whereas the strange quark is
included via the rational hybrid Monte Carlo (RHMC)
algorithm~\cite{Clark:2006fx}. The modifications of the target action
are corrected for by applying the appropriate reweighting, see
Refs.~\cite{Bruno:2014jqa,Mohler:2020txx,RQCD:2022xux} for further details.

\begin{figure*}
  \includegraphics[width=0.75\textwidth]{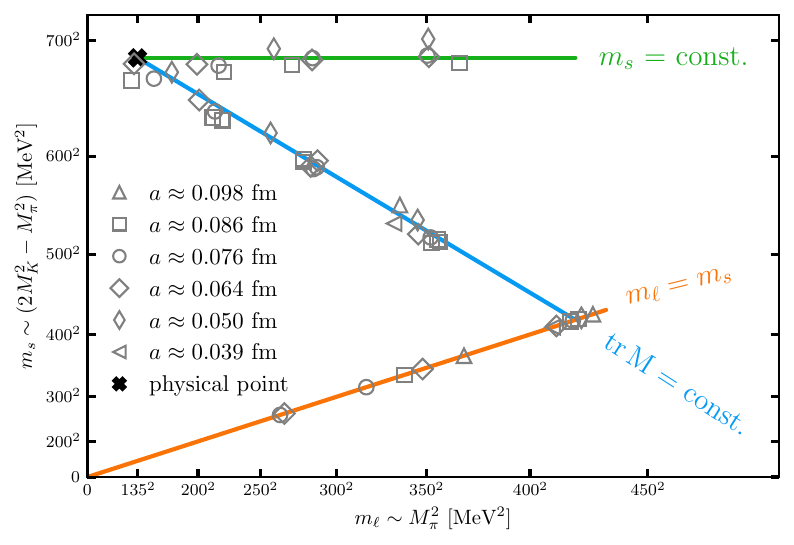}
  \caption{Position of the ensembles in the quark mass plane. The
    $\tr\,M = \const$ line (indicated as a blue line) and $m_s =
    \const$ line (green line) intersect close to the physical point
    (black cross). The symmetric line~(orange line), which approaches
    the SU(3) chiral limit, crosses the $\tr\,M = \const$ line around
    $M_\pi=411\,\MeV$. \label{fig:cls_trajectories}}
\end{figure*}

In total 47 ensembles were analysed spanning six lattice spacings $a$
in the range $0.039\,\fm\lesssim a\lesssim 0.098\,\fm$, with pion
masses between $430 \, \MeV$ and $130 \, \MeV$ (below the physical
pion mass), as shown in Fig.~\ref{fig:cls_ensembles}.  The lattice
spatial extent $L$ is kept sufficiently large, where $ L M_\pi \geq 4$
for the majority of the ensembles. A limited number of smaller volumes
are employed to enable finite volume effects to be investigated, with
the spatial extent varying across all the ensembles in the range $3.0
\leq L M_\pi \leq 6.5$. Further details are given in
Table~\ref{tab:cls_ensembles}.  The ensembles lie along three
trajectories in the quark mass plane, as displayed in
Fig.~\ref{fig:cls_trajectories}:
\begin{itemize}
  \item the symmetric line: the light and strange quark masses are
    degenerate ($m_\ell = m_s$) and SU(3) flavour symmetry is exact.
  \item The $\tr\,M = \const$ line: starting at the $m_\ell = m_s$
    flavour symmetric point, the trajectory approaches the physical
    point holding the trace of the quark mass matrix ($2m_\ell + m_s$,
    i.e., the flavour averaged quark mass) constant such that $2M_K^2
    + M_\pi^2$ is close to its physical value.
  \item The $m_s = \const$ line: the renormalized strange quark mass
    is kept near to its physical value~\cite{Bali:2016umi}.
\end{itemize}
The latter two trajectories intersect close to the physical point,
whereas the symmetric line approaches the SU(3) chiral limit. In
figures where data from different lines are shown, we will distinguish
these, employing the symbol shapes of Fig.~\ref{fig:cls_ensembles}
(triangle, circle, diamond).  The excellent coverage of the quark mass
plane enables the interpolation/extrapolation of the results for the
charges to the physical point to be tightly constrained. In addition,
considering the wide range of lattice spacings and spatial volumes and
the high statistics available for most ensembles, all sources of
systematic uncertainty associated with simulating at unphysical quark
mass, finite lattice spacing and finite volume can be
investigated. Our strategy for performing a simultaneous continuum,
quark mass and infinite volume extrapolation is given in
Sec.~\ref{sec:extrapol}.

\begin{table*}
  \caption{List of the gauge ensembles analysed in this work. The
    rqcd{\tt xyz} ensembles were generated by the RQCD group using the
    BQCD code~\cite{Nakamura:2010qh}, whereas all other ensembles were
    created within the CLS effort. Note that for H102 there are two
    replicas. These have the same parameters but were generated with
    slightly different algorithmic set-ups and, therefore, have to be
    analysed separately. N401 and N451 differ in terms of the boundary
    conditions (bc) imposed in the time direction~(open (o) and
    anti-periodic (p), respectively). The lattice spacings $a$ are
    determined in Ref.~\cite{RQCD:2022xux}. In the second to last
    column $t$ denotes the source-sink separation of the connected
    three-point functions. The subscript (superscript) given for each
    separation indicates the number of measurements performed using
    the sequential source (stochastic) method on each
    configuration~(see Secs.~\ref{sec:analysis}
    and~\ref{sec:stoch3pts}). The last column gives $N_{\text{cnfg}}$,
    the number of configurations analysed.\label{tab:cls_ensembles}}
  \centering
  \begin{ruledtabular}
  \begin{tabular}{llcccrcclcc}
    Ensemble  & $\beta$ & $ a [\fm] $ & trajectory & bc  & $N_t\cdot N_s^3$ & $M_\pi [\MeV]$ & $M_K [\MeV]$ & $LM_\pi$ & $t/a$ & $N_{\text{cnfg}}$ \\
    \cmidrule{1-11}
    A650     & 3.34 & 0.098 & sym       & p & $48 \cdot 24^3$  & 371 & 371 & 4.43 & $7_{3}^{}, 9_{3}^{}, 11_{3}^{}, 13_{4}^{}$ & 5062 \\ 
A653     &      &       & tr\,M/sym & p & $48 \cdot 24^3$  & 429 & 429 & 5.12 & $7_{3}^{}, 9_{3}^{}, 11_{3}^{}, 13_{4}^{}$ & 2525 \\ 
A654     &      &       & tr\,M     & p & $48 \cdot 24^3$  & 338 & 459 & 4.04 & $7_{3}^{2}, 9_{3}^{2}, 11_{3}^{2}, 13_{4}^{2}$ & 2533 \\ 
\cmidrule{1-11}
rqcd021  & 3.4  & 0.086 & sym       & p & $32 \cdot 32^3$  & 340 & 340 & 4.73 & $8_{2}^{}, 10_{2}^{}, 12_{4}^{}, 14_{4}^{}$ & 1541 \\
H101     &      &       & tr\,M/sym & o & $96 \cdot 32^3$  & 423 & 423 & 5.88 & $8_{2}^{}, 10_{2}^{}, 12_{2}^{}, 14_{2}^{}$ & 2000 \\ 
U103     &      &       & tr\,M/sym & o & $128 \cdot 24^3$ & 420 & 420 & 4.38 & $8_{1}^{}, 10_{2}^{}, 12_{3}^{}, 14_{4}^{}$ & 2470 \\ 
H102r001 &      &       & tr\,M     & o & $96 \cdot 32^3$  & 354 & 442 & 4.92 & $8_{1}^{4}, 10_{2}^{4}, 12_{3}^{4}, 14_{4}^{4}$ & 997  \\ 
H102r002 &      &       & tr\,M     & o & $96 \cdot 32^3$  & 359 & 444 & 4.99 & $8_{1}^{4}, 10_{2}^{4}, 12_{3}^{4}, 14_{4}^{4}$ & 1000 \\ 
U102     &      &       & tr\,M     & o & $128 \cdot 24^3$ & 357 & 445 & 3.72 & $8_{1}^{}, 10_{2}^{}, 12_{3}^{}, 14_{4}^{}$ & 2210 \\ 
N101     &      &       & tr\,M     & o & $128 \cdot 48^3$ & 281 & 467 & 5.86 & $8_{1}^{}, 10_{2}^{}, 12_{3}^{}, 14_{4}^{}$ & 1457 \\ 
H105     &      &       & tr\,M     & o & $96 \cdot 32^3$  & 281 & 468 & 3.91 & $8_{1}^{2}, 10_{2}^{2}, 12_{3}^{2}, 14_{4}^{2}$ & 2038 \\ 
D101     &      &       & tr\,M     & o & $128 \cdot 64^3$ & 222 & 476 & 6.18 & $8_{1}^{}, 10_{2}^{}, 12_{3}^{}, 14_{4}^{}$ & 608  \\ 
C101     &      &       & tr\,M     & o & $96 \cdot 48^3$  & 222 & 476 & 4.63 & $8_{1}^{2}, 10_{2}^{2}, 12_{3}^{2}, 14_{4}^{2}$ & 2000 \\ 
S100     &      &       & tr\,M     & o & $128 \cdot 32^3$ & 214 & 476 & 2.98 & $8_{1}^{}, 10_{2}^{}, 12_{3}^{}, 14_{4}^{}$ & 983  \\ 
D150     &      &       & tr\,M/ms     & p & $128 \cdot 64^3$ & 127 & 482 & 3.53 & $8_{1}^{}, 10_{2}^{}, 12_{3}^{}, 14_{4}^{}$ & 603  \\ 
H107     &      &       & ms        & o & $96 \cdot 32^3$  & 368 & 550 & 5.12 & $8_{2}^{2}, 10_{2}^{2}, 12_{3}^{2}, 14_{4}^{2}$ & 1564 \\ 
H106     &      &       & ms        & o & $96 \cdot 32^3$  & 273 & 520 & 3.80 & $8_{2}^{4}, 10_{2}^{4}, 12_{3}^{4}, 14_{4}^{4}$ & 1553 \\ 
C102     &      &       & ms        & o & $96 \cdot 48^3$  & 223 & 504 & 4.65 & $8_{2}^{4}, 10_{2}^{4}, 12_{3}^{4}, 14_{4}^{4}$ & 1500 \\ 
\cmidrule{1-11}
rqcd030  & 3.46 & 0.076 & sym       & p & $64 \cdot 32^3$  & 319 & 319 & 3.93 & $9_{4}^{}, 11_{4}^{}, 13_{8}^{}, 16_{8}^{}$ & 1224 \\ 
X450     &      &       & sym       & p & $64 \cdot 48^3$  & 265 & 265 & 4.90 & $9_{2}^{}, 11_{2}^{}, 13_{4}^{}, 16_{4}^{}$ & 400  \\ 
B450     &      &       & tr\,M/sym & p & $64 \cdot 32^3$  & 421 & 421 & 5.19 & $9_{3}^{}, 11_{3}^{}, 14_{3}^{}, 16_{4}^{}$ & 1612 \\ 
S400     &      &       & tr\,M     & o & $128 \cdot 32^3$ & 354 & 445 & 4.36 & $9_{1}^{2}, 11_{2}^{2}, 13_{3}^{}, 14_{}^{2}, 16_{4}^{2}$ & 2872 \\ 
N451     &      &       & tr\,M     & p & $128 \cdot 48^3$ & 289 & 466 & 5.34 & $9_{4}^{4}, 11_{4}^{4}, 13_{4}^{}, 14_{}^{4}, 16_{4}^{4}$ & 1011 \\ 
N401     &      &       & tr\,M     & o & $128 \cdot 48^3$ & 287 & 464 & 5.30 & $9_{1}^{2}, 11_{2}^{2}, 13_{3}^{}, 14_{}^{2}, 16_{4}^{2}$ & 1086 \\ 
D450     &      &       & tr\,M     & p & $128 \cdot 64^3$ & 216 & 480 & 5.32 & $9_{4}^{2}, 11_{4}^{2}, 13_{4}^{}, 14_{}^{2}, 16_{4}^{2}$ & 621  \\ 
D452     &      &       & tr\,M     & p & $128 \cdot 64^3$ & 156 & 488 & 3.84 & $9_{}^{4}, 11_{}^{4}, 14_{}^{4}, 16_{}^{4}$ & 1000 \\ 
B452     &      &       & ms        & p & $64 \cdot 32^3$  & 352 & 548 & 4.34 & $9_{3}^{2}, 11_{3}^{2}, 13_{3}^{}, 14_{}^{2}, 16_{4}^{2}$ & 1944 \\ 
N450     &      &       & ms        & p & $128 \cdot 48^3$ & 287 & 528 & 5.30 & $9_{4}^{2}, 11_{4}^{2}, 13_{4}^{}, 14_{}^{2}, 16_{4}^{2}$ & 1132 \\ 
D451     &      &       & ms        & p & $128 \cdot 64^3$ & 219 & 507 & 5.39 & $9_{4}^{2}, 11_{4}^{2}, 13_{4}^{}, 14_{}^{2}, 16_{4}^{2}$ & 458  \\ 
\cmidrule{1-11}
X250     & 3.55 & 0.064 & sym       & p & $64 \cdot 48^3$  & 350 & 350 & 5.47 & $11_{2}^{}, 14_{2}^{}, 16_{4}^{}, 19_{4}^{}$ & 1493 \\ 
X251     &      &       & sym       & p & $64 \cdot 48^3$  & 268 & 268 & 4.19 & $11_{4}^{}, 14_{4}^{}, 16_{8}^{}, 19_{8}^{}$ & 1474 \\ 
N202     &      &       & tr\,M/sym & o & $128 \cdot 48^3$ & 414 & 414 & 6.47 & $11_{1}^{}, 14_{2}^{}, 16_{2}^{}, 19_{4}^{}$ & 883  \\ 
N203     &      &       & tr\,M     & o & $128 \cdot 48^3$ & 348 & 445 & 5.44 & $11_{1}^{4}, 14_{2}^{4}, 16_{3}^{4}, 19_{4}^{4}$ & 1543 \\ 
N200     &      &       & tr\,M     & o & $128 \cdot 48^3$ & 286 & 466 & 4.47 & $11_{1}^{4}, 14_{2}^{4}, 16_{3}^{4}, 19_{4}^{4}$ & 1712 \\ 
S201     &      &       & tr\,M     & o & $128 \cdot 32^3$ & 290 & 471 & 3.02 & $11_{1}^{}, 14_{2}^{}, 16_{3}^{}, 19_{4}^{}$ & 2092 \\ 
D200     &      &       & tr\,M     & o & $128 \cdot 64^3$ & 202 & 484 & 4.21 & $11_{1}^{2}, 14_{2}^{2}, 16_{3}^{2}, 19_{4}^{2}$ & 2001 \\ 
E250     &      &       & tr\,M/ms     & p & $192 \cdot 96^3$ & 131 & 493 & 4.10 & $11_{4}^{}, 14_{4}^{}, 16_{4}^{}, 19_{4}^{}$ & 490  \\ 
N204     &      &       & ms        & o & $128 \cdot 48^3$ & 353 & 549 & 5.52 & $11_{2}^{2}, 14_{2}^{2}, 16_{3}^{2}, 19_{4}^{2}$ & 1500 \\ 
N201     &      &       & ms        & o & $128 \cdot 48^3$ & 287 & 527 & 4.49 & $11_{2}^{2}, 14_{2}^{2}, 16_{3}^{2}, 19_{4}^{2}$ & 1522 \\ 
D201     &      &       & ms        & o & $128 \cdot 64^3$ & 200 & 504 & 4.17 & $11_{1}^{4}, 14_{2}^{4}, 16_{3}^{4}, 19_{4}^{4}$ & 1078 \\ 
\cmidrule{1-11}
N300     & 3.7  & 0.049  & tr\,M/sym & o & $128 \cdot 48^3$ & 425 & 425 & 5.15 & $14_{1}^{}, 17_{2}^{}, 21_{2}^{}, 24_{4}^{}$ & 1539 \\ 
N302     &      &       & tr\,M     & o & $128 \cdot 48^3$ & 348 & 455 & 4.21 & $14_{1}^{2}, 17_{2}^{2}, 21_{3}^{2}, 24_{4}^{2}$ & 1383 \\ 
J303     &      &       & tr\,M     & o & $192 \cdot 64^3$ & 259 & 479 & 4.18 & $14_{2}^{2}, 17_{4}^{2}, 21_{6}^{2}, 24_{8}^{2}$ & 998  \\ 
E300     &      &       & tr\,M     & o & $192 \cdot 96^3$ & 176 & 496 & 4.26 & $14_{3}^{2}, 17_{3}^{2}, 21_{6}^{2}, 24_{6}^{2}$ & 1038 \\ 
N304     &      &       & ms        & o & $128 \cdot 48^3$ & 353 & 558 & 4.27 & $14_{2}^{2}, 17_{2}^{2}, 21_{3}^{2}, 24_{4}^{2}$ & 1652 \\ 
J304     &      &       & ms        & o & $192 \cdot 64^3$ & 261 & 527 & 4.21 & $14_{3}^{2}, 17_{3}^{2}, 21_{3}^{2}, 24_{4}^{2}$ & 1630 \\ 
\cmidrule{1-11}
J500     & 3.85 & 0.039 & tr\,M/sym & o & $192 \cdot 64^3$ & 413 & 413 & 5.24 & $17_{1}^{}, 22_{2}^{}, 27_{3}^{}, 32_{4}^{}$ & 1837 \\ 
J501     &      &       & tr\,M     & o & $192 \cdot 64^3$ & 336 & 448 & 4.26 & $17_{1}^{2}, 22_{2}^{2}, 27_{3}^{2}, 32_{4}^{2}$ & 1018 

  \end{tabular}
  \end{ruledtabular}
\end{table*}

\subsection{Correlation functions\label{sec:analysis}}

The baryon octet charges are extracted from two- and three-point
correlation functions of the form
\begin{align}
  &C_{\text{2pt}}^{B}(t) = \mathcal{P}_+^{\alpha \beta} 
  \sum_{\mathbf{x}'} 
  \langle \mathcal{B}_\alpha(\mathbf{x}',t)\; \bar{\mathcal{B}}_\beta(\mathbf{0},0) \rangle, \label{eq:2pt} \\
  &C_{\text{3pt}}^{B}(t, \tau) =  \mathcal{P}^{\alpha \beta} 
  \sum_{\mathbf{x}',\mathbf{y}} \langle \mathcal{B}_\alpha(\mathbf{x}',t) \mathcal{O}_J(\mathbf{y},\tau) \bar{\mathcal{B}}_\beta(\mathbf{0},0) \rangle.\label{eq:3pt}
\end{align}
Spin-1/2 baryon states are created~(annihilated) using suitable
interpolators $\overline{\mathcal{B}}$~($\mathcal{B}$): for the
nucleon, $\Sigma$ and $\Xi$, we employ interpolators corresponding to
the proton, $\Sigma^+$ and $\Xi^0$, respectively,
\begin{align}
  N_\alpha &= \epsilon^{ijk} u^i_\alpha \left( u^{jT} C \gamma_5 d^k \right), \\
  \Sigma_\alpha &= \epsilon^{ijk} u^i_\alpha \left( s^{jT} C \gamma_5 u^k \right), \\
  \Xi_\alpha &= \epsilon^{ijk} s^i_\alpha \left( s^{jT} C \gamma_5 u^k \right),
\end{align}
with spin index $\alpha$, colour indices $i$, $j$, $k$ and $C$ being
the charge conjugation matrix. The $\Lambda$ baryon is not considered
here since three-point functions with the currents
$\mathcal{O}_J=\bar{u}\Gamma_J u-\bar{d}\Gamma_J d$ vanish in this
case.  Without loss of generality, we place the source space-time
position at the origin \mbox{$(\mathbf{0}, 0)$} and the sink at
$(\mathbf{x}', t)$ such that the source-sink separation in time
equals~$t$. The current is inserted at $(\mathbf{y}, \tau)$ with $0
\leq \tau \leq t$.\footnote{Note that in practice we only analyse data
with $2a\leq\tau\leq t-2a$.} The annihilation interpolators are
projected onto zero-momentum via the sums over the spatial sink
position, while momentum conservation (and the sum over $\mathbf{y}$
for the current) means the source is also at rest.
\begin{figure}[!t]
  \includegraphics[width=0.48\textwidth]{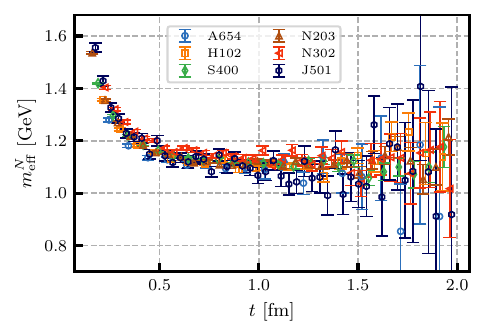}\\
  \includegraphics[width=0.48\textwidth]{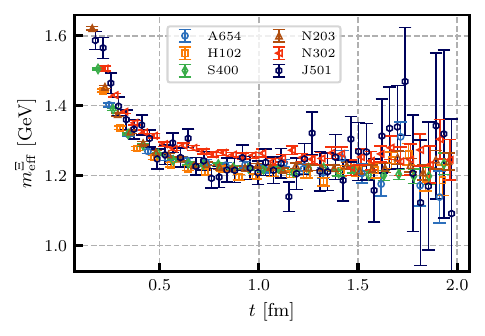}
  \caption{Effective masses $m_{\text{eff}}^B(t)=
    \ln[C_{\text{2pt}}^B(t-\tfrac{a}{2})/C_{\text{2pt}}^B(t+\tfrac{a}{2})]/a$ of the
    nucleon~(top) and $\Xi$~(bottom) determined on ensembles with
    $M_\pi\approx 340\,\MeV$ and $M_K\approx 450\,\MeV$ and lattice
    spacings ranging from $a= 0.098\,\fm$~(ensemble A654) down to $a=
    0.039\,\fm$~(ensemble J501). The errors of $m^B_{\text{eff}}(t)$
    are expected to increase in proportion to $a^{-1}$ but they also
    vary, e.g., with the number of effectively independent ensembles
    analysed.
    \label{fig:effmass}}
\end{figure}

\begin{figure}[!t]
  \includegraphics[width=0.48\textwidth]{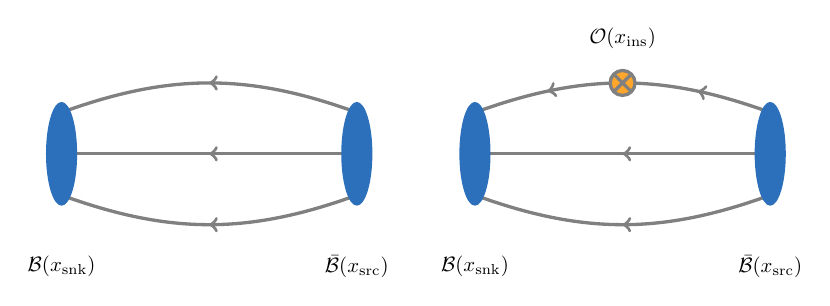}
  \caption{Quark-line diagrams of the two-point (left) and connected
    three-point (right) correlation functions where
    $x_{\text{src}}=(\mathbf{0},0)$,
    $x_{\text{ins}}=(\mathbf{y},\tau)$ and
    $x_{\text{snk}}=(\mathbf{x}',t)$.\label{fig:c3pt}}
\end{figure}

We ensure positive parity via the projection operator $\mathcal{P}_+ =
\frac{1}{2} (\mathds{1} + \gamma_4)$. For the three-point functions,
$\mathcal{P}=\mathcal{P}_+$ for $J=V,S$ and $\mathcal{P} = i\gamma_i
\gamma_5 \mathcal{P}_+$ for $J=A,T$. The latter corresponds to taking
the difference of the polarizations~(in the $i$~direction).  The
interpolators are constructed from spatially extended quark fields in
order to increase the overlap with the ground state of interest and
minimize contributions to the correlation functions from excited
states. Wuppertal smearing is employed~\cite{Gusken:1989ad}, together
with APE-smeared~\cite{Falcioni:1984ei} gauge transporters. The number
of smearing iterations is varied with the aim of ensuring that ground
state dominance sets in for moderate time separations. The root mean
squared light quark smearing radii range from about $0.6\,\fm$ (for
$M_\pi\approx 430\,\MeV$) up to about $0.8\,\fm$ (for $M_\pi\approx
130\,\MeV$), whereas the strange quark radii range from
about $0.5\,\fm$ (for the physical strange quark mass)
up to about $0.7\,\fm$ (for $M_K=M_{\pi}\approx 270\,\MeV$).
See Sec.~E.1 (and in particular Table~15) of
Ref.~\cite{RQCD:2022xux} for further details. Figure~\ref{fig:effmass}
demonstrates that, when keeping for similar pion and kaon masses the
smearing radii fixed in physical units, the ground state dominates the
two-point functions at similar physical times across different lattice
spacings.

Performing the Wick contractions for the two- and three-point
correlation functions leads to the connected quark-line diagrams
displayed in Fig.~\ref{fig:c3pt}. Note that there are no disconnected
quark-line diagrams for the three-point functions as these cancel when
forming the isovector flavour combination of the current. The
two-point functions are constructed in the standard way using
point-to-all propagators. For the three-point functions either a
stochastic approach~(described in the next subsection) or the
sequential source method~\cite{Maiani:1987by} (on some ensembles in
combination with the coherent sink technique~\cite{LHPC:2010jcs}) is
employed. The stochastic approach provides a computationally cost
efficient way of evaluating the three-point functions for the whole of
the baryon octet, however, additional noise is introduced. The
relevant measurements for the nucleon~(which has the worst signal-to-noise
ratio of the octet) have already been performed with the
sequential source method as part of other projects by our group, see,
e.g., Ref.~\cite{RQCD:2019jai}. We use these data in our analysis and
the stochastic approach for the correlation functions of the $\Sigma$
and the $\Xi$~baryons. Note that along the symmetric line
($m_\ell=m_s$) the hyperon three-point functions can be obtained as
linear combinations of the contractions carried out for the currents
$\bar{u}\Gamma_Ju$ and $\bar{d}\Gamma_Jd$ within the
proton. Therefore, no stochastic three-point functions are generated
in these cases.

We typically realize four source-sink separations with~$t/\fm \approx
\{0.7, 0.8, 1.0, 1.2\}$ in order to investigate excited state
contamination and reliably extract the ground state baryon octet
charges. Details of our fitting analysis are presented in
Sec.~\ref{sec:excited}. Multiple measurements are performed per
configuration, in particular for the larger source-sink separations to
improve the signal, see Table~\ref{tab:cls_ensembles}. The source
positions are chosen randomly on each configuration in order to reduce
autocorrelations. On ensembles with open boundary conditions in time
only the spatial positions are varied and the source and sink time
slices are restricted to the bulk of the lattice~(sufficiently away
from the boundaries), where translational symmetry is effectively
restored.

\subsection{Stochastic three-point correlation functions\label{sec:stoch3pts}}

\begin{figure}
  \begin{center}
    \includegraphics[width=0.48\textwidth]{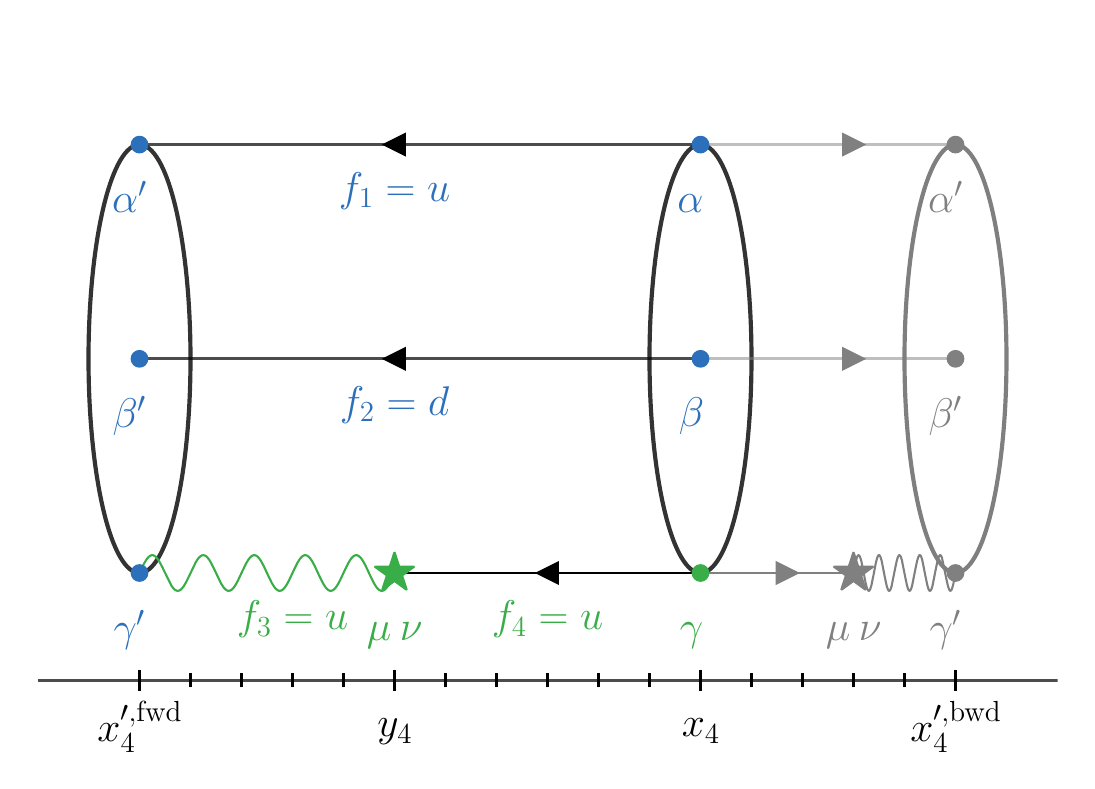}
    \caption{Schematic representation of the forward~(left) and
      backward~(right, shown in grey) propagating three-point
      correlation functions with open spin indices that are computed
      simultaneously with the stochastic approach. The indices
      corresponding to the spectator and insertion part from the
      factorization in Eq.~\eqref{eq:stoch3pt} are colour-coded in
      blue and green, respectively. The black solid lines correspond
      to the standard point-to-all propagators, whereas the green
      wiggly lines represent stochastic timeslice-to-all propagators. The
      temporal positions of the forward/backward sink, insertion and
      source are labelled as $x_4^{\prime,\text{fwd}|\text{bwd}}$,
      $y_4$ and $x_4$, respectively.  The flavour indices are chosen
      corresponding to a nucleon three-point function with a
      $\bar{u}\Gamma_J u$-current
      insertion. \label{fig:stoch3pt_diagram}}
  \end{center}
\end{figure}

\begin{figure}
  \begin{center}
    \includegraphics[width=0.45\textwidth]{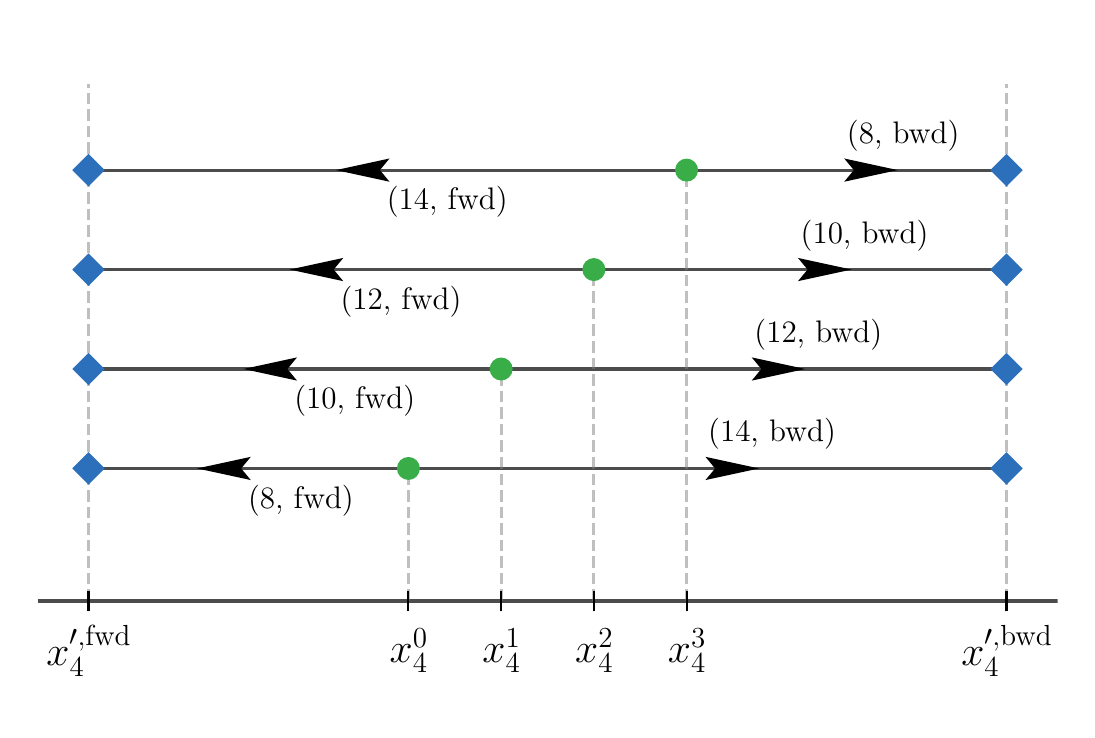}
    \caption{Sketch of the source and sink positions of the
      three-point functions realized using the stochastic
      approach. Blue diamonds depict the position of the forward
      ($x^{\prime,\text{fwd}}_x$) and backward
      ($x^{\prime,\text{bwd}}_x$) sink timeslices. Green points
      correspond to the source timeslices $x_4^k$ for $k =
      0,1,2,3$. Each three-point function measurement is labelled by
      the source-sink separation, where the values given correspond to
      the set-up for the ensembles at $\beta = 3.40$.
      \label{fig:src_snk_setup}}
  \end{center}
\end{figure}

In the following, we describe the construction of the connected
three-point correlation functions using a computationally efficient
stochastic approach. This method was introduced for computing meson
three-point functions in Ref.~\cite{Evans:2010tg} and utilized for
baryons in Refs.~\cite{Bali:2013gxx,Bali:2019svt} and also for mesons
in Refs.~\cite{Bali:2017mft,Loffler:2021afv}. Similar stochastic
approaches have been implemented by other groups, see, e.g.,
Refs.~\cite{Alexandrou:2013xon,Yang:2015zja}.

In Fig.~\ref{fig:c3pt} the quark-line diagram for the three-point
function contains an all-to-all quark propagator which connects the
current insertion at time $\tau$ with the baryon sink at time $t$. The
all-to-all propagator is too computationally expensive to evaluate
exactly. One commonly used approach avoids directly calculating the
propagator by constructing a sequential source~\cite{Maiani:1987by}
which depends on the baryon sink interpolator~(including its temporal
position and momentum). This method has the disadvantage that one
needs to compute a new quark propagator for each source-sink
separation, sink momentum and baryon sink interpolator. Alternatively,
one can estimate the all-to-all propagator stochastically. This
introduces additional noise on top of the gauge noise, however, the
quark-line diagram can be computed in a very efficient way. The
stochastic approach allows one to factorize the three-point
correlation function into a ``spectator'' and an ``insertion'' part
which can be computed and stored independently with all spin indices
and one colour index open. This is illustrated in
Fig.~\ref{fig:stoch3pt_diagram} and explained in more detail below. The
two parts can be contracted at a later (post-processing) stage with
the appropriate spin and polarization matrices, such that arbitrary
baryonic interpolators can be realized, making this method ideal for
SU(3) flavour symmetry studies. Furthermore, no additional inversions
are needed for different sink momenta.

\begin{figure*} 
  \begin{center}
    \includegraphics[width=0.8\textwidth]{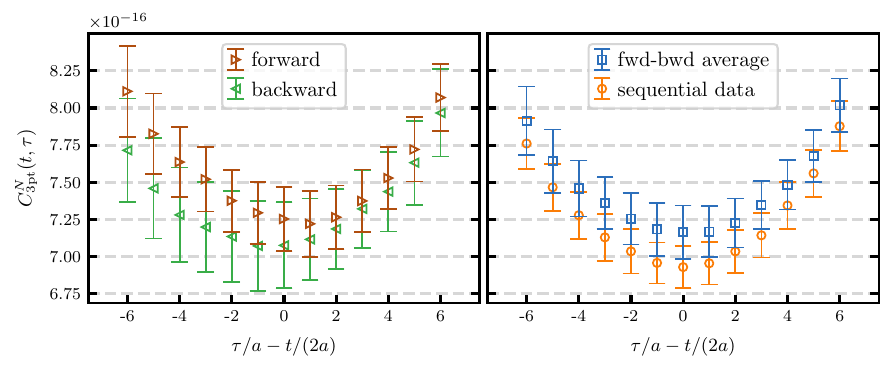}
    \caption{Left: polarized nucleon three-point correlation function
      propagating in the forward and backward directions for a
      $\bar{u} \gamma_y\gamma_z u$ current insertion obtained from one
      stochastic measurement on ensemble N200 ($a=0.064\,\fm$). Note
      that two different source positions were needed to obtain the
      same source-sink separation $t=14a$ in the two directions, see
      Fig.~\ref{fig:src_snk_setup}. Right: forward-backward average of
      the stochastic three-point function shown on the left compared
      to that obtained using the sequential source method~(with one
      source position, propagating in the forward
      direction).\label{fig:stoch_vs_seq_c3pts}}
  \end{center}
\end{figure*}

As depicted in Fig.~\ref{fig:src_snk_setup}, we simultaneously compute
the three-point functions for a baryon propagating (forwards) from
source timeslice $x_4$ to sink timeslice $x_4^{\prime, \text{fwd}}$
and propagating (backwards) from $x_4$ to $x_4^{\prime,
  \text{bwd}}$. We start with the definition of the stochastic source
and solution vectors which can be used to construct the timeslice-to-all
propagator~(shown as a green wiggly line in
Fig.~\ref{fig:stoch3pt_diagram}). In the following $i \in \{1, \dots,
N_{\text{sto}}\}$ is the ``stochastic index'', we denote spin indices
with Greek letters, colour indices with Latin letters (other than $f$
or $i$) and we use flavour indices $f_n \in \{u, d, s\}$. We introduce
(time partitioned) complex $\mathbb{Z}_2$ noise
vectors~\cite{Dong:1993pk,Bernardson:1993he}
 \begin{align}
  \eta_i(x)^\alpha_a = \left\{\begin{array}{ccl}
    \left(\mathbb{Z}_2 \otimes \mathrm{i} \mathbb{Z}_2\right) / \sqrt{2}&\text{if}&x_4= x_4^{\prime, \text{fwd}|\text{bwd}}\\
    \\
    0&&\text{otherwise} \ , \end{array}\right.,
 \end{align}
 where the noise vector has support on timeslices $x_4^{\prime,
   \text{fwd}}$ and $x_4^{\prime, \text{bwd}}$.  The noise vectors
 have the properties
\begin{align}
  \frac{1}{N_{\text{sto}}} \sum_{i=1}^{N_{\text{sto}}} \eta_i(x)^\alpha_a &= \mathcal{O}\left( \frac{1}{\sqrt{N_{\text{sto}}}}\right), \label{eq:st1}\\
  \frac{1}{N_{\text{sto}}} \sum_{i=1}^{N_{\text{sto}}} \eta_i(x)^\alpha_a  \eta^*_i(y)^\beta_b &=
  \delta_{xy}\delta_{\alpha\beta}\delta_{ab} + 
  \mathcal{O}\left( \frac{1}{\sqrt{N_{\text{sto}}}}\right).\label{eq:st2}
\end{align}
The solution vectors~$s_{f,i}(y)$ are defined through the linear system
\begin{align}
  D_f(x,y)^{\alpha\beta}_{ab} s_{f,i}(y)^\beta_b = \eta_i(x)^\alpha_a ,
  \label{eq:linsys}
\end{align}
where we sum over repeated indices~(other than $f$) and
$D_f(x,y)^{\alpha\beta}_{ab}$ is the Wilson-Dirac operator for the quark flavour~$f$.
Note that $s_{u,i}=s_{d,i}$ since our light quarks are mass-degenerate.

Using $\gamma_5$-Hermiticity ($\gamma_5D_f\gamma_5=D_f^\dagger$)
and the properties given in Eqs.~\eqref{eq:st1}
and~\eqref{eq:st2}, the timeslice-to-all propagator connecting
all points of the (forward and backward) sink timeslices~$x_4 ^\prime$
to all points of any insertion timeslice~$y_4$  can be estimated as
\begin{align}
  G_{f_3} (x^\prime, y)^{\gamma^\prime \mu}_{c^\prime d} \approx 
  \frac{1}{N_{\text{sto}}} \sum_{i=1}^{N_{\text{sto}}}
  \left[\gamma_5\eta_i(x^\prime)\right]^{\gamma^\prime}_{c^\prime}
  \left[ \gamma_5 s^*_{f_3}(y) \right]^{\mu}_d.
\end{align}

Combining this timeslice-to-all propagator with point-to-all
propagators for the source position $x_4$, the baryonic three-point
correlation functions Eq.~\eqref{eq:3pt} can be factorized as
visualized in Fig.~\ref{fig:stoch3pt_diagram} into a spectator part
(S) and an insertion part (I), leaving all flavour and spin indices
open:
\begin{align}
  C_{\text{3pt}} &(
    \mathbf{p}^\prime, \mathbf{q}| x_4^\prime, y_4, x_4
    )^{
      \alpha^\prime \alpha \beta^\prime \beta \gamma^\prime \mu \nu \gamma 
    }_{
    f_1 f_2 f_3 f_4
    } \nonumber \\ 
    &\approx \frac{1}{N_{\text{sto}}} \sum_{i=1}^{N_{\text{sto}}} \sum_{c=1}^{3}   
    \Big( 
    S_{f_1 f_2} (\mathbf{p}^\prime, x_4^\prime, x_4)^{\alpha^\prime \alpha \beta^\prime \beta \gamma^\prime}_{i c}  \nonumber \\ 
    &\qquad\qquad\qquad\qquad\qquad\times I_{f_3 f_4} (\mathbf{q}, y_4, x_4)^{\mu \nu \gamma }_{i c}
    \Big). \label{eq:stoch3pt}
\end{align}
The spectator and insertion parts are defined as
\begin{align}
  S_{f_1 f_2} &(\mathbf{p}^\prime, x_4^\prime, x_4)^{\alpha^\prime \alpha \beta^\prime \beta \gamma^\prime}_{i c} \nonumber \\ 
  \coloneqq & \sum_{\mathbf{x}^\prime} \Big( \epsilon_{a^\prime b^\prime c^\prime} \epsilon_{a b c}
  G_{f_1} (x^\prime, x)^{\alpha^\prime \alpha}_{a^\prime a}
  G_{f_2} (x^\prime, x)^{\beta^\prime \beta}_{b^\prime b} \nonumber \\ 
  &\qquad\qquad\qquad\qquad \times\big[ \gamma_5\eta_{i}(x^\prime) \big]^{\gamma^\prime}_{c^\prime}
  \; e^{-i \mathbf{p}^\prime \cdot (\mathbf{x}^\prime - \mathbf{x})} \Big), \label{eq:spe}\\
  I_{f_3 f_4} &(\mathbf{q}, y_4, x_4)^{\mu \nu \gamma }_{i c} \nonumber \\ 
  &\coloneqq \sum_{\textbf{y}} \left[\gamma_5 s^{*}_{f_3, i}(y)\right]^{\mu}_{d}
  G_{f_4}(y,x)^{\nu\gamma}_{dc} \; e^{+i \mathbf{q} \cdot (\mathbf{y} - \mathbf{x})}. \label{eq:ins}
\end{align}
Using these building blocks, three-point functions for given baryon
interpolators and currents for any momentum combination can be
constructed. Note that in this article we restrict ourselves to the
case $\mathbf{q}=\mathbf{p}'={\bf 0}$. The point-to-all propagators
within the spectator part are smeared at the source and at the sink,
whereas $G_{f_4}$ is only smeared at the source. The stochastic source
is smeared too, however, this is carried out after solving
Eq.~\eqref{eq:linsys}. In principle, the spectator part also depends
on $f_3$ because for $f_3=s$ and $f_3\in\{u, d\}$ different smearing
parameters are used. We ignore the dependence of the spectator part on
$f_3$ since here we restrict ourselves to $f_3,f_4\in\{u,d\}$. For
details on the smearing see the previous subsection.  Using the same set
of timeslice-to-all propagators, we compute point-to-all propagators for a
number of different source positions at timeslices $x_4$ in-between
$x_4^{\prime,\text{bwd}}$ and $x_4^{\prime,\text{fwd}}$ which allows
us to vary the source-sink distances, see
Figs.~\ref{fig:stoch3pt_diagram} and~\ref{fig:src_snk_setup}.

The number of stochastic estimates $N_{\text{sto}}$ is chosen by
balancing the computational cost against the size of the stochastic
noise introduced. We find that for $N_{\text{sto}} \gtrsim 100$ the
stochastic noise becomes relatively small compared to the gauge noise
and we employ 100 estimates across all the ensembles. In some channels
the signal obtained for the three-point function, after averaging over
the forward and backward directions, is comparable to that obtained
from the traditional sequential source method~(for a single source,
computed in the forward direction), as shown in
Fig.~\ref{fig:stoch_vs_seq_c3pts}. Nonetheless, when taking the ratio
of the three-point function with the two-point function for the
fitting analysis, discussed in the next subsection, a significant part
of the gauge noise cancels, while the stochastic noise remains. This
results in larger statistical errors in the ratio for the stochastic
approach. This is a particular problem in the vector channel. A more
detailed comparison of the two methods is given in
Appendix~\ref{sec:details3pt}.

As mentioned above, only flavour conserving currents and zero momentum
transfer are considered, however, the data to construct three-point
functions with flavour changing currents containing up to one
derivative for various different momenta is also available, enabling
an extensive investigation of (generalized) form factors in the
future. Similarly, meson three-point functions can be constructed by
computing the relatively inexpensive meson spectator part and
(re-)using the insertion part, see Ref.~\cite{Loffler:2021afv} for
first results.

\subsection{Fitting and excited state analysis\label{sec:excited}}

The spectral decompositions of the two- and three-point correlation
functions read
\begin{align}
  &C_{\text{2pt}}^{B}(t) = \sum_{n} |Z^B_n|^2 e^{-E_{n}^{B}
    t}\ , \label{eq:spect2pt}\\ &C_{\text{3pt}}^{B}(t,\tau;\mathcal{O}_J) =
  \sum_{n,m} Z^B_n Z^{B*}_m \langle n | \mathcal{O}_J | m \rangle
  e^{-E_{n}^{B} (t-\tau)} e^{E_{m}^{B} \tau},\label{eq:spect3pt}
\end{align}
where $E_{n}^{B}$ is the energy of state $|n\rangle$ ($n=0,1,\ldots$),
created when applying the baryon interpolator $\bar{\mathcal{B}}$ to
the vacuum state $|\Omega\rangle$ and $Z_n^B$ is the associated overlap
factor \mbox{$Z_n^B \propto \langle n | \bar{\mathcal{B}}| \Omega
  \rangle$}. The ground state matrix elements of interest $\langle 0 |
\mathcal{O}_J | 0 \rangle=g_J^{B,\latt}$ can be obtained in the limit
of large time separations from the ratio of the three-point and
two-point functions
\begin{align}
  R^B_J(t,\tau) &=
  \frac{C^B_{\text{3pt}}(t,\tau;\mathcal{O}_J)}{C^B_{\text{2pt}}(t)}\stackrel{t,\tau\rightarrow
    \infty}{\longrightarrow}g_J^{B,\latt}. \label{eq:rat}
\end{align}
However, the signal-to-noise ratio of the correlation functions
deteriorates exponentially with the time separation and with current
techniques it is not possible to achieve a reasonable signal for
separations that are large enough to ensure ground state dominance. At
moderate $t$ and $\tau$, one observes significant excited state
contributions to the ratio. All states with the same quantum numbers
as the baryon interpolator contribute to the sums in
Eqs.~\eqref{eq:spect2pt} and~\eqref{eq:spect3pt}, including
multi-particle excitations such as $B\pi$ P-wave and $B\pi\pi$ S-wave
scattering states. The spectrum of states becomes increasingly dense
as one decreases the pion mass while keeping the spatial extent of the
lattice sufficiently large, where the lowest lying excitations are
multi-particle states.

\begin{table}
  \caption{Summary of the fits performed. We vary the combinations of
    channels~$J$ that are fitted simultaneously as well as the number
    of excited states (ES) included in the fit and the fit interval
    $\tau \in [\delta t, t - \delta t]$ with $\delta t\in\{\delta t_1,
    \delta t_2\}$, where $\delta t_1\approx 0.15\,\fm$, $\delta
    t_2\approx 0.25\,\fm$.  The last two columns indicate which
    parameters in Eq.~\eqref{eq:ratfit} are constrained by a prior or set to
    zero. All other parameters are determined in the
    fit.\label{tab:rat_fits}}
  \begin{ruledtabular}
  \begin{tabular}{llcccl}
  Fit & $J$ & $\delta t$ & ES & prior & set to zero\\
  \cmidrule{1-6}
   1  & $A,S,T$   &  $\delta t_1$  & 1 & -            & $b_{2}^J$, $b_{3}^J$, $b_{4}^J$, $\Delta E_2$ \\ 
   2  & $A,S,T$   &  $\delta t_2$  & 1 & -            & $b_{2}^J$, $b_{3}^J$, $b_{4}^J$, $\Delta E_2$ \\  
   3  & $A,S,T,V$ &  $\delta t_1$  & 1 & -            & $b_{2}^J$, $b_{3}^J$, $b_{4}^J$, $\Delta E_2$ \\ 
   4  & $A,S,T,V$ &  $\delta t_2$  & 1 & -            & $b_{2}^J$, $b_{3}^J$, $b_{4}^J$, $\Delta E_2$ \\ 
   5  & $A,S,T$   &  $\delta t_1$  & 2 & $\Delta E_1$ & $b_{1}^{T}$, $b_{2}^{J}$, $b_{4}^J$ \\ 
   6  & $A,S,T$   &  $\delta t_2$  & 2 & $\Delta E_1$ & $b_{1}^{T}$, $b_{2}^{J}$, $b_{4}^J$ \\  
   7  & $A,S,T,V$ &  $\delta t_1$  & 2 & $\Delta E_1$ & $b_{1}^{T,V}$, $b_{2}^{J}$, $b_{4}^J$ \\ 
   8  & $A,S,T,V$ &  $\delta t_2$  & 2 & $\Delta E_1$ & $b_{1}^{T,V}$, $b_{2}^{J}$, $b_{4}^J$ \\ 
   9  & $A,S,T$   &  $\delta t_1$  & 2 & $\Delta E_1$ & $b_{2}^{J}$, $b_{4}^J$ \\ 
   10 & $A,S,T$   &  $\delta t_2$  & 2 & $\Delta E_1$ & $b_{2}^{J}$, $b_{4}^J$ \\  
   11 & $A,S,T,V$ &  $\delta t_1$  & 2 & $\Delta E_1$ & $b_{2}^{J}$, $b_{4}^J$ \\ 
   12 & $A,S,T,V$ &  $\delta t_2$  & 2 & $\Delta E_1$ & $b_{2}^{J}$, $b_{4}^J$ \\ 
   13 & $A,S,T$   &  $\delta t_1$  & 2 & $\Delta E_2$ & $b_{1}^{T}$, $b_{2}^{J}$, $b_{4}^J$ \\ 
   14 & $A,S,T$   &  $\delta t_2$  & 2 & $\Delta E_2$ & $b_{1}^{T}$, $b_{2}^{J}$, $b_{4}^J$ \\  
   15 & $A,S,T,V$ &  $\delta t_1$  & 2 & $\Delta E_2$ & $b_{1}^{T,V}$, $b_{2}^{J}$, $b_{4}^J$ \\ 
   16 & $A,S,T,V$ &  $\delta t_2$  & 2 & $\Delta E_2$ & $b_{1}^{T,V}$, $b_{2}^{J}$, $b_{4}^J$ \\ 
   17 & $A,S,T$   &  $\delta t_1$  & 2 & $\Delta E_2$ & $b_{2}^{J}$, $b_{4}^J$ \\ 
   18 & $A,S,T$   &  $\delta t_2$  & 2 & $\Delta E_2$ & $b_{2}^{J}$, $b_{4}^J$ \\  
   19 & $A,S,T,V$ &  $\delta t_1$  & 2 & $\Delta E_2$ & $b_{2}^{J}$, $b_{4}^J$ \\ 
   20 & $A,S,T,V$ &  $\delta t_2$  & 2 & $\Delta E_2$ & $b_{2}^{J}$, $b_{4}^J$
  \end{tabular}
  \end{ruledtabular}
\end{table}

\begin{figure*}[tb]
  \centering
  \includegraphics[width=0.75\textwidth]{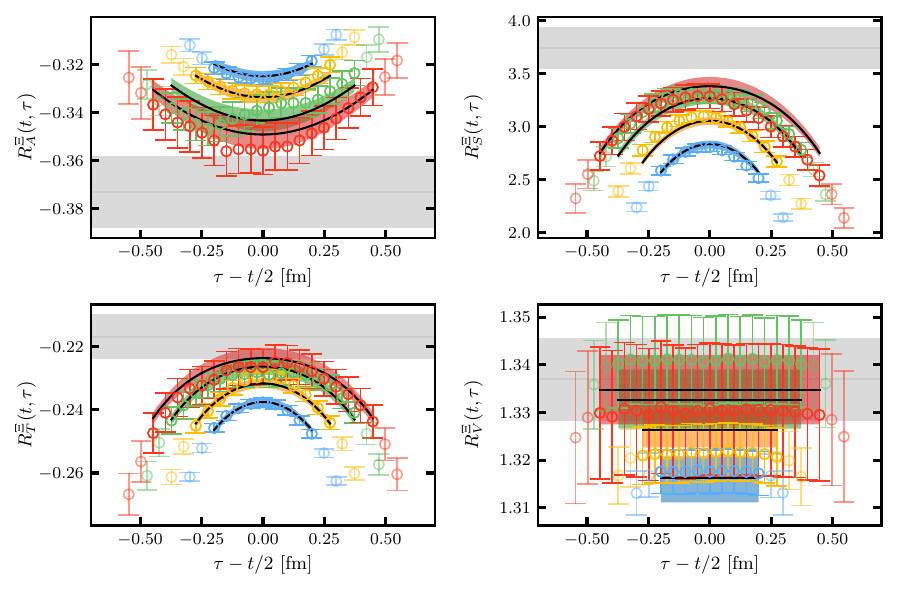}
  \caption{Unrenormalized ratios~$R_J^\Xi(t,\tau)$, $J\in \{A, S, T,
    V\}$~(defined in Eq.~\eqref{eq:rat}) for the cascade baryon on
    ensemble N302~($M_\pi=348\,\MeV$ and $a=0.049\,\fm$), where $t
    \approx \{0.7, 0.8,1.0, 1.2\}\, \fm$. The grey horizontal lines
    and bands show the results for the ground state matrix elements
    $\langle 0 | \bar{u} \Gamma_J u - \bar{d} \Gamma_J d| 0
    \rangle=g_J^{\Xi,\latt}$, obtained from a simultaneous fit to the
    ratios for all channels and source-sink separations using
    parametrization 7 (see Eq.~\eqref{eq:ratfit} and
    Table~\ref{tab:rat_fits}). The data points with $\tau \in
    \left[\delta t, t-\delta t\right]$, where $\delta t = 2a$, are
    included in the fit~(the faded data points are omitted), which is
    the maximum fit range possible for our action.  The coloured
    curves show the expectation from the fit for each source-sink
    separation.\label{fig:ratio_fit}}
\end{figure*}
\begin{figure*}[bt]
  \centering
  \includegraphics[width=\textwidth]{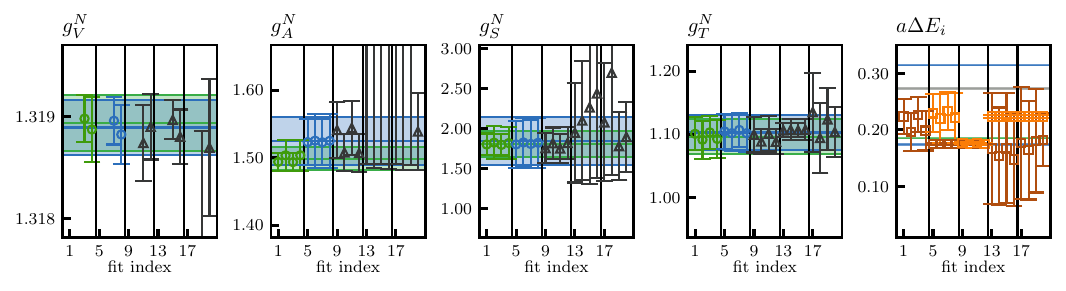}\\
  \includegraphics[width=\textwidth]{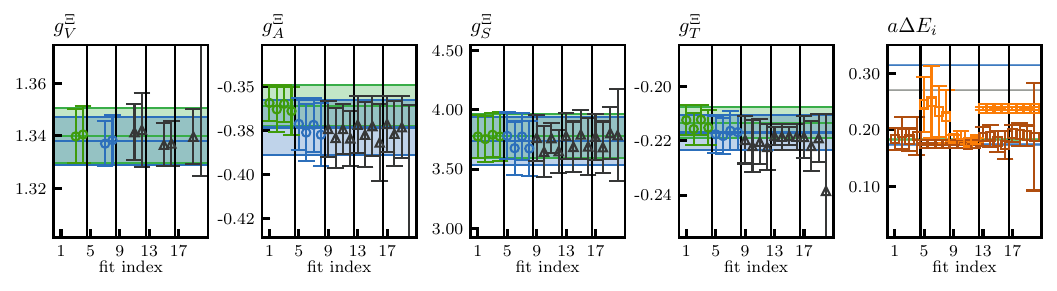}
  \caption{Results for the four unrenormalized charges of the
    nucleon~(top) and cascade baryon~(bottom) obtained from the fits
    listed in Table~\ref{tab:rat_fits} for ensemble
    N302~($M_\pi=348\,\MeV$ and $a=0.049\,\fm$). The green~(blue)
    horizontal lines and bands indicate the final results and errors
    obtained from the median and 68\% confidence level interval of the
    combined bootstrap distributions determined from the fits
    indicated by the green~(blue) data points which include one~(two)
    excited state(s). On the right the energy gaps determined in the
    fits and those corresponding to the lowest lying multi-particle
    states are displayed using the same colour coding as in
    Fig.~\ref{fig:ex_spec}. \label{fig:fits_nucleon}}
\end{figure*}

One possible strategy is to first determine the energies of the ground
state and lowest lying excitations by fitting to the two-point
function (which is statistically more precise than the three-point
function) with a suitable functional form. The energies can then be
used in a fit to the three-point function~(or the ratio $R_J^B$) to
extract the charge $g_J^B$.\footnote{Given the precision of the
two-point function relative to that of the three-point function, this
strategy is very similar to fitting $C_{\text{2pt}}^{B}$ and
$C_{\text{3pt}}^{B}$ simultaneously.} However, the three-quark baryon
interpolators we use by design have only a small overlap with the
multi-particle states containing five or more quarks and antiquarks
and it is difficult to extract the lower lying excited state spectrum
from the two-point function. Nonetheless, multi-particle states can
significantly contribute to the three-point function if the transition
matrix elements $\langle n|{\cal O}_J| 0\rangle$ are
large. Furthermore, depending on the current, different matrix
elements, and hence excited state contributions, will dominate. In
particular, one would expect the axial and scalar currents to couple
to the $B\pi$ P-wave and $B\pi\pi$ S-wave states, respectively, while
the tensor and vector currents may enhance transitions between $B$ and
$B\pi\pi$ states when $\pi\pi$ is in a P-wave.

The summation method~\cite{Maiani:1987by} is an alternative approach,
which involves summing the ratio over the operator insertion time
$S_J^B(t) = \sum_{\tau = \tau_0 }^{t-\tau_0} R^B_J(t,\tau)$, where one
can show that the leading excited state contributions to $S_J^B(t)$
only depend on $t$~(rather than also on $t-\tau$ and $\tau$ as for
$R^B_J(t,\tau)$). However, one needs a large number of source-sink
separations~(more than the four values of $t$ that are realized in this
study) in order to extract reliable results from this approach.

These considerations motivate us to extract the charges by fitting to
the ratio of correlation functions using a fit form which takes into
account contributions from up to two excited states,
\begin{align}
  R^B_J &(t,\tau) = b_0^J \nonumber \\
    &+ b_1^J \left(e^{-\Delta E_1 (t-\tau)}+e^{-\Delta E_1 \tau} \right) + b_2^J e^{-\Delta E_1 t} \nonumber \\
    &+ b_3^J \left(e^{-\Delta E_2 (t-\tau)}+e^{-\Delta E_2 \tau} \right) + b_4^J e^{-\Delta E_2 t}.\label{eq:ratfit}
\end{align}
where $\Delta E_n = E_n^B - E_0^B$ denotes the energy gap between the
ground state and the $n^{th}$ excited state of baryon~$B$ and we have
not included transitions between the first and the second excited
state. The amplitude $b_0^J = g_J^{B,\latt}$ gives the charge, while
$b_{1,3}^J$ and $b_{2,4}^J$ are related to the ground state to excited
state and excited state to excited state transition matrix elements,
respectively. In practice, even when simultaneously fitting to all
available source-sink separations, it is difficult to determine the
energy gaps~(and amplitudes) for a particular channel~$J$. Similar to
the strategy pursued in Ref.~\cite{Harris:2019bih}, we simultaneously
fit to all four channels $J\in \{V, A, S, T\}$ for a given baryon. As
the same energy gaps are present, the overall number of fit parameters
is reduced and the fits are further constrained.

To ensure that the excited state contributions are sufficiently under
control, we carry out a variety of different fits, summarized in
Table~\ref{tab:rat_fits}. We vary
\begin{itemize}
\item the data sets included in the fit: simultaneous fits are
  performed to the data for $J\in\{A,S,T,V\}$ and $J\in\{A,S,T\}$. As
  the axial, scalar and tensor channels are the main focus of this
  study, we only consider excluding the vector channel data.
\item the parametrization: either one~(`ES=1') or two~(`ES=2') excited
  states are included in the fits. In the latter case, in order to
  stabilize the fit, we use a prior for $\Delta E_1$ corresponding to
  the energy gap for the lowest lying multi-particle state. As a
  cross-check we repeat these fits using the average result obtained for
  $\Delta E_2$ in fits 5--8 as a prior and leaving $\Delta E_1$ as a free
  parameter~(fits 13--20). The widths of the
  priors are set to $E_1/100$ and to $E_2/100$, respectively. In general, the
  contributions from excited state to excited state transitions could
  not be resolved and the parameters $b^J_{2,4}$ are set to zero. We
  also found that the tensor and vector currents couple more strongly
  to the second excited state, consistent with the expectations
  mentioned above, and the first excited state contributions are
  omitted for these channels in the `ES=2' fits. Furthermore, due to
  the large statistical error of the stochastic three-point functions
  for the $\Sigma$ and $\Xi$ baryons in the vector channel (see
  Fig.~\ref{fig:ratio_fit} and the discussion in
  Appendix~\ref{sec:details3pt}), we are not able to resolve
  $b_1^V$~(and analogously $b_3^V$). For these baryons we also set
  $b_{1,3}^V = 0$ in all the fits.
\item the fit range: two fit intervals $\tau \in [\delta t_j, t -
  \delta t_j]$ are used with $\delta t_1=n_1 a\approx 0.15\,\fm$ and
  $\delta t_2=n_2a \approx 0.25\,\fm$.\footnote{Due to $n_j\geq 2$ and its
quantization, $\delta t_1$ and $\delta t_2$ depend slightly
on the lattice spacing:
$\delta t_j\approx 0.20\,\fm, 0.29\,\fm$ ($\beta=3.34$),
$\delta t_j\approx 0.17\,\fm, 0.26\,\fm$ ($\beta=3.40$),
$\delta t_j\approx 0.15\,\fm, 0.23\,\fm$ ($\beta=3.46$),
$\delta t_j\approx 0.13\,\fm, 0.26\,\fm$ ($\beta=3.55$),
$\delta t_j\approx 0.15\,\fm, 0.25\,\fm$ ($\beta=3.70$),
$\delta t_j\approx 0.16\,\fm, 0.27\,\fm$ ($\beta=3.85$).
}
\end{itemize}
A typical fit to the ratios for the cascade baryon is shown in
Fig.~\ref{fig:ratio_fit} for ensemble N302~($M_\pi=348\,\MeV$ and
$a=0.049\,\fm$). The variation in the ground state matrix elements
extracted from the 20 different fits is shown in
Fig.~\ref{fig:fits_nucleon}, also for the nucleon on the same
ensemble.  See Appendix~\ref{sec:plots} for the analogous plot for the
$\Sigma$ baryon. Overall, the results are consistent within errors,
however, some trends in the results can be seen across the different
ensembles. In the axial channel, in particular the results for the
fits involving a single excited state~(fits 1--4), tend to be lower
than those involving two excited states~(fits 5--20). The former are,
in general, statistically more precise than the latter due to the
smaller number of parameters in the fit.

In order to study the systematics arising from any residual excited
state contamination in the final results at the physical point (in the
continuum limit at infinite volume), the extrapolations, detailed in
Sec.~\ref{sec:extrapol}, are performed for the results obtained from
fits 1--4~(`ES=1') and fits 5--8~(`ES=2'), separately. For each set of
fits, 500 samples are drawn from the combined bootstrap distributions
of the four fit variations. The final result and error, shown as the
green and blue bands in Fig.~\ref{fig:fits_nucleon}, correspond to the
median and the 68\% confidence interval, respectively.
Note that we take the same 500 bootstrap samples for all the baryons
to preserve correlations. The final results for all the ensembles are
listed in Tables~\ref{tab:data_nucleon}, ~\ref{tab:data_sigma}
and~\ref{tab:data_xi} of Appendix~\ref{sec:lattice_data} for the
nucleon, sigma and cascade baryons, respectively.

\begin{figure}[tb!]
  \centering
  \includegraphics[width=0.4\textwidth]{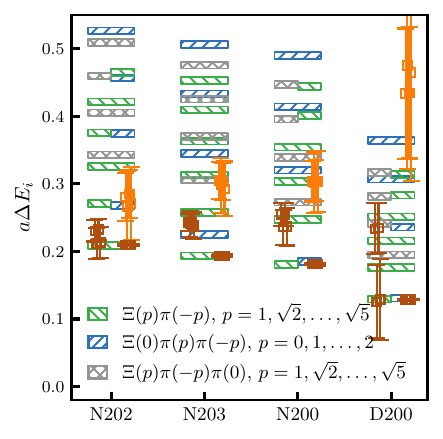}
  \caption{Results for the first and second excited state energy gaps
    of the cascade baryon, $\Delta E_1$~(brown data points) and
    $\Delta E_2$~(orange data points), respectively, determined on
    ensembles lying on the $\tr\,M = \const$ trajectory with
    $a=0.064\,\fm$. The pion mass decreases from left to right with
    $M_\pi=414$~MeV for ensemble N202 and $M_\pi=202$~MeV for ensemble
    D200, see Table~\ref{tab:cls_ensembles}. For each ensemble, the
    $\Delta E_1$ obtained using fits 1--4 of Table~\ref{tab:rat_fits}
    are shown on the left and the $\Delta E_1$~(fixed with a prior to
    the lowest multi-particle energy gap) and $\Delta E_2$ resulting
    from fits 5--8 are displayed on the right. For comparison, the
    energy gaps of the lower lying non-interacting multi-particle
    states with the quantum numbers of the cascade baryon are shown as
    horizontal lines, where the momenta utilized for each hadron are
    indicated in lattice units. \label{fig:ex_spec}}
\end{figure}

In terms of the energy gaps extracted, Fig.~\ref{fig:fits_nucleon}
shows that we find consistency across variations in the fit range and
whether the vector channel data is included or not. However, the first
excited energy gap $\Delta E_1$ obtained from the single excited state
fits tends to be higher than the lowest multi-particle level, in
particular, as the pion mass is decreased, suggesting that
contributions from higher excited states are significant. This can be
seen in Fig.~\ref{fig:ex_spec}, where we compare the results for the
energy gaps for the cascade baryon with the lower lying
non-interacting $\Xi\pi$ and $\Xi\pi\pi$ states for four ensembles
with $a=0.064\,\fm$ and pion masses ranging from $414\,\MeV$ down to
$202\,\MeV$. Note that the multi-particle levels are modified in a
finite volume, although the corresponding energy shifts may be small
for the large volumes realized here. There are a number of levels
within roughly $500\,\MeV$ of the first excited state. Some levels lie
close to each other and one would not expect that the difference can
be resolved by fits with one or two excited states. The $\Delta E_2$ energy
gaps from the two excited state fits (with the first excited state
fixed with a prior to the lowest multi-particle level) are consistent
with the next level that is significantly above the first excited
state, although for ensemble D200 the errors are too large to draw a
conclusion. Given that more than one excited state is contributing
significantly, we expect that the latter fits isolate the ground state
contribution more reliably. We remark that within present statistics,
two-exponential fits to the two-point functions alone give energy gaps
$a\Delta E=0.390(37), 0.371(34), 0.430(37)$ and $0.312(46)$ for
N202, N203, N200 and D200, respectively, that are all larger than
$a\Delta E_2$, with the exception of D200, where the
two gaps agree within errors.

\subsection{Non-perturbative renormalization and improvement\label{sec:renorm}} 

The isovector lattice charges, $g_J^{B,\latt}$, extracted in the
previous subsection need to be matched to the continuum
$\MS$ scheme. The renormalized matrix elements suffer
from discretization effects, however, the leading order effects are
reduced to $O(a^2)$ when implementing full $O(a)$ improvement. In the
forward limit, in addition to using a non-perturbatively $O(a)$
improved fermion action, this involves taking mass dependent terms
into account. The following multiplicative factors are applied,
\begin{align}
  g_J^B = Z^k_J \left(1 + a m_\ell b_J + 3 a \bar{m} \tilde{b}_J \right) g_J^{B,\latt} +O(a^2) \ , \label{eq:renorm}
\end{align}
for $J \in \{V, A, S, T\}$, where $Z_J$ are the renormalization
factors and $b_J$ and $\tilde{b}_J$ are the $O(a)$ improvement
coefficients. Note that the renormalization factors for the scalar and
tensor currents depend on the scale, $Z_{S,T}=Z_{S,T}(\mu)$, where we
take $\mu=2\,\GeV$. The vector Ward identity lattice quark mass~$am_q$
is obtained from the hopping parameter $\kappa_q$ ($q = \ell, s$) and
the critical hopping parameter $\kappa_{\text{crit}}$ via \mbox{$am_q =
  \left( 1/\kappa_q - 1/\kappa_{\text{crit}} \right)/2$}.  $\bar{m} =
(2m_\ell + m_s)/ 3$ denotes the flavour averaged quark mass. The
hopping parameters for all ensembles used within this work are
tabulated in Table~\ref{tab:data_lat} of
Appendix~\ref{sec:lattice_data}. For $\kappa_{\text{crit}}$ we utilize
the interpolation formula~\cite{RQCD:2022xux}
\begin{align}
  \frac{1}{\kappa_{\text{crit}}} = 
  8 - 0.402454 g^2 \frac{1 + 0.28955 g^2 - 0.1660 g^6}{1 + 0.22770 g^2 - 0.2540 g^4}.
\end{align}

The improvement coefficients $b_J$ and $\tilde{b}_J$ are determined
non-perturbatively in Ref.~\cite{Korcyl:2016cmx}. We make use of
updated preliminary values, which will appear in a future
publication~\cite{inprep2}. These are listed in
Tables~\ref{tab:improvement_b} and~\ref{tab:improvement_btilde},
respectively. Note that no estimates of $\tilde{b}_J$ are available
for $\beta=3.85$. Considering the size of the statistical errors, the
general reduction of the $|\tilde{b}_J|$ values with increasing
$\beta$ (and the decreasing $a$), at this lattice spacing we set
$\tilde{b}_J=0$ for all~$J$.

\begin{table}
  \caption{Improvement coefficients $b_J$ for $J \in \{A, S, T, V \}$ from Refs.~\cite{Korcyl:2016cmx,inprep2}.
    \label{tab:improvement_b}}
  \begin{ruledtabular}
  \begin{tabular}{lllll}
    $\beta$  &  $b_A$ &  $b_S$  &  $b_T$  &  $b_V$ \\ 
    \cmidrule{1-5}
    3.34 & 1.249(16)    & 1.622(47)    & 1.471(11)    & 1.456(11)    \\ 
    3.4  & 1.244(16)    & 1.583(62)    & 1.4155(48)   & 1.428(11)    \\ 
    3.46 & 1.239(15)    & 1.567(74)    & 1.367(12)    & 1.410(13)    \\ 
    3.55 & 1.232(15)    & 1.606(98)    & 1.283(14)    & 1.388(17)    \\ 
    3.7  & 1.221(13)    & 1.49(11)     & 1.125(15)    & 1.309(22)    \\ 
    3.85 & 1.211(12)    & 1.33(16)     & 0.977(38)    & 1.247(26)
  \end{tabular}
  \end{ruledtabular}
\end{table}
\begin{table}
  \caption{Improvement coefficients $\tilde{b}_J$ for $J \in \{A, S,
    T, V \}$ from Refs.~\cite{Korcyl:2016cmx,inprep2}. Note that no
    results are available for $\beta=3.85$.\label{tab:improvement_btilde}}
  \begin{ruledtabular}
  \begin{tabular}{lllll}
    $\beta$  &  $\tilde{b}_A$ &  $\tilde{b}_S$  &  $\tilde{b}_T$  &  $\tilde{b}_V$ \\
    \cmidrule{1-5}
    3.34 & -0.06(28)           & -0.24(55)           & \phantom{-}1.02(16) & \phantom{-}1.05(13) \\ 
    3.4  & -0.11(13)           & -0.36(23)           & \phantom{-}0.49(17) & \phantom{-}0.41(11) \\ 
    3.46 & \phantom{-}0.08(11) & -0.421(83)          & \phantom{-}0.115(19) & \phantom{-}0.158(28) \\ 
    3.55 & -0.03(13)           & -0.25(12)           & \phantom{-}0.000(37) & \phantom{-}0.069(42) \\ 
    3.7  & -0.047(75)          & -0.274(65)          & -0.0382(60)         & -0.031(18)
  \end{tabular}
  \end{ruledtabular}
\end{table}

For the renormalization factors, we employ the values obtained in
Ref.~\cite{RQCD:2020kuu}.  The factors are determined
non-perturbatively in the RI$^\prime$-SMOM
scheme~\cite{Martinelli:1994ty,Sturm:2009kb} and then~(for $Z_S$ and
$Z_T$) converted to the $\MS$ scheme using three-loop
matching~\cite{Kniehl:2020sgo,Bednyakov:2020ugu,Kniehl:2020nhw}. We
remark that the techniques for implementing the Rome-Southampton
method were extended in Ref.~\cite{RQCD:2020kuu} to ensembles with
open boundary conditions in time. This development enables us to
utilize ensembles with $a<0.06\,\fm$, where only open boundary
conditions in time are available due to the need to maintain
ergodicity.
\begin{table}
  \caption{Set of renormalization factors taken from
    Ref.~\cite{RQCD:2020kuu}, denoted as $Z_J^1$ in the text. The
    factors are determined using the RI$^\prime$-SMOM scheme and the
    `fixed-scale method' with the perturbative subtraction of lattice
    artefacts. For $Z_A$ and $Z_V$, the values correspond to those
    listed under $Z^\prime_A$ and $Z^\prime_V$, respectively, which
    are obtained using renormalization conditions consistent with the
    respective Ward identities. The statistical and systematic errors
    have been added in quadrature. \label{tab:ZtabXVI}}
  \begin{ruledtabular}
    \begin{tabular}{lllll}
      $\beta$  & $Z_A$ & $Z_S^{\MS}(2 \,\GeV)$  & $Z_T^{\MS}(2 \,\GeV)$  & $Z_V$ \\
      \cmidrule{1-5}
      3.34 & 0.77610(58) & 0.6072(26)  & 0.8443(35)  & 0.72690(71) \\ 
      3.4  & 0.77940(36) & 0.6027(25)  & 0.8560(35)  & 0.73290(67) \\ 
      3.46 & 0.78240(32) & 0.5985(25)  & 0.8665(36)  & 0.73870(71) \\ 
      3.55 & 0.78740(22) & 0.5930(25)  & 0.8820(37)  & 0.74740(82) \\ 
      3.7  & 0.79560(98) & 0.5846(24)  & 0.9055(42)  & 0.76150(94) \\ 
      3.85 & 0.8040(13)  & 0.5764(25)  & 0.9276(42)  & 0.77430(76)
    \end{tabular}
  \end{ruledtabular}
\end{table}
\begin{table}
  \caption{Set of renormalization factors denoted as $Z_J^2$ in the text. These are determined as in Table~\ref{tab:ZtabXVI} but now using the `fit method'.  \label{tab:ZtabXIII}}
  \begin{ruledtabular}
    \begin{tabular}{lllll}
      $\beta$  & $Z_A$ & $Z_S^{\MS}(2 \,\GeV)$  & $Z_T^{\MS}(2 \,\GeV)$  & $Z_V$ \\
      \cmidrule{1-5}
      3.34 & 0.7579(42)  & 0.6115(93)  & 0.8321(95)  & 0.7072(60)  \\ 
      3.4  & 0.7641(35)  & 0.6068(86)  & 0.8462(88)  & 0.7168(49)  \\ 
      3.46 & 0.7695(36)  & 0.6025(79)  & 0.8585(84)  & 0.7250(43)  \\ 
      3.55 & 0.7774(36)  & 0.5968(66)  & 0.8756(76)  & 0.7367(37)  \\ 
      3.7  & 0.7895(32)  & 0.5880(45)  & 0.9010(63)  & 0.7544(30)  \\ 
      3.85 & 0.8006(25)  & 0.5793(35)  & 0.9243(55)  & 0.7699(38)  \\ 
    \end{tabular}
  \end{ruledtabular}
\end{table}
\begin{table}
\caption{Renormalization factors $Z_A$ and $Z_V$ obtained from the
  interpolation formulas in Eqs.~(C.7) and~(C.8) in
  Ref.~\cite{DallaBrida:2018tpn}, denoted as $Z_J^3$ in the
  text.\label{tab:Zextrapol}}
\begin{minipage}{0.35\textwidth}
  \begin{ruledtabular}
    \begin{tabular}{lll}
      $\beta$  & $Z_A$ & $Z_V$ \\
      \cmidrule{1-3}
      3.34 & 0.7510(11)  & 0.7154(11)  \\ 
      3.4  & 0.75629(65) & 0.72221(65) \\ 
      3.46 & 0.76172(39) & 0.72898(39) \\ 
      3.55 & 0.76994(34) & 0.73905(35) \\ 
      3.7  & 0.78356(32) & 0.75538(33) \\ 
      3.85 & 0.79675(45) & 0.77089(47)
    \end{tabular}
  \end{ruledtabular}
\end{minipage}
\end{table}
A number of different methods are employed in Ref.~\cite{RQCD:2020kuu}
to determine the renormalization factors. In order to assess the
systematic uncertainty arising from the matching in the final results
for the charges at the physical point in the continuum limit, we make
use of two sets of results, collected in Tables~\ref{tab:ZtabXVI}
and~\ref{tab:ZtabXIII} and referred to as $Z_J^1$ and $Z_J^2$,
respectively, in the following. The first set of results are extracted
using the `fixed-scale method', where the RI$^\prime$-SMOM factors are
determined at a fixed scale~(ignoring discretization effects), while
the second set are obtained by fitting the factors as a function of
the scale and the lattice spacing, the `fit method'. See
Ref.~\cite{RQCD:2020kuu} for further details. In both cases, lattice
artefacts are reduced by subtracting the perturbative one-loop
expectation. For the axial and vector currents, we also consider a
third set of renormalization factors, $Z_J^3$, listed in
Table~\ref{tab:Zextrapol}, that are obtained with the chirally rotated
Schrödinger functional approach~\cite{Sint:2010eh}, see
Ref.~\cite{DallaBrida:2018tpn}. We emphasize that employing the
different sets of renormalization factors should lead to consistent
results for the charges in the continuum limit.

\subsection{Extrapolation strategy\label{sec:extrapol}}

In the final step of the analysis the renormalized charges $g_J^B$
determined at unphysical quark masses and finite lattice spacing and 
spatial volume are extrapolated to the physical point in the continuum
and infinite volume limits. We employ a similar strategy to the one
outlined in Ref.~\cite{Bali:2022qja} and choose continuum fit
functions of the form
\begin{equation}
\begin{aligned}
  g_J^B(M_\pi, M_K&, L, a=0) =  \\ 
   &c_0 + c_{\pi} M_\pi^2 + c_{K} M_K^2+ c_{V} M_\pi^2 \frac{e^{-L M_\pi}}{\sqrt{L M_\pi}}   \ ,
  \label{eq:g_chpt}
\end{aligned}
\end{equation}
to parameterize the quark mass and finite volume dependence, where $L$
is the spatial lattice extent and the coefficients $c_X$,
$X\in\{0,\pi,K,V\}$ are understood to depend on the baryon $B$ and the
current $J$. The leading order coefficients $c_0$ give the charges in
the SU(3) chiral limit, which can be expressed in terms of two LECs,
e.g., $F$ and $D$, for the axial charges, see
Eq.~\eqref{eq:FD1}.

Equation~\eqref{eq:g_chpt} is a phenomenological fit form based on the
SU(3) ChPT expressions for the axial charge. It contains the expected
$O(p^2)$ terms for the quark mass dependence and the dominant finite
volume corrections. The $O(p^3)$ expressions for
$g_A^B$~\cite{Jenkins:1990jv,Bijnens:1985kj,Ledwig:2014rfa} contain
log terms with coefficients completely determined by the LECs $F$ and
$D$. In an earlier study of the axial charges on the $m_s=m_{\ell}$
subset of the ensembles used here~\cite{Bali:2022qja}, we found that
including these terms did not provide a satisfactory description of
the data. When terms arising from loop corrections that contain
decuplet baryons are taken into account, additional LECs enter that
are difficult to resolve. If the coefficient of the log term is left
as a free parameter, one finds that the coefficient has the opposite
sign to the ChPT expectation without decuplet loops. We made similar
observations in this study and this is also consistent with the
findings of previous works, see, e.g.,
Refs.~\cite{Chang:2018uxx,Gupta:2018qil,Lutz:2020dfi}. Finite volume
effects appear at $O(p^3)$ with no additional LECs appearing in the
coefficients. Again the signs of the corrections are the opposite to
the trend seen in the data and, when included, it is difficult to
resolve the effects of the decuplet baryons. As is shown in
Sec.~\ref{sec:results}, the data for all the charges are well
described when the fit form is restricted to the dominant terms, with
free coefficients $c_0$, $c_{\pi}$, $c_K$ and $c_V$.

We remark that the same set of LECs appear in the $O(p^2)$ SU(3) ChPT
expressions for the three different octet baryons~(for a particular
charge). Ideally, one would carry out a simultaneous fit to the whole
baryon octet~(taking the correlations between the $g_J^B$ determined
on the same ensemble into account). However, we obtain very similar
results when fitting the $g_J^B$ individually compared to fitting the
results for all the octet baryons simultaneously. For simplicity, we
choose to do the former, such that the coefficients $c_X$ for the
different baryons are independent of one another.

Lattice spacing effects also need to be taken into account and we add both mass independent and mass dependent terms to the continuum fit ansatz to give
\begin{align}
    g_J^B(M_\pi, M_K, L, a) &=  g_J^B(\mathcal{M}_\pi, \mathcal{M}_K, L,0) \nonumber\\ 
    &+c_a\; \mathbbm{a}^2 + \bar{c}_{a} \overline{\mathcal{M}}^2 \;\mathbbm{a}^2  + \delta c_{a} \delta\mathcal{M}^2 \;\mathbbm{a}^2 \nonumber\\
    &+c_{a,3}\; \mathbbm{a}^3, \label{eq:g_cont}
\end{align}
where $\overline{\mathcal{M}}^2=(2\mathcal{M}_K^2 +
\mathcal{M}_\pi^2)/3$ and $\delta\mathcal{M}^2 = \mathcal{M}_K^2 -
\mathcal{M}_\pi^2$. The meson masses are rescaled with the Wilson flow
scale $t_0$~\cite{Luscher:2010iy}, $\mathcal{M}_{\pi,K}=\sqrt{8 t_0}
M_{\pi,K}$ to form dimensionless combinations. This rescaling is
required to implement full $O(a)$ improvement~(along with employing a
fermion action and isovector currents that are non-perturbatively
$O(a)$ improved) when simulating at fixed bare lattice coupling
instead of at fixed lattice spacing, see Sec. 4.1 of
Ref.~\cite{RQCD:2022xux} for a detailed discussion of this issue. The
values of $t_0/a^2$ and the pion and kaon masses in lattice units for
our set of ensembles are given in Table~\ref{tab:data_lat} of
Appendix~\ref{sec:lattice_data}.  We translate between different lattice
spacings using $t_0^\star$, the value of $t_0$ along the symmetric
line where $12t_0^*M_\pi^2=1.110$~\cite{Bruno:2016plf},
i.e., $\mathbbm{a}={a}/{\sqrt{8t_0^\star}}$. The values, determined in
Ref.~\cite{RQCD:2022xux}, are listed in Table~\ref{tab:t0star}. Note
that we include a term that is cubic in the lattice spacing in the fit
form, however, this term is only utilized in the analysis of the
vector charge, for which we have the most precise data.

\begin{table}
  \footnotesize
  \caption{Values for $t_0^\star/a^2$ at each $\beta$-value as determined in Ref.~\cite{RQCD:2022xux}. \label{tab:t0star}}
  \def\arraystretch{1.6}
  \addtolength{\tabcolsep}{-1pt}
  \begin{ruledtabular}
  \begin{tabular}{ccccccc}
    $\beta$                  & 3.34 & 3.4 & 3.46 & 3.55 & 3.7 & 3.85 \\ \cmidrule{1-7}
    $\tfrac{t_0^\star}{a^2}$ & 2.204(6) & 2.888(8) & 3.686(11) & 5.157(15) & 8.617(22) & 13.988(34)
  \end{tabular}
  \end{ruledtabular}
  \addtolength{\tabcolsep}{1pt}
\end{table}

To obtain results at the physical quark mass point, we make use of the
scale setting parameter
\begin{align}
  \sqrt{8t_{0,\text{phys}}} = 0.4098^{(20)}_{(25)} \, \fm ,
\end{align}
determined in Ref.~\cite{RQCD:2022xux} and take the isospin corrected pion and kaon masses quoted in the FLAG~16 review~\cite{Aoki:2016frl} to define the physical point in the quark mass plane,
\begin{align}
  M_\pi^{\text{phys}} &= 134.8(3)\, \MeV  , \\
  M_K^{\text{phys}} &= 494.2(3)\, \MeV  .
\end{align}

In practice, we choose to fit to the bare lattice charges
$g_J^{B,\latt}$ rather than the renormalized ones as this enables us to
include the uncertainties of the renormalization and improvement
factors (which are the same for all ensembles at fixed $\beta$)
consistently. Therefore, our final fit form reads
\begin{align}
  g_J^{B,\latt} = \frac{g_J^B(M_\pi, M_K, L, a)}{Z^k_J(\beta) \left(1 + a m_\ell b_J(\beta) + 3 a \bar{m} \tilde{b}_J(\beta) \right)},
  \label{eq:g_fit}
\end{align}
where the dependence of the factors on the $\beta$-value is made
explicit and the superscript $k$ of $Z_J^k$ refers to the different
determinations of the renormalization factors that we consider,
$k=1,2,3$ for $J \in \{A, V\}$ and $k=1,2$ for $J \in \{S, T\}$~(see
Tables~\ref{tab:ZtabXVI}, \ref{tab:ZtabXIII} and~\ref{tab:Zextrapol}
in the previous subsection). We introduce a separate parameter for
$Z^k_J$, $b_J$ and $\tilde{b}_J$ for each $\beta$-value and add
corresponding ``prior'' terms to the $\chi^2$~function.  The
statistical uncertainties of these quantities are incorporated by
generating pseudo-bootstrap distributions.

The systematic uncertainty in the determination of the charges at the
physical point is investigated by varying the fit model and by
employing different cuts on the ensembles that enter the fits. For the
latter we consider
\begin{enumerate}
  \item[1] no cut: including all the available data points, denoted as
    data set 0, DS$(0)$,
  \item[2] pion mass cut: excluding all ensembles with \mbox{$M_\pi >
    400\, \MeV$}, DS($M_\pi^{\footnotesize{<400\, \MeV}}$),
  \item[3] pion mass cut: excluding all ensembles with \mbox{$M_\pi >
    300\, \MeV$}, DS($M_\pi^{\footnotesize{<300\,\MeV}}$),
  \item[4] a lattice spacing cut: excluding the coarsest lattice
    spacing, i.e., the ensembles with $a\approx 0.098\,\fm$,
    DS($a^{\footnotesize{<0.1\,\fm}}$),
  \item[5] a volume cut: excluding all ensembles with \mbox{$LM_\pi < 4$},
    DS($LM_\pi^{\footnotesize{> 4}}$).
\end{enumerate}
In some cases, more than one cut is applied, e.g., cut~2 and~4, with
the data set denoted DS($M_\pi^{\footnotesize{<400\,\MeV}}$,
$a^{\footnotesize{<0.1\,\fm}}$), etc..  Our final results are obtained
by carrying out the averaging procedure described in Appendix~B of
Ref.~\cite{Bali:2022qja} which gives an average and error that
incorporates both the statistical and systematic uncertainties.

\section{Extrapolations to the continuum, infinite volume, physical quark mass limit\label{sec:results}}

We present the extrapolations to the physical point in the continuum
and infinite volume limits of the isovector vector~($V$), axial~($A$),
scalar~($S$) and tensor~($T$) charges for the nucleon~($N$),
sigma~($\Sigma$) and cascade~($\Xi$) octet baryons.

\subsection{Vector charges\label{sec:gv}}

The isovector vector charges for the nucleon, cascade and sigma
baryons are $g_V^N = g_V^\Xi = 1$ and $g_V^\Sigma = 2$, up to second
order isospin breaking corrections~\cite{Ademollo:1964sr}. These 
values also apply to our isospin symmetric lattice results in the continuum limit 
for any quark mass combination and volume. A determination of the
vector charges provides an important cross-check of our analysis
methods and allows us to demonstrate that all systematics are under
control.

\begin{figure}
  \begin{center}
    \includegraphics[width=0.48\textwidth]{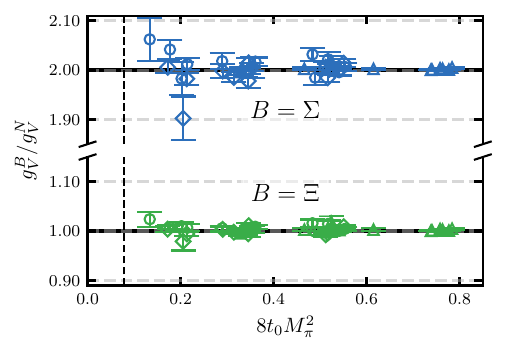}
    \caption{Ratio of the hyperon ($B = \Sigma, \Xi$) vector charges
      over the nucleon charge, $ g_V^B / g_V^N$, as a function of the
      rescaled pion mass
      squared~($8t_0M_\pi^2=\mathcal{M}_\pi^2$). The data were
      extracted using two excited states in the fitting analysis, see
      Sec.~\ref{sec:excited}, and not corrected for lattice spacing
      or volume effects. Circles (diamonds) correspond to the $\tr M=\const$
      ($m_s=\const$) trajectories, the triangles to the $m_s=m_{\ell}$ line.\label{fig:hyperons_gv_ratio}}
  \end{center}
\end{figure}

To start with, we display the ratios of the hyperon charges over the
nucleon charge in Fig.~\ref{fig:hyperons_gv_ratio}. The
renormalization factors drop out in the ratio and lattice spacing
effects are expected to cancel to some extent. As one can see, the
results align very well with the expected values.

For the individual
charges, we perform a continuum extrapolation of the data using the
fit form
\begin{align}
  g_V = 
  &\; c_0 
  + c_a \;\mathbbm{a}^2 + \bar{c}_{a} \overline{\mathcal{M}}^2 \;\mathbbm{a}^2 + \delta c_{a} \delta\mathcal{M}^2 \;\mathbbm{a}^2 +c_{a,3}\; \mathbbm{a}^3.
  \label{eq:gv_extrapol} 
\end{align}
Note that there is no dependence on the pion or kaon mass nor on the
spatial volume in the continuum limit. $\mathcal{M}^2$ and $\delta
\mathcal{M}^2$ represent the flavour average and difference of the
kaon and pion masses squared, rescaled with the scale parameter $t_0$,
while the lattice spacing $\mathbbm{a}=a/\sqrt{8t_0^*}$. See the
previous section for further details of the extrapolation
procedure. We implement full $O(a)$ improvement and leading
discretization effects are quadratic in the lattice spacing. However,
the data for the nucleon vector charge are statistically very precise
and higher order effects can be resolved. This motivates the addition
of the cubic term in Eq.~\eqref{eq:gv_extrapol}. The data for
$g_V^\Sigma$ and $g_V^\Xi$ are less precise as they are determined
employing the stochastic approach outlined in Sec.~\ref{sec:stoch3pts}
which introduces additional noise, see Appendix~\ref{sec:details3pt}
for further discussion.

\begin{figure}
  \begin{center}
    \includegraphics[width=0.48\textwidth]{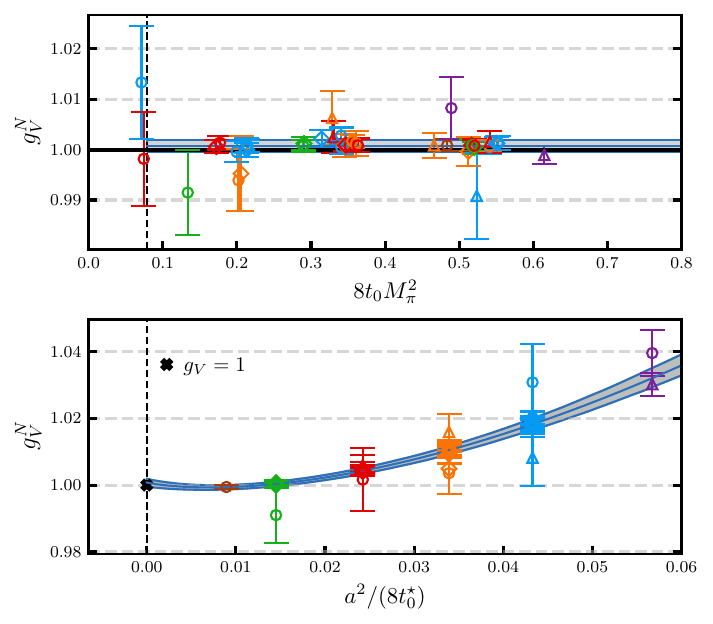}
    \caption{Continuum limit extrapolation of the nucleon isovector
      vector charge $g_V^N$ for a five parameter
      fit~(Eq.~\eqref{eq:gv_extrapol}) using the renormalization
      factors $Z_V^3$~(see Table~\ref{tab:Zextrapol}) and imposing the
      cut $M_\pi < 400\,\MeV$. The data were extracted including two
      excited states in the fitting analysis, see
      Sec.~\ref{sec:excited}. The upper panel shows the data points
      corrected for discretization effects according to the fit. They
      are consistent with $g_V^N=1$. The bottom panel shows the
      lattice spacing dependence at the physical point. The blue lines
      and grey bands indicate the expectations from the fit.  For
      better visibility, the data point for ensemble D452, which has a
      relatively large error~(see Table~\ref{tab:data_nucleon}), is
      not displayed. Circles (diamonds) correspond to the $\tr
      M=\const$ ($m_s=\const$) trajectories, the triangles to the
      $m_s=m_{\ell}$ line.
      \label{fig:extrapol_gv_nucleon}}
  \end{center}
\end{figure}

The data are well described by Eq.~\eqref{eq:gv_extrapol}, as
demonstrated by the fit, shown in Fig.~\ref{fig:extrapol_gv_nucleon},
for $g_V^N$ which has a goodness of fit of $\chi^2/N_{\text{dof}}=0.92$. The
data are extracted using two excited states in the fitting
analysis~(see Sec.~\ref{sec:excited}) and we employ the most precise
determination of the renormalization factors~($Z_V^3$, see
Table~\ref{tab:Zextrapol}). A cut of $M_\pi<400$~MeV is imposed on the
ensembles entering the fit, however, fits including all data points
are also performed, as detailed below. When the data are corrected for
the discretization effects according to the fit, we see consistency
with $g_V^N=1$, for all pion and kaon masses. Using the fit to shift
the data points to the physical point, we observe that the lattice
spacing dependence is moderate but statistically significant, with a
3--4\% deviation from the continuum value at the coarsest lattice
spacing (lower panel of Fig.~\ref{fig:extrapol_gv_nucleon}).

\begin{figure}
  \begin{center}
    \includegraphics[width=0.48\textwidth]{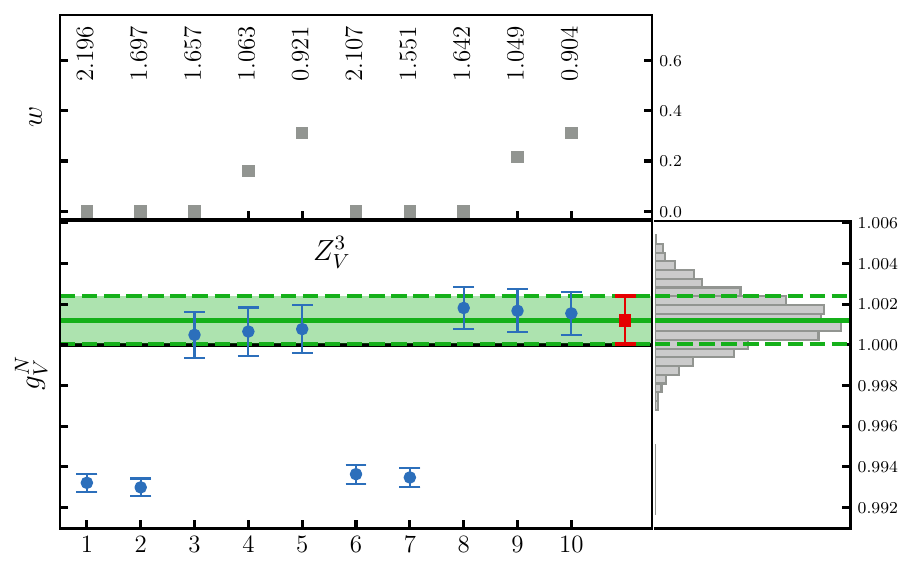}
    \caption{Results for the nucleon vector charge $g_V^N$ in the
      continuum limit at the physical point obtained using
      $Z_V^3$~(see Table~\ref{tab:Zextrapol}) and five different
      parametrizations applied to data set
      DS($M_\pi^{\footnotesize{<400\, \MeV}}$)~(fits to
      Eq.~\eqref{eq:g_cont} with different coefficients set to zero
      labelled 1,...,5, see the text) and
      DS($M_\pi^{\footnotesize{<400\, \MeV}}$,
      $a^{\footnotesize{<0.1\,\fm}}$)~(6,...,10). See
      Sec.~\ref{sec:extrapol} for the definitions of the data
      sets. The data were extracted including two excited states in
      the fitting analysis, see Sec.~\ref{sec:excited}. The model
      average is shown as the red data point and the green horizontal
      line and band. On the right the model averaged distribution is
      displayed as a histogram where also the median and the 68\%
      confidence level interval, which form the final result, are
      indicated (green lines). The top panel shows the weights (grey
      points) assigned to the individual fits, with the corresponding
      $\chi^2/N_{\text{dof}}$ values given
      above. \label{fig:modelavg_hist_gv}}
  \end{center}
\end{figure}

In order to investigate the uncertainty arising from the choice of
parametrization and the importance of the different terms, we repeat
the extrapolations employing five different parametrizations~(listed
in terms of the coefficients of the terms entering the fit):
$(1,\{c_0, c_a\})$, $(2,\{c_0, c_a, \delta c_a\})$, $(3,\{c_0, c_a,
c_{a,3}\})$, $(4,\{c_0, c_a, c_{a,3}, \delta c_a\})$ and $(5,\{c_0,
c_a, c_{a,3}, \bar{c}_{a}, \delta c_{a} \})$.\footnote{We also
investigated the possibility of residual $\mathcal{O}(a)$ effects, in
spite of the non-perturbative improvement of the current and the
action. Indeed, the coefficients of additional terms $\propto a$ were
found to be consistent with zero.}  Regarding the lattice spacing
dependence, the mass independent term $c_a$ is always included as the
other terms are formally at a higher order. These five fits are
performed on two data sets. The first set contains ensembles with
$M_\pi<400\,\MeV$~(data set DS($M_\pi^{\footnotesize{<400\,\MeV}}$)),
while in the second set the ensembles with the coarsest lattice
spacing are also excluded~(DS($M_\pi^{\footnotesize{<400\,\MeV}}$,
$a^{\footnotesize{<0.1\,\fm}}$), $+5$ is added to the fit number). See
the end of Sec.~\ref{sec:extrapol} for the definitions of the data
sets.

The results for $g_V^N$, displayed in Fig.~\ref{fig:modelavg_hist_gv},
show that the cubic term and at least one mass dependent term are
needed to obtain a reasonable description of the data in terms of the
$\chi^2/N_{\text{dof}}$.  Two of the fit forms with large $\chi^2/N_{\text{dof}}$
values (corresponding to 1, 2, 6, 7, with negligible weight in the
model averaging procedure) give values that are inconsistent with the
continuum expectation. The results are stable under the removal of the
coarsest ensembles. Performing the model averaging procedure, the
final result for the nucleon, given in the last row of the first
column of Table~\ref{tab:gv_final}, agrees with the expectation
$g_V^N=1$ within a combined statistical and systematic uncertainty of
about 1\textperthousand.

\begin{figure*}
  \begin{center}
    \includegraphics[width=0.7\textwidth]{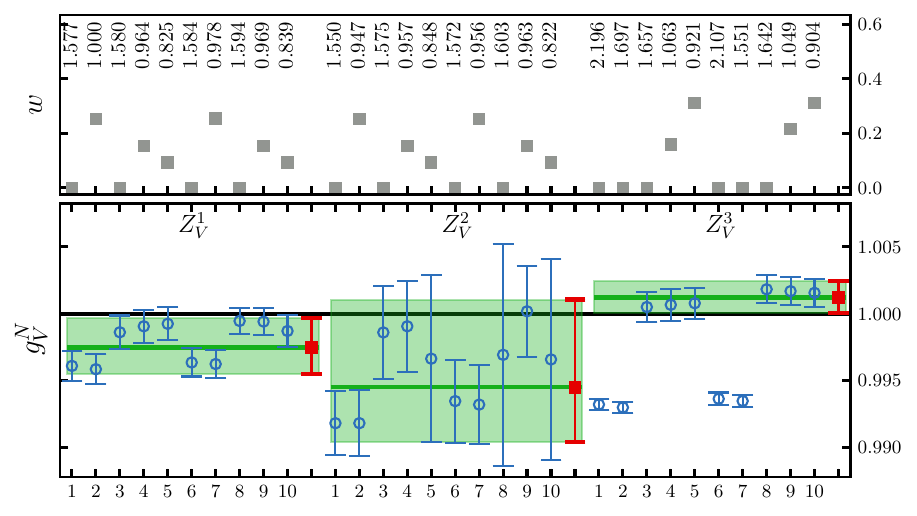}
    \caption{Results for the nucleon vector charge $g_V^N$ as in
      Fig.~\ref{fig:modelavg_hist_gv} but now also including those
      obtained employing $Z_V^1$ and $Z_V^2$. \label{fig:modelavg_gv}}
  \end{center}
\end{figure*}

The above analysis is also performed utilizing the sets of
renormalization factors $Z_V^1$ and $Z_V^2$, determined via the
RI$^\prime$-SMOM scheme~\cite{RQCD:2020kuu}. The results for the
nucleon vector charge are compared in Fig.~\ref{fig:modelavg_gv}. The
uncertainties on these factors are larger, in particular for $Z_V^2$,
than those of set $Z_V^3$, which is derived using the chirally rotated
Schrödinger functional approach~\cite{DallaBrida:2018tpn}. This
translates into larger errors for $g_V^N$ for those fits. The lattice
spacing dependence is somewhat different: the first quadratic mass
dependent term in Eq.~\eqref{eq:gv_extrapol} and the cubic term can no
longer be fully resolved and also parametrization $(2,\{c_0, c_a,
\delta c_a\})$ gives a $\chi^2/N_{\text{dof}}=1.00$~(0.95) when employing
$Z_V^1$~($Z_V^2$).

\begin{figure*}
  \begin{center}
    \includegraphics[width=0.32\textwidth]{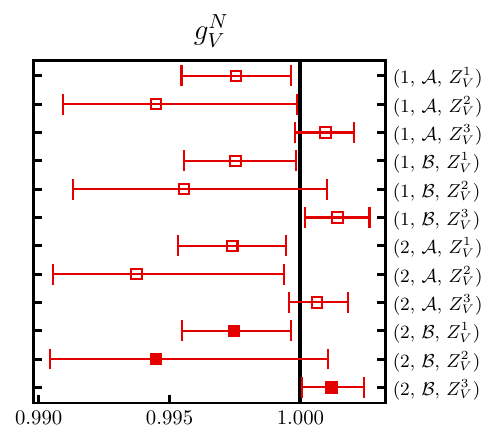}
    \includegraphics[width=0.32\textwidth]{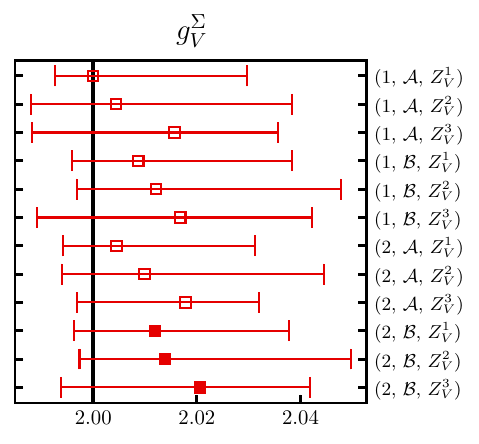}
    \includegraphics[width=0.32\textwidth]{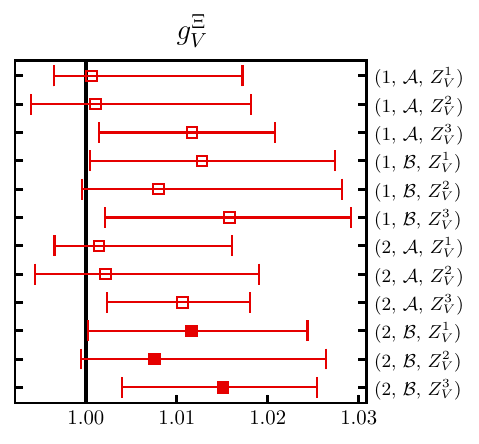}
    \caption{Overview of the results for the vector charges $g_V^B$, $B\in\{N,\Sigma,\Xi\}$,
      obtained from different data sets. These are labelled by the
      number of excited states used in the fitting analysis (ES=1, 2),
      the pion mass cut imposed (denoted $\mathcal{A}$ or
      $\mathcal{B}$) and the set of renormalization factors employed
      ($Z_V^k$, $k=1,2,3$). Each data point represents a model
      averaged result.  The label $\mathcal{A}$ indicates that 15 fits
      (5 fit variations applied to three data sets, DS(0),
      DS($M_\pi^{\footnotesize{<400\, \MeV}}$) and
      DS($a^{\footnotesize{<0.1\,\fm}}$)) are averaged, while
      the results labelled with $\mathcal{B}$ are based on the set of
      10 fits utilized in Fig.~\ref{fig:modelavg_hist_gv}~(5 fit
      variations applied to two data sets,
      DS($M_\pi^{\footnotesize{<400\, \MeV}}$) and
      DS($M_\pi^{\footnotesize{<400\, \MeV}}$,
      $a^{\footnotesize{<0.1\,\fm}}$)). See
      Sec.~\ref{sec:extrapol} for the definitions of the data
      sets. The final results for each $Z_V^k$~(filled squares) are
      listed in
      Table~\ref{tab:gv_final}.\label{fig:modelavg_overview_gv}}
  \end{center}
\end{figure*}

The systematic uncertainty of the results due to residual excited state
contamination and the range of pion masses employed in the
extrapolations is also
considered. Figure~\ref{fig:modelavg_overview_gv} shows the model
averaged results discussed so far, displayed as filled squares, and
also those obtained using several other sets of fits. These are
labelled in terms of the number of excited states~(one or two)
included in the fitting analysis, the cuts imposed on the pion
mass~($\mathcal{A}$ or $\mathcal{B}$) and the renormalization factors
utilized. For the results from pion mass cut $\mathcal{A}$, 15 fits
enter the model average, the five different parametrizations are
applied to three data sets DS(0), DS($M_\pi^{\footnotesize{<400\,
    \MeV}}$) and DS($a^{\footnotesize{<0.1\,\fm}}$). Note that
the first and third data set include ensembles with pion masses up to
$430\,\MeV$. For mass cut $\mathcal{B}$, data sets
DS($M_\pi^{\footnotesize{<400\, \MeV}}$) and
DS($M_\pi^{\footnotesize{<400\, \MeV}}$,
$a^{\footnotesize{<0.1\,\fm}}$)) are used, giving 10 fits in
total. The results only depend on the choice of renormalization
factors, suggesting that the systematic uncertainties due to excited
state contamination and the cut made on the pion mass are very
small.

Repeating the whole procedure for the sigma and the cascade baryons
gives vector charges which are also consistent with the expected
values to within $1.5\,\sigma$, as shown in
Fig.~\ref{fig:modelavg_overview_gv}~(see
Figs.~\ref{fig:extrapol_gv_hyperon} and~\ref{fig:modelavg_gv_hyperon}
in Appendix~\ref{sec:plots} for the individual fits for mass cut
$\mathcal{B}$). The statistical noise introduced by the stochastic
approach dominates, leading to much less precise values and very
little variation between the results for the different hyperon data
sets. We take the values obtained from the data sets (2,
$\mathcal{B}$, $Z_V^k$), listed in Table~\ref{tab:gv_final}, as our
estimates of the vector charges as these data sets give the most
reliable determinations of the charges across the different
channels~(as discussed in the following subsections).

\begin{table}
  \caption{Results for $g_V^B$, $B\in\{N,\Sigma,\Xi\}$,  obtained with three different sets of renormalization factors. The errors include the statistical and all the systematic uncertainties. \label{tab:gv_final}}
  \begin{ruledtabular}
    \def\arraystretch{1.7}
    \begin{tabular}{lccc}
      Renormalization                       & $g_V^N$                & $g_V^\Sigma$          & $g_V^\Xi$  \\\hline 
      $Z_V^{1}$ (Table~\ref{tab:ZtabXVI})   & $0.9975^{(22)}_{(20)}$ & $2.012^{(26)}_{(16)}$ & $1.012^{(13)}_{(11)}$ \\
      $Z_V^{2}$ (Table~\ref{tab:ZtabXIII})  & $0.9945^{(66)}_{(41)}$ & $2.014^{(36)}_{(16)}$ & $1.008^{(19)}_{(8)}$ \\
      $Z_V^{3}$ (Table~\ref{tab:Zextrapol}) & $1.0012^{(12)}_{(11)}$ & $2.021^{(21)}_{(27)}$ & $1.015^{(10)}_{(11)}$ 
    \end{tabular}
  \end{ruledtabular}
\end{table}

Overall, the results demonstrate that the systematics arising from
excited state contamination, renormalization and finite lattice
spacing are under control in our analysis in this channel~(to within
an error of 1\textperthousand\ for the nucleon).

\subsection{Axial charges\label{sec:ga}}

\begin{figure}
  \begin{center}
    \includegraphics[width=0.48\textwidth]{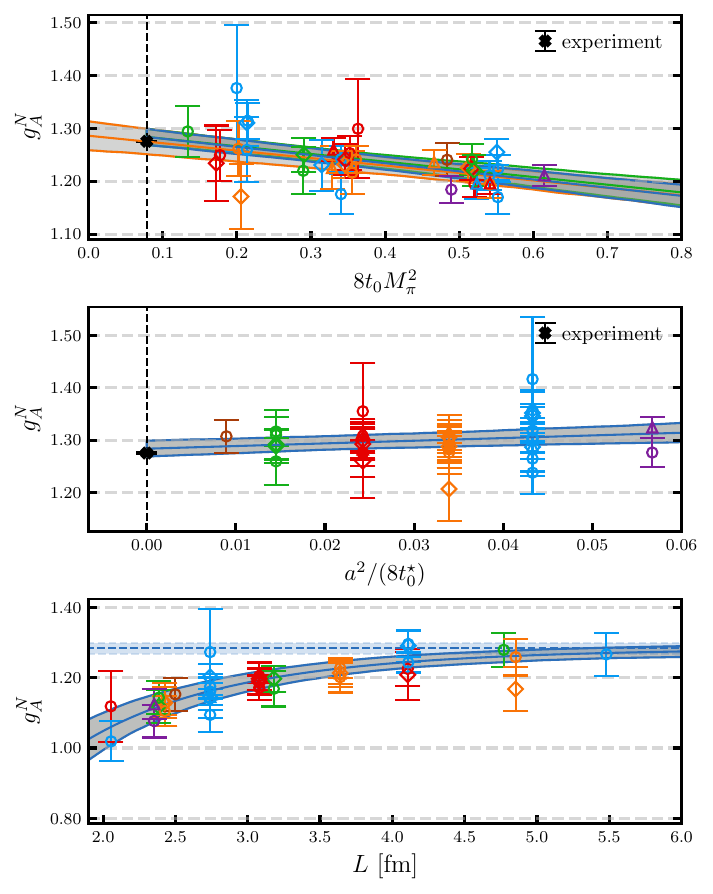}
    \caption{Simultaneous quark mass, continuum and finite volume
      extrapolation of the nucleon isovector axial charge $g_A^N$
      extracted on ensembles with $M_\pi<400\,\MeV$ using two excited
      states in the fitting analysis~(see Sec.~\ref{sec:excited}) and
      renormalization factors $Z_A^3$~(see
      Table~\ref{tab:Zextrapol}). A five parameter fit form is
      employed, see the text. (Top) Pion mass dependence of $g_A^N$,
      where the data points are corrected, using the fit, for finite
      volume and discretization effects and shifted~(depending on the
      ensemble) to kaon masses corresponding to the $\tr\,M = \const$
      and $m_s = \const$ trajectories. The fit is shown as a grey band
      with the three trajectories distinguished by blue ($\tr
      M=\const$, circles), green ($m_s=\const$, diamonds) and orange
      ($m_s=m_{\ell}$, triangles) lines, respectively. The vertical
      dashed line indicates the physical point. (Middle) Lattice
      spacing dependence at the physical point in the infinite volume
      limit. (Bottom) Finite volume dependence at the physical point
      in the continuum limit. The dashed blue line (band) indicates
      the infinite volume result. For better visibility, the data
      points for ensembles D150, E250 and D452, which have relatively
      large errors~(see Table~\ref{tab:data_nucleon}), are not
      displayed.  The black cross at the physical point indicates the
      experimental value~\cite{Workman:2022ynf}.
      \label{fig:extrapol_ga_nucleon}}
  \end{center}
\end{figure}

 \begin{figure*}
  \begin{center}
    \includegraphics[width=0.48\textwidth]{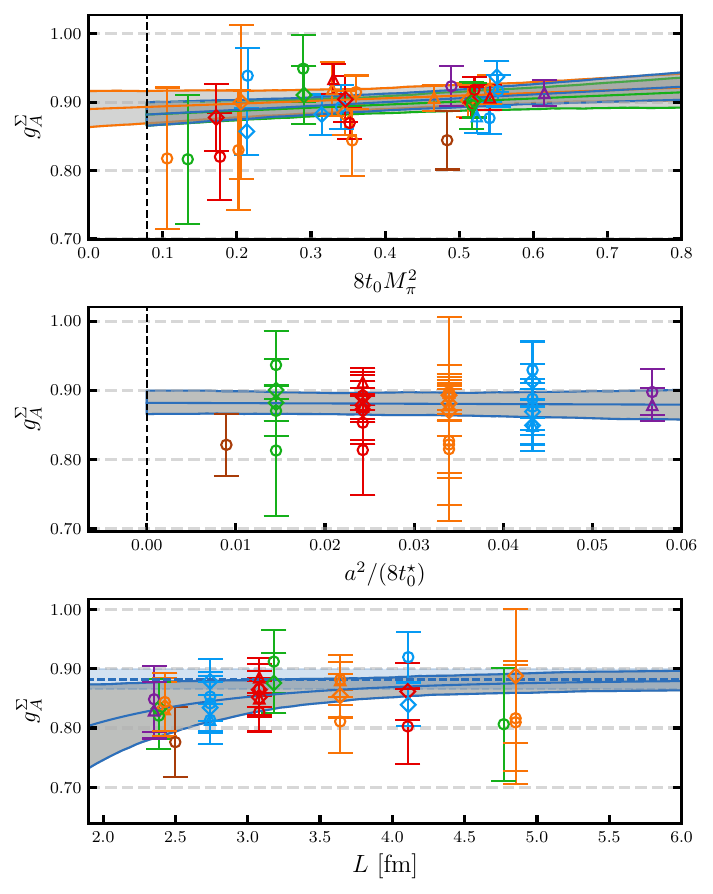}
    \includegraphics[width=0.48\textwidth]{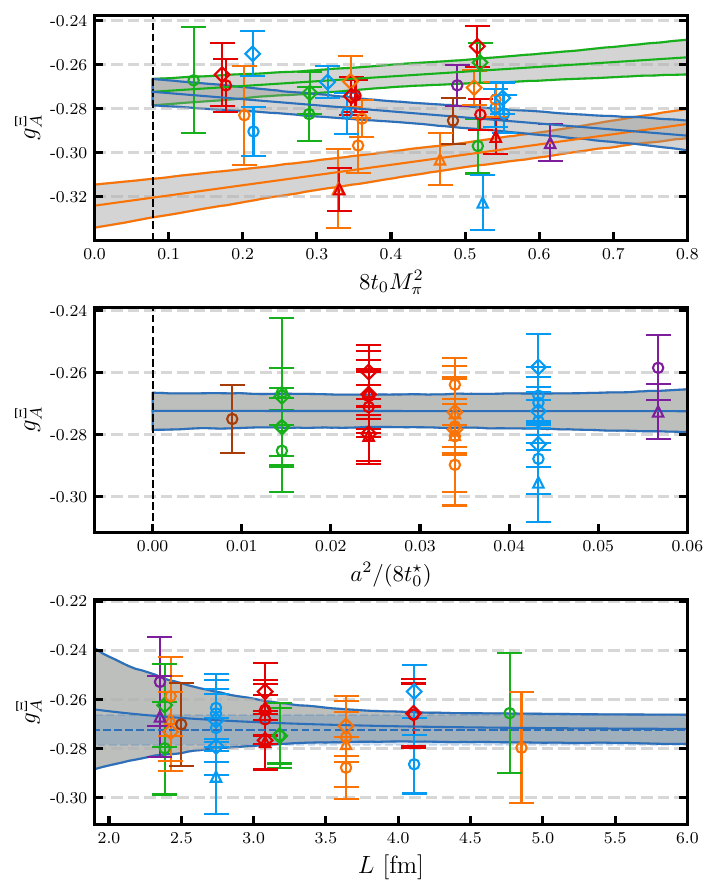}
    \caption{The same as Fig.~\ref{fig:extrapol_ga_nucleon} for the
      isovector axial charges~$g_A^B$ of the sigma baryon (left) and
      the cascade baryon (right).  For better visibility, the data
      points for ensembles D452 and D451, which have relatively
      large errors~(see Table~\ref{tab:data_xi}), are not
      displayed for the cascade baryon. Compared to the nucleon, for
      the analysis of the hyperon charges a reduced set of ensembles
      is employed, see Tables~\ref{tab:data_sigma} and
      \ref{tab:data_xi} for a complete list of ensembles.
      \label{fig:extrapol_ga_hyperons}}
  \end{center}
\end{figure*}

In the following we present our results for the nucleon, sigma and
cascade isovector axial charges~$g_A^B$, $B\in\{N,\Sigma,\Xi\}$. The
nucleon axial charge is very precisely measured in experiment,
$\lambda=g_A^N/g_V^N = 1.2754(13)$~\cite{Workman:2022ynf}, and serves
as another benchmark quantity when assessing the size of the
systematics of the final results. Note, however, that possible
differences of up to 2\%, due to radiative corrections, between
$\lambda$ computed in QCD and an effective $\lambda$ measured in
experiment have been discussed
recently~\cite{Hayen:2020cxh,Cirigliano:2022hob}. Lattice
determinations of $g_A^N$ are known to be sensitive to excited state
contributions, finite volume effects and other systematics. Whereas
there is a long history of lattice QCD calculations of $g_A^N$, see,
e.g., the FLAG~21 review~\cite{Aoki:2021kgd}, there are very few
lattice computations of hyperon axial
charges~\cite{Lin:2007ap,Erkol:2009ev,Alexandrou:2016xok,Savanur:2018jrb,Smail:2023eyk} and only
few phenomenological estimates exist from measurements of
semileptonic hyperon decay rates.

 \begin{figure*}[b]
  \begin{center}
    \includegraphics[width=0.7\textwidth]{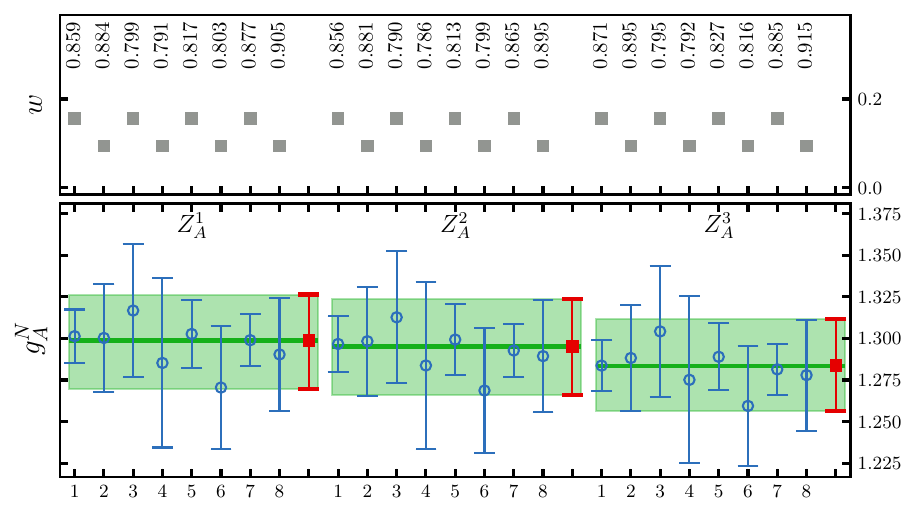}
    \caption{The same as Fig.~\ref{fig:modelavg_gv} for the nucleon
      axial charge $g_A^N$. The eight fits correspond to two fit
      variations, see the text, applied to four data sets,
      DS($M_\pi^{\footnotesize{<400\, \MeV}}$),
      DS($M_\pi^{\footnotesize{<300\, \MeV}}$),
      DS($M_\pi^{\footnotesize{<400\, \MeV}}$,
      $a^{\footnotesize{<0.1\,\fm}}$) and
      DS($M_\pi^{\footnotesize{<400\, \MeV}}$,
      $LM_\pi^{\footnotesize{> 4}}$). The data are extracted using two
      excited states in the fitting analysis, see
      Sec.~\ref{sec:excited}.  \label{fig:modelavg_ga_nucleon}}
  \end{center}
\end{figure*}

We carry out simultaneous continuum, quark mass and finite volume fits
to the individual baryon charges employing the parametrization in
Eq.~\eqref{eq:g_cont}~(with the continuum form in
Eq.~\eqref{eq:g_chpt}). The discretization effects are found to be
fairly mild and we are not able to resolve the quadratic mass
dependent terms or a cubic term. These terms are omitted
throughout. As already mentioned in Sec.~\ref{sec:extrapol} we are
also not able to resolve any higher order ChPT terms in the continuum
parametrization.

A five parameter fit, with free coefficients $\{c_0, c_\pi, c_K, c_V,
c_a\}$, describes the data well, as demonstrated in
Fig.~\ref{fig:extrapol_ga_nucleon} for the nucleon~(with
$\chi^2/N_{\text{dof}}=0.86$) and Fig.~\ref{fig:extrapol_ga_hyperons} for the
sigma and cascade baryons~(with $\chi^2/N_{\text{dof}}=0.85$ and 1.25,
respectively). The data are extracted using two excited
states~(`ES=2') in the fitting analysis~(see Sec.~\ref{sec:excited})
and renormalized with factors $Z_A^3$~(that are the most precise of
the three determinations considered, see
Table~\ref{tab:Zextrapol}). For the cascade baryon, with two strange
quarks, the data on the three quark mass trajectories~($\tr\,M =
\const$, $m_s = \const$ and $m_\ell=m_s$) are clearly delineated,
however, note the different scale on the right of
Fig.~\ref{fig:extrapol_ga_hyperons}. The availability of ensembles on
two trajectories which intersect at the physical point helps to
constrain the physical value of the axial charge. In terms of the
finite volume effects, only the nucleon shows a significant dependence
on the spatial extent. The quark mass dependence is
also pronounced in this case.

As in the vector case, we quantify the systematics associated with the
extraction of the charges at the physical point (in the continuum and
infinite volume limits) by varying the parametrization and the set of
ensembles that are included in the fit. We consider two fit forms
$(1,\{c_0, c_\pi, c_K, c_V, c_a\})$ and $(2,\{c_0, c_\pi, c_K, c_V,
c_a, \delta c_a\}$) and four data sets,
DS($M_\pi^{\footnotesize{<400\, \MeV}}$),
DS($M_\pi^{\footnotesize{<300\, \MeV}}$),
DS($M_\pi^{\footnotesize{<400\, \MeV}}$,
$a^{\footnotesize{<0.1\,\fm}}$) and DS($M_\pi^{\footnotesize{<400\,
    \MeV}}$, $LM_\pi^{\footnotesize{> 4}}$), see
Sec.~\ref{sec:extrapol} for their definitions.

The results of the eight fits and their model averages for the three
different determinations of the renormalization factors are shown in
Fig.~\ref{fig:modelavg_ga_nucleon} for the nucleon and in
Fig.~\ref{fig:modelavg_ga_hyperons} of Appendix~\ref{sec:plots} for the
hyperon axial charges.  In all cases, we find consistent results
across the different fits and choice of renormalization factor
suggesting that the statistical errors dominate. The additional
lattice spacing term is not really resolved with the goodness of fit
only changing slightly, while the errors on the coefficients
increase. For the nucleon and sigma baryon, all fits have a
$\chi^2/N_{\text{dof}}<1$ and are given a similar weight in the model
average, while for the cascade baryon, the cut
$M_\pi^{\footnotesize{<300\, \MeV}}$ is needed to achieve a goodness
of fit around 1 and these fits have the highest weight factors.

In order to further explore the systematics, additional data sets are
considered. We assess the sensitivity of the results to excited state
contributions by performing extrapolations of the data extracted using
only one excited state~(`ES=1') in the fitting analysis. In addition,
as only the $O(p^2)$ ChPT terms are included in the continuum
parametrization, we test the description of the quark mass dependence
by performing 10 fits, involving the two parametrization variations
above, applied to five data sets, DS(0),
DS($M_\pi^{\footnotesize{<400\, \MeV}}$),
DS($M_\pi^{\footnotesize{<300\, \MeV}}$),
DS($a^{\footnotesize{<0.1\,\fm}}$) and DS($LM_\pi^{\footnotesize{>
    4}}$). The first, fourth and fifth data sets include ensembles
with pion masses up to $430\,\MeV$.
\begin{figure*}
  \begin{center}
    \includegraphics[width=0.3\textwidth]{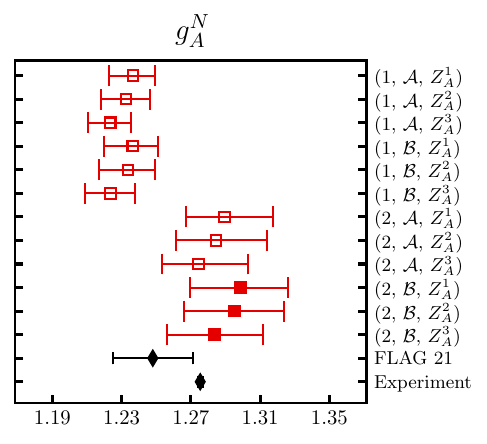}
    \includegraphics[width=0.3\textwidth]{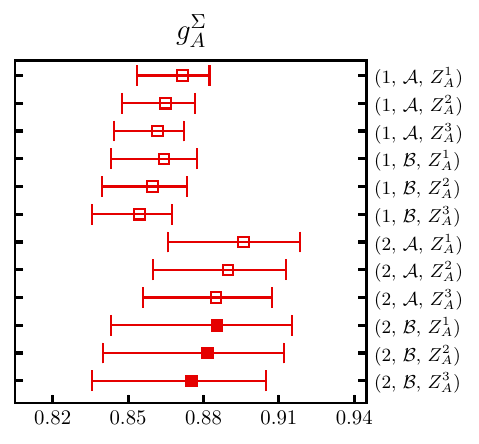}
    \includegraphics[width=0.3\textwidth]{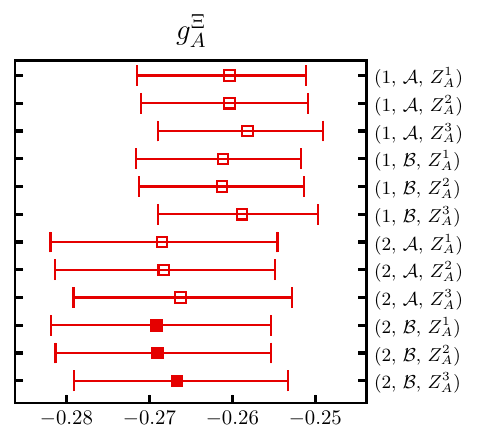}
    \caption{The same as Fig.~\ref{fig:modelavg_overview_gv} for the
      nucleon, sigma and cascade axial charges. The label
      $\mathcal{A}$ indicates that 10 fits enter the model average
      corresponding to two fit variations, see the text, applied to 5
      data sets, DS(0), DS($M_\pi^{\footnotesize{<400\, \MeV}}$),
      DS($M_\pi^{\footnotesize{<300\, \MeV}}$),
      DS($a^{\footnotesize{<0.1\,\fm}}$) and
      DS($LM_\pi^{\footnotesize{> 4}}$). For the data points labelled
      with $\mathcal{B}$, the results of the 8 fits employed in
      Fig.~\ref{fig:modelavg_ga_nucleon} are averaged. For the
      nucleon, the FLAG~21 average for $N_f=2+1$~\cite{Liang:2018pis,Harris:2019bih}
      and the experimental value~\cite{Workman:2022ynf} are indicated~(black
      diamonds) \label{fig:modelavg_overview_ga}}
  \end{center}
\end{figure*}

\begin{table}[b]
  \caption{Results for $g_A^B$, $B\in\{N,\Sigma,\Xi\}$, obtained with three different sets of renormalization factors. The errors include the statistical and all the systematic uncertainties.\label{tab:ga_final}}
  \begin{ruledtabular}
    \def\arraystretch{1.7}
    \begin{tabular}{lccc}
      Renormalization                       & $g_A^N$               & $g_A^\Sigma$           & $g_A^\Xi$  \\\hline 
      $Z_A^{1}$ (Table~\ref{tab:ZtabXVI})   & $1.299^{(28)}_{(29)}$ & $0.885^{(30)}_{(42)}$ & $-0.269^{(14)}_{(13)}$ \\
      $Z_A^{2}$ (Table~\ref{tab:ZtabXIII})  & $1.295^{(28)}_{(29)}$ & $0.882^{(30)}_{(42)}$ & $-0.269^{(14)}_{(12)}$ \\
      $Z_A^{3}$ (Table~\ref{tab:Zextrapol}) & $1.284^{(28)}_{(27)}$ & $0.875^{(30)}_{(39)}$ & $-0.267^{(13)}_{(12)}$ 
    \end{tabular}
  \end{ruledtabular}
\end{table}

The results for the axial charges from model averaging the 10
fits~(denoted $\mathcal{A}$) employing the 5 data sets and also from
the 8 fits~(denoted $\mathcal{B}$) using the 4 data sets given above,
for the `ES=1' and `ES=2' data and the different renormalization
factors are displayed in Fig.~\ref{fig:modelavg_overview_ga}. Very
little variation is seen in the results in terms of the range of pion
masses included and, as before, the renormalization factors employed,
suggesting the associated systematics are accounted for within the
combined statistical and systematic error~(which includes the
uncertainty due to lattice spacing and finite volume
effects). However, the results are sensitive to the number of excited
states included in the fitting analysis. This is only a significant
effect for the nucleon, for which the `ES=1' results lie around
$2.5\,\sigma$ below experiment. Similar underestimates of $g_A^N$ have
been observed in many earlier lattice studies~\cite{Aoki:2021kgd}.

As detailed in Sec.~\ref{sec:excited}, more than one excited state is
contributing significantly to the ratio of three-point over two-point
correlation functions and including two excited states in the fitting
analysis enables the ground state matrix element to be isolated more
reliably. Considering the pion mass cuts, to be conservative we take
the results of the model averages of the $\mathcal{B}$ data
sets~(where all the ensembles have $M_\pi < 400\,\MeV$) as only the
dominant mass dependent terms are included in the continuum
parametrization. Our estimates, corresponding to the (`ES=2',
$\mathcal{B}$, $Z_A^k$) results in
Fig.~\ref{fig:modelavg_overview_ga}, are listed in
Table~\ref{tab:ga_final}.

\subsection{Scalar charges\label{sec:gs}}

As there is no isovector scalar current interaction at tree-level in
the Standard Model, the scalar charges cannot be measured directly in
experiment. However, the conserved vector current (CVC) relation can
be used to estimate the charges from determinations of the up and down
quark mass difference, $\delta_m=m_u-m_d$, and the QCD contribution to
baryon mass isospin splittings, e.g.,  between the mass of the proton
and the neutron, $\Delta m^{\QCD}_N$,~(for $g_S^N$ see
Eq.~\eqref{eq:massssh} below).
Reference~\cite{Gonzalez-Alonso:2013ura} finds $g_S^N = 1.02(11)$
employing lattice estimates for $\delta_m$ and an average of lattice
and phenomenological values for $\Delta m^{\QCD}_N$, which is
consistent with the FLAG~21~\cite{Aoki:2021kgd} $N_f=2+1$ result of $g_S^N =
1.13(14)$~\cite{Harris:2019bih}.  Estimates can also be made of the
isovector scalar charges of the other octet baryons, see the
discussion in Sec.~\ref{sec:cmp}. Conversely, direct determinations of
the scalar charges can be used to predict $\delta_m$, as presented in
Sec.~\ref{sec:ud_diff}. So far, there has been only one previous study of
the hyperon scalar charges~\cite{Smail:2023eyk}.

\begin{figure}
  \begin{center}
    \includegraphics[width=0.48\textwidth]{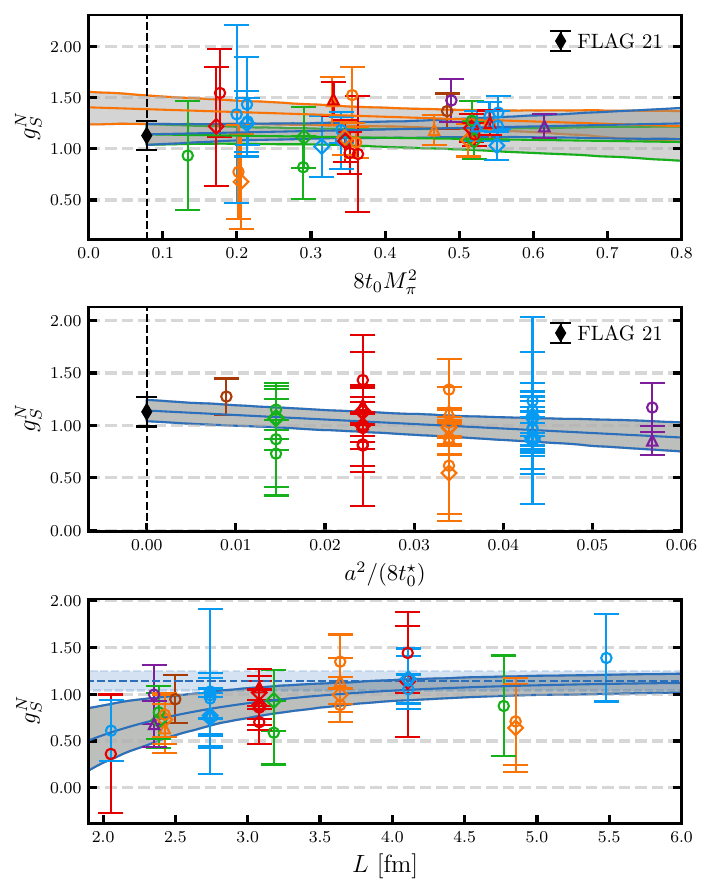}
    \caption{The same as Fig.~\ref{fig:extrapol_ga_nucleon} for the
      nucleon scalar charge~$g_S^N$. The factors $Z_S^1$ are used for
      the matching~(see Table~\ref{tab:ZtabXVI}). For orientation, the
      FLAG~21 result for $N_f=2+1$~\cite{Harris:2019bih} is
      indicated~(black diamond) at the physical point.  For better
      visibility, the data points for ensembles D150, E250 and
      D452, which have relatively large errors~(see
      Table~\ref{tab:data_nucleon}), are not displayed.
      \label{fig:extrapol_gs_nucleon}}
  \end{center}
\end{figure}

For the extrapolation of the scalar charges and the extraction of the
value at the physical point, we follow the same procedures as for the
axial channel, presented in the previous subsection. The five
parameter fit~(with coefficients $\{c_0, c_\pi, c_K, c_V, c_a\}$) can
again account for the observed quark mass, lattice spacing and volume
dependence as illustrated in Fig.~\ref{fig:extrapol_gs_nucleon} for
the nucleon~(with $\chi^2/N_{\text{dof}}=0.56$) and
Fig.~\ref{fig:extrapol_gs_hyperon} of Appendix~\ref{sec:plots} for the
sigma and cascade baryons~(with $\chi^2/N_{\text{dof}}=0.97$ and 1.14,
respectively). The data are extracted using two excited states in the
fitting analysis. For both hyperons, the quark mass and lattice
spacing effects can be resolved, in contrast to the nucleon, while for
all baryons the dependence on the spatial volume is marginal. When
investigating the systematics in the estimates of the charges at the
physical point, we perform model averages of the results of
($\mathcal{A}$): 8 fits from the two fit variations~(as for the axial
case) and the four data sets, DS(0), DS($M_\pi^{\footnotesize{<400\,
    \MeV}}$), DS($a^{\footnotesize{<0.1\,\fm}}$) and
DS($LM_\pi^{\footnotesize{> 4}}$), ($\mathcal{B}$): 6 fits from the
two fit variations to the three data sets
DS($M_\pi^{\footnotesize{<400\, \MeV}}$),
DS($M_\pi^{\footnotesize{<400\, \MeV}}$,
$a^{\footnotesize{<0.1\,\fm}}$) and DS($M_\pi^{\footnotesize{<400\,
    \MeV}}$, $LM_\pi^{\footnotesize{> 4}}$). Note that a cut on the
pion mass $M_\pi < 300\,\MeV$ is not considered.  The scalar matrix
elements are generally less precise than the axial ones and utilizing
such a reduced data set leads to instabilities in the extrapolation
and spurious values of the coefficients.

\begin{table}[b]
  \caption{Results for $g_S^B$, $B\in\{N,\Sigma,\Xi\}$, obtained with two different sets of renormalization factors. The errors include the statistical and all the systematic uncertainties. \label{tab:gs_final}}
  \begin{ruledtabular}
    \def\arraystretch{1.7}
    \begin{tabular}{lccc}
      Renormalization                      & $g_S^N$               & $g_S^\Sigma$           & $g_S^\Xi$  \\\hline 
      $Z_S^{1}$ (Table~\ref{tab:ZtabXVI})  & $1.11^{(14)}_{(16)}$ & $3.98^{(22)}_{(24)}$ & $2.57^{(11)}_{(11)}$ \\
      $Z_S^{2}$ (Table~\ref{tab:ZtabXIII}) & $1.12^{(14)}_{(17)}$ & $4.00^{(23)}_{(24)}$ & $2.57^{(11)}_{(11)}$ \\
    \end{tabular}
  \end{ruledtabular}
\end{table}

\begin{figure*}
  \begin{center}
    \includegraphics[width=0.3\textwidth]{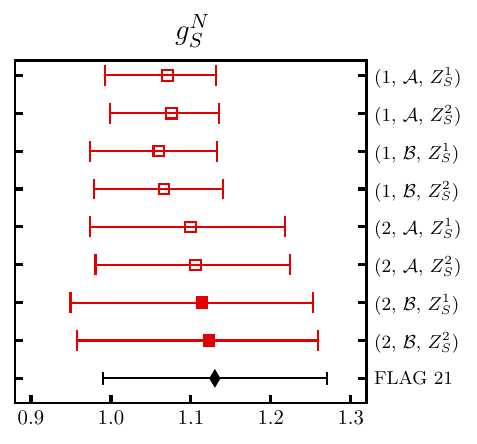}
    \includegraphics[width=0.3\textwidth]{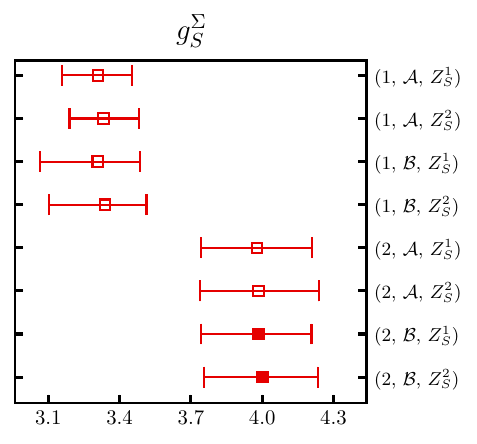}
    \includegraphics[width=0.3\textwidth]{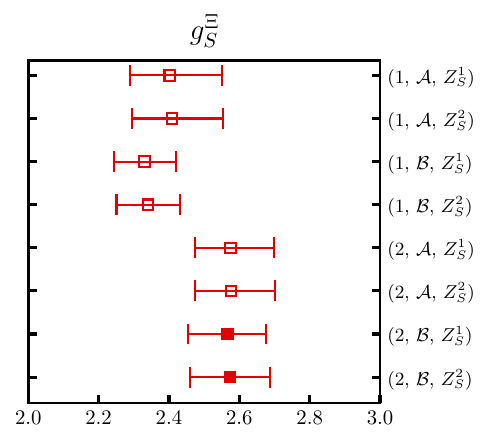}
    \caption{The same as Fig.~\ref{fig:modelavg_overview_gv} for the
      nucleon, sigma and cascade scalar charges. The label
      $\mathcal{A}$ indicates that 8 fits enter the model average
      corresponding to two fit variations, see the text, applied to 4
      data sets, DS(0), DS($M_\pi^{\footnotesize{<400\, \MeV}}$),
      DS($a^{\footnotesize{<0.1\,\fm}}$) and
      DS($LM_\pi^{\footnotesize{> 4}}$). For the data points labelled
      with $\mathcal{B}$, the two fit variations are performed on 3
      data sets, DS($M_\pi^{\footnotesize{<400\, \MeV}}$),
      DS($M_\pi^{\footnotesize{<400\, \MeV}}$,
      $a^{\footnotesize{<0.1\,\fm}}$) and
      DS($M_\pi^{\footnotesize{<400\, \MeV}}$,
      $LM_\pi^{\footnotesize{> 4}}$), giving a total of 6 fits for the
      average. For the nucleon, the FLAG~21 result for
      $N_f=2+1$~\cite{Harris:2019bih} is also shown~(black
      diamond). \label{fig:modelavg_overview_gs}}
  \end{center}
\end{figure*}

For illustration, the values from the individual fits and the model
averages over the $\mathcal{B}$ data sets for the two different
determinations of the renormalization factors are given in
Fig.~\ref{fig:modelavg_gs} in Appendix~\ref{sec:plots}. The results are
consistent across the different fits, although the weights vary. The
values of the scalar charges for all the model averages performed are
compiled in Fig.~\ref{fig:modelavg_overview_gs}. There are no
significant variations in the results obtained using the different
renormalization factors and data sets~($\mathcal{A}$ or
$\mathcal{B}$). For the nucleon, there is also agreement between the
values for the data extracted including one~(`ES=1') or two~(`ES=2')
excited states in the fitting analysis and consistency with the
current FLAG~21 result. For the sigma baryon, and to a lesser extent for
the cascade baryon, there is a tension between the `ES=1' and `ES=2'
determinations. As discussed previously, the (`ES=2', $\mathcal{B}$,
$Z_S^k$) values are considered the most reliable. These are listed in
Table~\ref{tab:gs_final}.

\subsection{Tensor charges\label{sec:gt}}

\begin{figure}
  \begin{center}
    \includegraphics[width=0.48\textwidth]{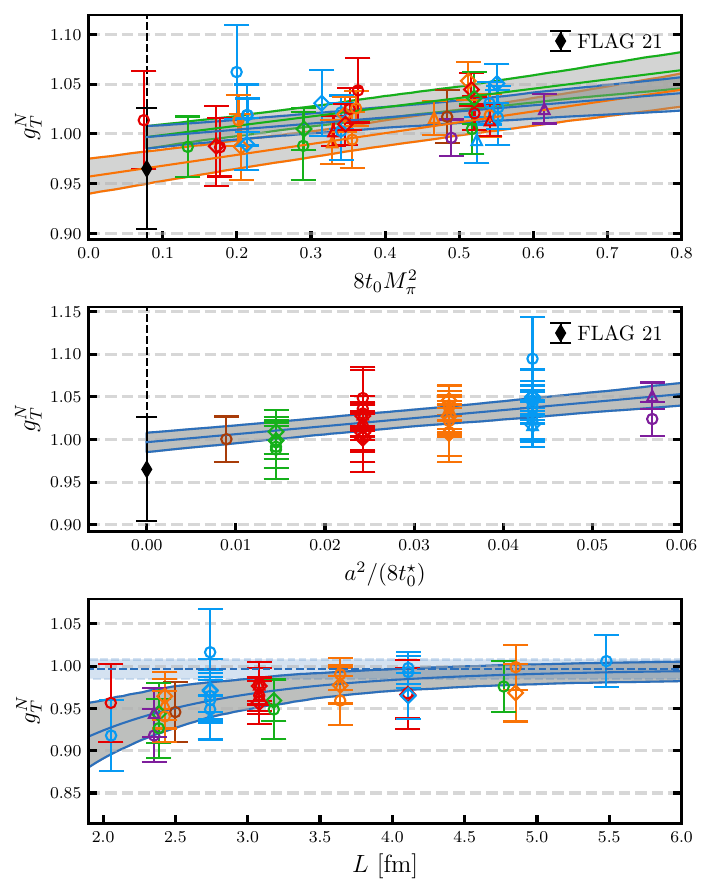}
    \caption{The same as Fig.~\ref{fig:extrapol_ga_nucleon} for the
      nucleon tensor charge~$g_T^N$.  The factors $Z_T^1$ are used for
      the matching~(see Table~\ref{tab:ZtabXVI}). For orientation, the
      FLAG~21 result for $N_f=2+1$~\cite{Harris:2019bih} is
      indicated~(black diamond) at the physical point.  For better
      visibility, the data points for ensembles D150 and D452,
      which have relatively large errors~(see
      Table~\ref{tab:data_nucleon}), are not displayed.
      \label{fig:extrapol_gt_nucleon}}
  \end{center}
\end{figure}
\begin{figure*}
  \begin{center}
    \includegraphics[width=0.3\textwidth]{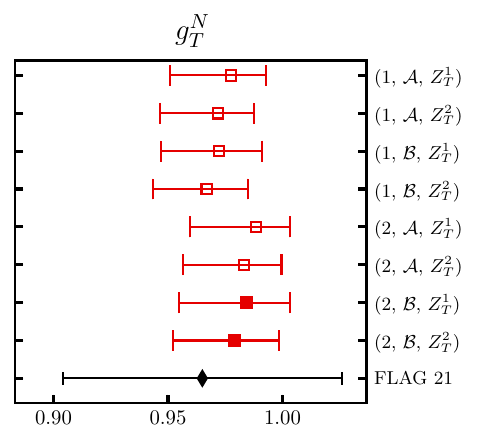}
    \includegraphics[width=0.3\textwidth]{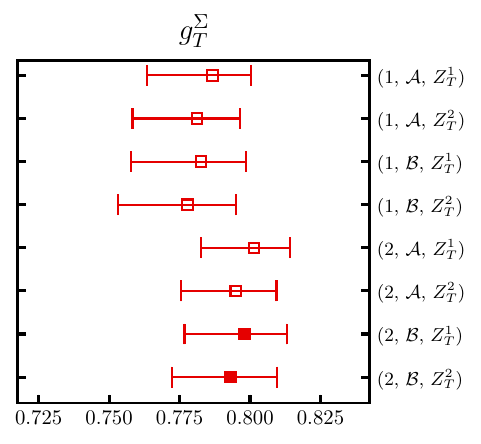}
    \includegraphics[width=0.3\textwidth]{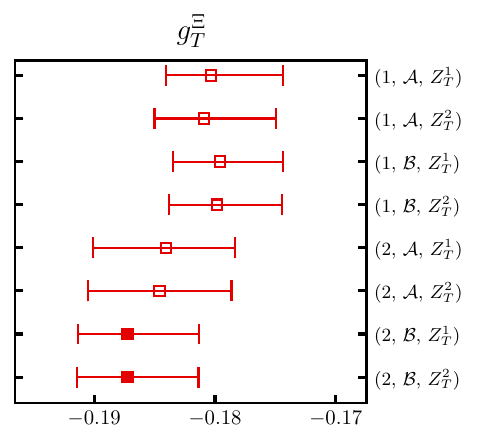}
    \caption{The same as Fig.~\ref{fig:modelavg_overview_gv} for the
      nucleon, sigma and cascade tensor charges.  For the nucleon, the FLAG~21 result for
      $N_f=2+1$~\cite{Harris:2019bih} is also shown~(black
      diamond).  \label{fig:modelavg_overview_gt}}
  \end{center}
\end{figure*}

In the isosymmetric limit, the nucleon tensor charge is equal to the first
moment of the nucleon isovector transversity parton distribution
function. Due to the lack of experimental data, estimates of
$g_T^N$ from phenomenological fits have very large uncertainties,
unless some assumptions are made. In fact, in some analyses, the fit
is constrained to reproduce the lattice results for the isovector
charge, see Refs.~\cite{Lin:2017stx,Gamberg:2022kdb}. The FLAG~21
review~\cite{Aoki:2021kgd} gives as the $N_f = 2+1$ value for the
nucleon tensor charge the result of Ref.~\cite{Harris:2019bih}, $g_T^N
= 0.965(61)$, whereas, as far as we know, there is only one previous study of
the hyperon tensor charges~\cite{Smail:2023eyk}.

The extraction of the octet baryon tensor charges at the physical
point follows the analysis of the axial charges in
Sec.~\ref{sec:ga}. In particular, the parametrizations employed and
the data sets considered are the
same. Figure~\ref{fig:extrapol_gt_nucleon} displays a typical example
of an extrapolation for the nucleon tensor charge for a five parameter
fit with a $\chi^2/N_{\text{dof}}=0.63$. See
Fig.~\ref{fig:extrapol_gt_hyperon} in Appendix~\ref{sec:plots} for the
analogous figures for the sigma and cascade baryons. The variation of
the fits with the parametrization and the data sets
utilized and the corresponding model averages, for the data sets with
pion mass cut $\mathcal{B}$~(see Sec.~\ref{sec:ga}), are shown in
Fig.~\ref{fig:modelavg_gt}.

\begin{table}
  \caption{Results for $g_T^B$, $B\in\{N,\Sigma,\Xi\}$, obtained with two different sets of renormalization factors. The errors include the statistical and all the systematic uncertainties. \label{tab:gt_final}}
  \begin{ruledtabular}
    \def\arraystretch{1.7}
    \begin{tabular}{lccc}
      Renormalization                      & $g_T^N$               & $g_T^\Sigma$           & $g_T^\Xi$  \\\hline 
      $Z_T^{1}$ (Table~\ref{tab:ZtabXVI})  & $0.984^{(19)}_{(29)}$ & $0.798^{(15)}_{(21)}$ & $-0.1872^{(59)}_{(41)}$ \\
      $Z_T^{2}$ (Table~\ref{tab:ZtabXIII}) & $0.979^{(19)}_{(27)}$ & $0.793^{(17)}_{(21)}$ & $-0.1872^{(59)}_{(42)}$ \\
    \end{tabular}
  \end{ruledtabular}
\end{table}

An overview of the model averaged results for all variations of the
input data is given in Fig.~\ref{fig:modelavg_overview_gt}. The
agreement between the different determinations suggests the
systematics associated with the extrapolation are under
control. Although the results utilizing data extracted with two
excited states~(`ES=2') in the fitting analysis are consistently above
or below those extracted from the `ES=1' data, considering the size of
the errors of the model averages~(which combine the statistical and
systematic uncertainties), the differences are not significant.  Our
estimates for the tensor charges, corresponding to the (`ES=2',
$\mathcal{B}$, $Z_T^k$) values, are listed in Table~\ref{tab:gt_final}.

\section{Discussion of the results\label{sec:discussion}}

Our values for the vector, axial, scalar and tensor charges of the
nucleon, sigma and cascade baryons are given in
Tables~\ref{tab:gv_final}, \ref{tab:ga_final}, \ref{tab:gs_final}
and~\ref{tab:gt_final}, respectively. In each case, we take the most
precise value as our final result, i.e., the one obtained using $Z_V^3$ and
$Z_A^3$ for the vector and axial channels, respectively, and $Z_S^1$ and $Z_T^1$ for
the scalar and the tensor.  In the following we compare with previous
determinations of the charges taken from the literature and discuss
the SU(3) flavour symmetry breaking effects in the different channels.
We use the conserved vector current relation and our result for the
scalar charge of the sigma baryon to determine the up and down quark
mass difference. We compute the QCD contributions to baryon isospin
mass splittings and evaluate the isospin breaking effects on the
pion baryon $\sigma$~terms.

\subsection{Individual charges\label{sec:cmp}}

\begin{figure}
  \centering
  \includegraphics[width=0.48\textwidth]{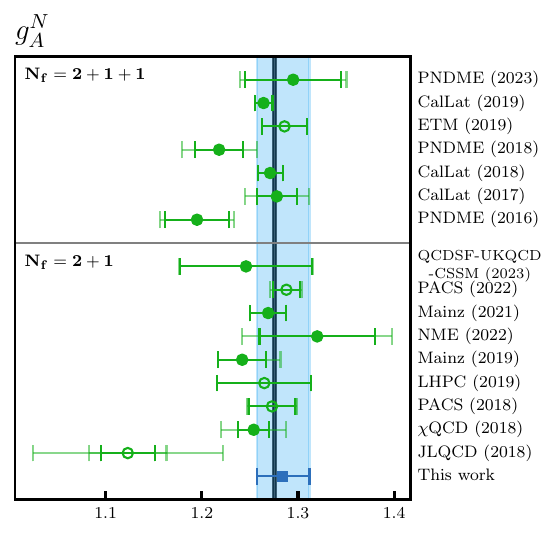}
  \caption{Compilation of recent lattice determinations of the nucleon
    axial charge $g_A^N$ with $N_f =
    2+1$~\cite{Yamanaka:2018uud,Liang:2018pis,Shintani:2018ozy,Hasan:2019noy,Harris:2019bih,Park:2021ypf,Ottnad:2021tlx,Tsuji:2022ric,Smail:2023eyk}
    and
    $N_f=2+1+1$~\cite{Bhattacharya:2016zcn,Berkowitz:2017gql,Chang:2018uxx,Gupta:2018qil,Alexandrou:2019brg,Walker-Loud:2019cif,Jang:2023zts}
    dynamical fermions. Values with filled symbols were obtained via a
    chiral, continuum and finite volume extrapolation.  The vertical
    black line gives the experimental
    result~\cite{Workman:2022ynf}.\label{fig:cmp_nucleon_ga}}
\end{figure}

We first consider the axial charges. Our final values read
\begin{align}
  g_A^N = 1.284^{(28)}_{(27)},\quad
  g_A^\Sigma = 0.875^{(30)}_{(39)},\quad
  g_A^\Xi = -0.267^{(13)}_{(12)}.
\end{align}
The result for the nucleon compares favourably with the experimental
value $g_A^N / g_V^N = 1.2754(13)$~\cite{Workman:2022ynf} and the
FLAG~21~\cite{Aoki:2021kgd} average for $N_f=2+1$,
$g_A^N=1.248(23)$. The latter is based on the determinations in
Refs.~\cite{Liang:2018pis,Harris:2019bih}. All sources of systematic
uncertainty must be reasonably under control to be included in the FLAG
average and a number of more recent studies incorporate continuum,
quark mass and finite volume extrapolations. A compilation of results
for $g_A^N$ is displayed in Fig.~\ref{fig:cmp_nucleon_ga}. Although
the determinations are separated in terms of the number of dynamical
fermions employed, including charm quarks in the sea is not expected
to lead to a discernible effect.

Regarding the hyperon axial charges, far fewer works exist. Lin
\textit{et al.}~\cite{Lin:2007gv,Lin:2007ap} performed the first
study, utilizing $N_f=2+1$ ensembles with pion masses ranging between
$350\,\MeV$ and $750\,\MeV$ and a single lattice spacing of
$0.12\,\fm$. After an extrapolation to the physical pion mass they
obtain $g_A^\Sigma = 0.900(42)_{\text{stat}}(54)_{\text{sys}}$ and
$g_A^\Xi = -0.277(15)_{\text{stat}}(19)_{\text{sys}}$, where estimates
of finite volume and discretization effects are included in the
systematic uncertainty. Note that we have multiplied their result for
$g_A^{\Sigma}$ by a factor of two to match our normalization
convention.  In Refs.~\cite{Alexandrou:2014vya,Alexandrou:2016xok}
ETMC determined all octet and decuplet (i.e., nucleon, hyperon and
$\Delta$) axial couplings employing $N_f = 2 + 1 + 1$ ensembles with
pion masses between $210\,\MeV$ and $430\,\MeV$ and two lattice
spacings $a \in \{0.065\,\fm, 0.082\,\fm\}$. Using a simple linear
ansatz for the quark mass extrapolation, they quote $g_A^\Sigma =
0.7629(218)_{\text{stat}}$ and $g_A^\Xi = -0.2479(87)_{\text{stat}}$,
where the errors are purely statistical.

More recently, Savanur \textit{et al.}~\cite{Savanur:2018jrb}
extracted the axial charges on $N_f = 2 + 1 + 1$ ensembles with three
different lattice spacings $a \in \{0.06\,\fm, 0.09\,\fm,
0.12\,\fm\}$, pion masses between $135\,\MeV$ and $310\,\MeV$ and
volumes in the range $3.3 \leq LM_\pi \leq 5.5$. The ratios
$g_A^\Sigma/g_A^N$ and $g_A^\Xi/g_A^N$ are extrapolated taking the
quark mass dependence and lattice spacing and finite volume effects
into account. The experimental value of $g^N_A$ is then used to obtain
$g_A^\Sigma = 0.891(11)_{\text{stat}}(13)_{\text{sys}}$ (again
multiplied by a factor of two to meet our conventions) and $g_A^\Xi =
-0.2703(47)_{\text{stat}}(13)_{\text{sys}}$.  Finally,
QCDSF-UKQCD-CSSM presented results for the isovector axial, scalar and
tensor charges in Ref.~\cite{Smail:2023eyk}. They employ $N_f=2+1$ ensembles lying
on a $\tr M = \const$ trajectory with pion masses ranging between
$220\,\MeV$ and $470\,\MeV$ and five different values of the lattice
spacing in the range $(0.052 - 0.082)\,\fm$. The Feynman-Hellmann theorem is used to calculate the baryon
matrix elements. Performing an extrapolation to the physical mass
point including lattice spacing and finite volume effects, they find
$g_A^\Sigma = 0.876(26)_{\text{stat}}(09)_{\text{sys}}$ and $g_A^\Xi =
-0.206(22)_{\text{stat}}(19)_{\text{sys}}$.

We also mention the earlier studies of Erkol \textit{et
  al.}~\cite{Erkol:2009ev}~($N_f=2$), utilizing pion masses above
$500\,\MeV$, and QCDSF-UKQCD~($N_f=2+1$) carried out at a single lattice
spacing~\cite{Gockeler:2011ze}.

\begin{figure}
  \centering
  \includegraphics[width=0.48\textwidth]{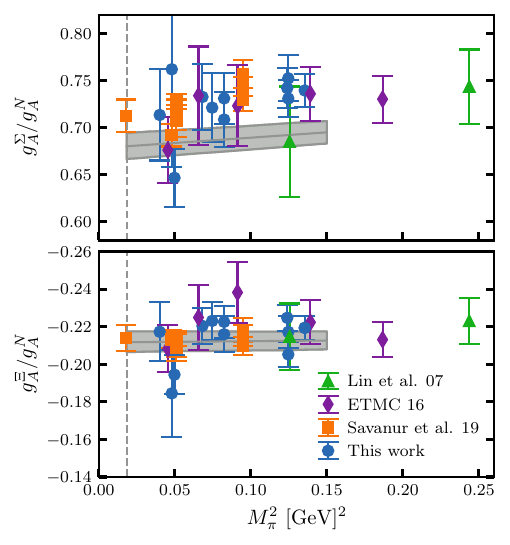}
  \caption{Comparison of lattice determinations~\cite{Lin:2007ap,Alexandrou:2016xok,Savanur:2018jrb}
    of the hyperon axial
    charges for the $\Sigma$ and $\Xi$ baryons normalized to the
    nucleon axial charge. Some of the results from Lin \textit{et
      al.}\ at heavier pion masses are not shown.  The grey vertical
    dashed line indicates the physical pion mass point. Only our
    results from the $m_s=\const$ trajectory are shown.  The grey
    bands indicate the (continuum limit, infinite volume) quark mass
    behaviour according to the fits displayed in
    Figs.~\ref{fig:extrapol_ga_nucleon}
    and~\ref{fig:extrapol_ga_hyperons}. Note that the data points are
    not corrected for finite volume or discretization effects. All
    data are converted to our phase and normalization conventions, Eqs.~\eqref{eq:decomp}--\eqref{eq:ccc3}. The Lin \textit{et al.}\ and 
    ETMC results are obtained by taking the ratio of the individual
    charges and employing error propagation.
    \label{fig:ga_cmp}}
\end{figure}

\begin{figure}
  \centering
  \includegraphics[width=0.48\textwidth]{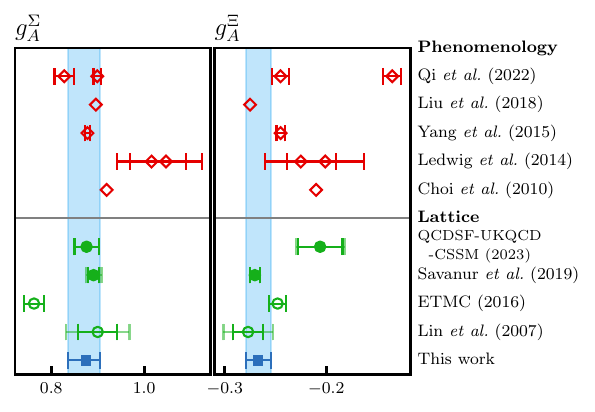}
  \caption{Comparison of our results for the axial charges
    $g_A^\Sigma$ and $g_A^\Xi$ (blue symbols and error bands) with
    other lattice
    determinations~\cite{Lin:2007ap,Alexandrou:2016xok,Savanur:2018jrb,Smail:2023eyk}
    and phenomenological
    estimates~\cite{Choi:2010ty,Ledwig:2014rfa,Yang:2015era,Liu:2018jiu,Qi:2022sus}.
    Values with filled symbols were obtained via a chiral, continuum
    and finite volume extrapolation. All results are converted to our phase and normalization conventions, \mbox{Eqs.~\eqref{eq:decomp}--\eqref{eq:ccc3}}. \label{fig:cmp_ga}}
\end{figure}

In Fig.~\ref{fig:ga_cmp} we compare the ratios of the hyperon axial
charges to the nucleon axial charge, $g_A^B/g_A^N$, from
Refs.~\cite{Lin:2007ap,Alexandrou:2016xok,Savanur:2018jrb}, obtained on
individual ensembles to our results. A comparison of the charges
themselves cannot be made since, as mentioned above, Savanur
\textit{et al.} only present results for the ratio. As the strange
quark mass is held approximately constant in these works, only our
results from the $m_s = \const$ trajectory are displayed. Similarly, the
QCDSF-UKQCD-CSSM values are omitted as the ensembles utilized lie on
a  $\tr M = \const$ trajectory. We observe reasonable
agreement between the data.  Note that our continuum, infinite volume
limit result (the grey band in the figure) for $g_A^\Sigma/g_A^N$ lies
slightly below the central values of most of our $m_s=\const$ data
points.

The individual hyperon axial charges at the physical point are shown
in Fig.~\ref{fig:cmp_ga}, along with a number of phenomenological
determinations employing a variety of quark
models~\cite{Choi:2010ty,Liu:2018jiu,Qi:2022sus}, the chiral soliton
model~\cite{Yang:2015era} and SU(3) covariant baryon
ChPT~\cite{Ledwig:2014rfa}. Within errors, the lattice results are
consistent apart from the rather low value for $g_A^\Sigma$ from
ETMC~\cite{Alexandrou:2016xok} and the rather high value for $g_A^\Xi$
from QCDSF-UKQCD-CSSM~\cite{Smail:2023eyk}. The phenomenological
estimates for $g_A^\Sigma$ are in reasonable agreement with our value,
while there is a large spread in the expectations for $g_A^\Xi$.

We remark that, in analogy to the CVC relation~(discussed in
Sec.~\ref{sec:ud_diff} below), the axial Ward identity, $\partial_\mu
(\bar{u}\gamma_\mu \gamma_5 d) = i (m_d + m_u) \bar{u}\gamma_5 d$,
connects the axial and pseudoscalar charges,
\begin{align} g^B_P = \frac{m_B}{m_{\ell}} g_A^B,\end{align}
where $m_B$ and $m_{\ell}$ correspond to the baryon and the light
quark mass, respectively. This relation was employed in
Ref.~\cite{Gonzalez-Alonso:2013ura} to determine the pseudoscalar
charge of the nucleon, which is defined as the pseudoscalar form
factor in the forward limit.  Taking the baryon masses of
isosymmetric QCD from Table~14 of Ref.~\cite{RQCD:2022xux} and the
isospin averaged light quark mass $m_{\ell} = 3.381(40)\,\MeV$
in the $N_f=4$ flavour $\MS$ scheme at $\mu=2\,\GeV$ from
the FLAG~21 review~\cite{Aoki:2021kgd}, we
find
\begin{align}
  g_{P,N_f=4}^N &= 356^{(9)}_{(9)},\quad
  g_{P,N_f=4}^\Sigma =  308^{(11)}_{(14)}, \nonumber\\
  g_{P,N_f=4}^\Xi &=  -104^{(5)}_{(5)}.
\end{align}

\begin{figure}
  \centering
  \includegraphics[width=0.465\textwidth]{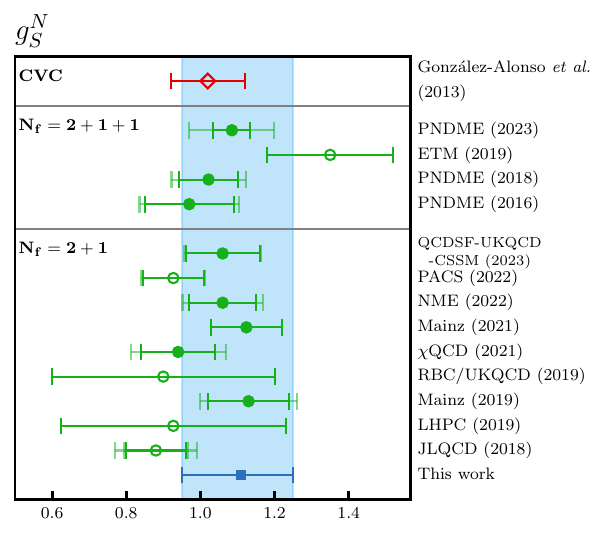}
  \caption{As in Fig.~\ref{fig:cmp_nucleon_ga} for the nucleon scalar
    charge $g_S^N$ with $N_f =
    2+1$~\cite{Yamanaka:2018uud,Hasan:2019noy,Harris:2019bih,Abramczyk:2019fnf,Liu:2021irg,Ottnad:2021tlx,Park:2021ypf,Tsuji:2022ric,Smail:2023eyk}
    and
    $N_f=2+1+1$~\cite{Bhattacharya:2016zcn,Gupta:2018qil,Alexandrou:2019brg,Jang:2023zts}
    dynamical fermions. González-Alonso~\textit{et~al.}\ estimate the
    scalar charge via the conserved vector current (CVC)
    relation~\cite{Gonzalez-Alonso:2013ura}.\label{fig:cmp_nucleon_gs}}
\end{figure}

\begin{figure}
  \centering
  \includegraphics[width=0.465\textwidth]{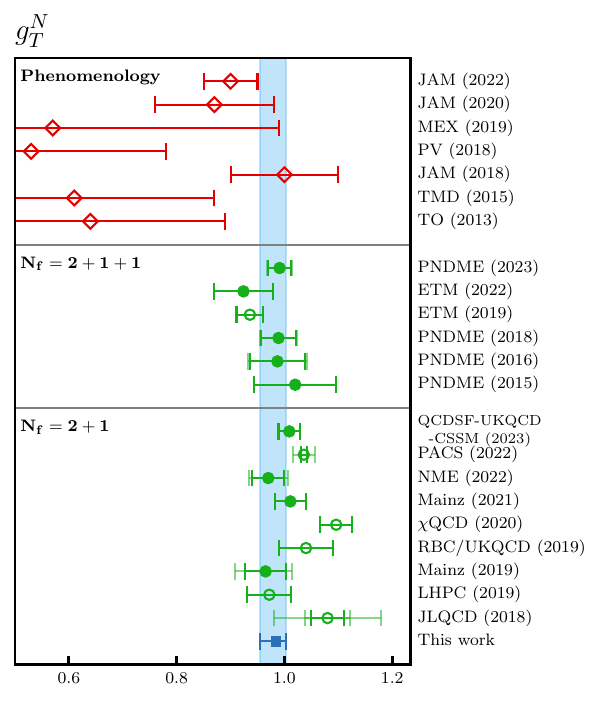}
  \caption{As in Fig.~\ref{fig:cmp_nucleon_ga} for the nucleon tensor
    charge $g_T^N$ with $N_f =
    2+1$~\cite{Yamanaka:2018uud,Hasan:2019noy,Harris:2019bih,Abramczyk:2019fnf,Horkel:2020hpi,Ottnad:2021tlx,Park:2021ypf,Tsuji:2022ric,Smail:2023eyk}
    and
    $N_f=2+1+1$~\cite{Bhattacharya:2015wna,Bhattacharya:2016zcn,Gupta:2018qil,Alexandrou:2019brg,Alexandrou:2022dtc,Jang:2023zts}
    dynamical fermions. Recent phenomenological estimates are also
    displayed for
    comparison~\cite{Anselmino:2013vqa,Kang:2015msa,Lin:2017stx,Radici:2018iag,Benel:2019mcq,Cammarota:2020qcw,Gamberg:2022kdb}.
    Values with filled symbols were obtained via a quark mass, continuum and 
    finite volume extrapolation. In addition, the filled ETM (2022)~\cite{Alexandrou:2022dtc} 
    point is obtained from a continuum limit extrapolation of results 
    determined on three physical point ensembles with large spatial volumes.
    \label{fig:cmp_nucleon_gt}}
\end{figure}

Turning to the scalar charges, our final results in the three flavour
$\MS$ scheme at $\mu=2\,\GeV$ read\footnote{
Using Version~3 of {\sc RunDec}~\cite{Herren:2017osy}, we compute the
conversion factor from $N_f=3$ to $N_f=4$:
$1.00082(2)_{\Lambda}(1)_{\text{pert}}(56)_{m_c}=1.0008(6)$.  The
errors reflect the uncertainty of the
$\Lambda$-parameter~\cite{Bruno:2017gxd}, the difference between
5-loop running~\cite{Baikov:2016tgj,Baikov:2017ujl}/4-loop
decoupling~\cite{Schroder:2005hy,Liu:2015fxa,Marquard:2016dcn} and
4-loop running/3-loop decoupling and a $200\,\MeV$ uncertainty in the
charm quark on-shell mass, respectively: at $\mu=2\,\GeV$ there is no
noteworthy difference between $N_f=3$ and $N_f=4$ $\MS$ pseudo(scalar)
charges.}
\begin{align}
  g_S^N = 1.11^{(14)}_{(16)},\quad
  g_S^\Sigma=  3.98^{(22)}_{(24)},\quad
  g_S^\Xi =  2.57^{(11)}_{(11)}.
\end{align}
For the nucleon, our result for $g_S^N$ agrees with the FLAG~21 value
$g_S^N = 1.13(14)$ for $N_f=2+1$~\cite{Aoki:2021kgd}~(taken from
Ref.~\cite{Harris:2019bih}) and more recent lattice determinations,
see Fig.~\ref{fig:cmp_nucleon_gs}. There is only one previous lattice
determination of the hyperon scalar couplings by
QCDSF-UKQCD-CSSM~\cite{Smail:2023eyk}, who obtain $g_S^\Sigma =
2.80(24)_{\text{stat}}(05)_{\text{sys}}$ and $g_S^\Xi =
1.59(11)_{\text{stat}}(04)_{\text{sys}}$. These values are much smaller than
ours.

One can also employ the CVC relation and estimates of the QCD
contribution to the isospin mass splittings and the light quark mass
difference to determine the scalar charges. For a detailed discussion
see Sec.~\ref{sec:ud_diff} below. Ref.~\cite{Gonzalez-Alonso:2013ura} obtains
$g_S^N = 1.02(11)$ assuming $\Delta
m_N^{\QCD}=m^{\QCD}_p-m^{\QCD}_n=-2.58(18)\,\MeV$ and the quark mass
difference $\delta_m=m_u-m_d=-2.52(19)\,\MeV$. Similarly, using the
results by BMWc on the light quark mass splitting~\cite{Fodor:2016bgu}
and their QCD contributions to the baryon mass
splittings~\cite{Borsanyi:2014jba}, we obtain
\begin{align}
  g_S^N = 1.05(13),\quad
  g_S^\Sigma= 3.35(19),\quad
  g_S^\Xi = 2.29(15),
\end{align}
which agree with our results to within two standard deviations. Note
that a smaller value for $|\delta_m|$~(see Sec.~\ref{sec:ud_diff})
would uniformly increase these charges.

\begin{figure}
  \centering
  \includegraphics[width=0.48\textwidth]{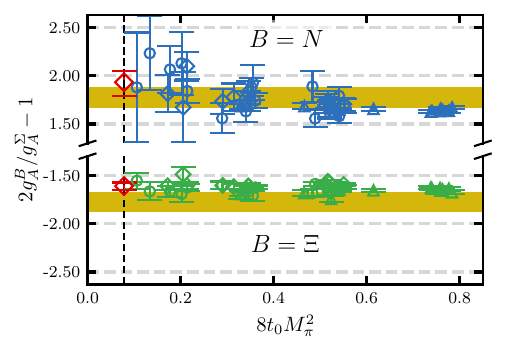}
  \caption{The ratios $2g_A^B / g_A^\Sigma - 1$ for $B\in\{N,\Xi\}$ as
    a function of the pion mass squared. The latter are rescaled with
    the Wilson flow scale $t_0$.  The red diamonds are the continuum
    and infinite volume limit results at the physical point~(indicated
    by the dashed vertical line) obtained from our extrapolations of
    the individual charges. The yellow bands depict the corresponding
    $m_s=m_{\ell}=0$ predictions $\pm D/F$.  Circles (diamonds)
    correspond to the $\tr M=\const$ ($m_s=\const$) trajectories, the
    triangles to the $m_s=m_{\ell}$ line.  The data are not corrected
    for lattice spacing or volume
    effects.\label{fig:hyperons_ga_ratio}}
\end{figure}

\begin{figure}
  \centering
  \includegraphics[width=0.48\textwidth]{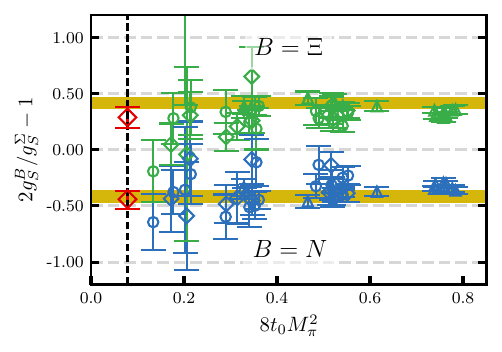}
  \caption{The same as Fig.~\ref{fig:hyperons_ga_ratio} for the scalar
    channel. The yellow bands depict the $m_s=m_{\ell}=0$ predictions
    $\pm D_S(0)/F_S(0)$ obtained from the extrapolations of the
    individual charges.\label{fig:hyperons_gs_ratio}}
\end{figure}
\begin{figure}
  \centering
  \includegraphics[width=0.48\textwidth]{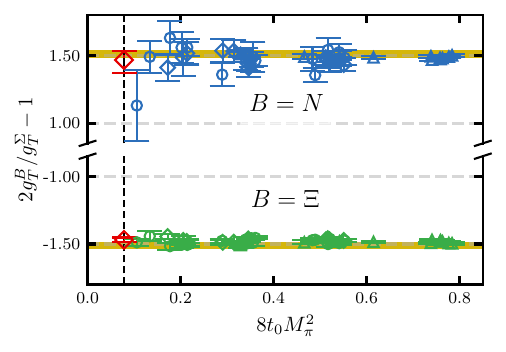}
  \caption{The same as Fig.~\ref{fig:hyperons_ga_ratio} for the tensor
    channel.\label{fig:hyperons_gt_ratio}}
\end{figure}

Regarding the tensor charges we find in the $N_f=3$ $\MS$ scheme at $\mu=2\,\GeV$
\begin{align}
  g_T^N = 0.984^{(19)}_{(29)},\quad
  g_T^\Sigma =  0.798^{(15)}_{(21)},\quad
  g_T^\Xi =  -0.1872^{(59)}_{(41)}. \label{eq:gtfinal}
\end{align}
Since the anomalous dimension of the tensor bilinear is smaller than
for the scalar case, we would expect no statistically relevant
difference between the $N_f=3$ and $N_f=4$ schemes at
$\mu=2\,\GeV$. The nucleon charge agrees with the
FLAG~21~\cite{Aoki:2021kgd} value of $g_T^N =
0.965(61)$~\cite{Harris:2019bih} for $N_f = 2+1$ and other recent
lattice studies. These are shown in Fig.~\ref{fig:cmp_nucleon_gt} along with
 determinations from phenomenology. The large uncertainties of the latter
reflect the lack of experimental data. In particular, in
Refs.~\cite{Lin:2017stx,Gamberg:2022kdb} the JAM collaboration
constrain the first Mellin moment of the isovector combination of the
transverse parton distribution functions to reproduce a lattice result
for $g_T^N$. QCDSF-UKQCD-CSSM also determined the hyperon tensor
charges~\cite{Smail:2023eyk}. Their results $g_T^\Sigma =
0.805(15)_{\text{stat}}(02)_{\text{sys}}$ and $g_T^\Xi =
-0.1952(74)_{\text{stat}}(10)_{\text{sys}}$ are in good agreement with
ours.

\subsection{SU(3) flavour symmetry breaking\label{sec:su3}}

On the SU(3) flavour symmetric line, i.e., for $m=m_\ell = m_s$, the
baryon charges~$g_J^B(m)$ can be decomposed into two functions,
$F_J(m)$ and $D_J(m)$, see Sec.~\ref{sec:hyperon} and
Eqs.~\eqref{eq:symm1}--\eqref{eq:symm3}.  For the axial charges
$g_A^B$, the values of these functions in the SU(3) chiral limit
correspond to the LECs $F=F_A(0)$ and $D=D_A(0)$. We will not consider
the vector channel~($J=V$) here since $F_V(m)=1$ and $D_V(m)=0$, which
holds even at $m_s\neq m_{\ell}$, due to charge conservation.

Estimates of baryon structure observables often rely on SU(3) flavour
symmetry arguments, however, it is not known {\em a priori} to what
extent this symmetry is broken for $m_s \neq m_\ell$ and, in
particular, at the physical point. Since within this analysis, we only
determined three isovector charges ($B\in\{N,\Sigma,\Xi\}$) for each
channel ($J\in\{A,S,T\}$), we cannot follow the systematic approach to
investigate SU(3) flavour symmetry breaking of matrix elements
proposed in Ref.~\cite{Bickerton:2019nyz}. Nevertheless, constructing
appropriate ratios from the individual charges will provide us with
estimates of the flavour symmetry breaking effects for each channel.

Using Eqs.~\eqref{eq:symm1}--\eqref{eq:symm3}, we obtain for
$m=m_s=m_{\ell}$
\begin{equation}
  \frac{2g_J^B(m)}{g_J^\Sigma(m)} - 1 = \pm \frac{D_J(m)}{F_J(m)},\label{eq:foverd}
\end{equation}
where `$+$' and `$-$' corresponds to $B = N$ and $B = \Xi$,
respectively.  Figure~\ref{fig:hyperons_ga_ratio} shows these
combinations for the axial charges, as functions of the squared pion
mass, compared to the chiral, continuum limit expectations $\pm D/F$
(yellow bands) determined from our global fit (see
Fig.~\ref{fig:extrapol_ga_hyperons} and Table~\ref{tab:ratios_final}).
The chiral limit value agrees with our earlier result
$D/F=1.641^{(27)}_{(44)}$~\cite{Bali:2022qja} (model averaging recomputed
for $D/F$ instead of $F/D$) within 1.5 standard errors.
The data shown in the figure are not corrected for volume or lattice
spacing effects. Note that the renormalization factors and improvement
coefficients and, possibly, other systematics cancel from
Eq.~\eqref{eq:foverd}.  For the ratio of the $\Xi$ over the $\Sigma$
axial charge we see no significant difference between the physical
point value and that obtained for the same average quark mass at the
flavour symmetric point. The symmetry breaking effect of the
combination involving $g_A^N/g_A^{\Sigma}$ can be attributed to the
pion mass dependence of $g_A^N$, see
Fig.~\ref{fig:extrapol_ga_nucleon}. The red symbols at the physical
point (dashed vertical line) correspond to our continuum, infinite
volume limit extrapolated results, listed in
Table~\ref{tab:ratios_final} for the combinations
Eq.~\eqref{eq:foverd}.

\begin{table}
  \caption{The combinations $2g_J^{B}/g_J^{\Sigma}-1$ of
    Eq.~\eqref{eq:foverd} with $B\in\{N,\Xi\}$ at the physical point,
    in the continuum and infinite volume limit. These are computed
    from the individual charges in Tables~\ref{tab:ga_final} (for
    $Z_A^3$),~\ref{tab:gs_final} and~\ref{tab:gt_final} (for $Z_J^1$).
    The last row gives the combination $D_J(0)/F_J(0)$ in the chiral,
    continuum and infinite volume limit (yellow bands in
    Figs.~\ref{fig:hyperons_ga_ratio}--\ref{fig:hyperons_gt_ratio}),
    computed from the individual charges: $D_J/F_J = (g_J^N -
    g_J^\Xi)/g_J^\Sigma$.
\label{tab:ratios_final}}
  \begin{ruledtabular}
    \def\arraystretch{1.7}
    \begin{tabular}{llll}
      $B$   & $2g_A^{B}/g_A^{\Sigma}-1$ & $2g_S^{B}/g_S^{\Sigma}-1$ & $2g_T^{B}/g_T^{\Sigma}-1$  \\\hline 
      $N$   & $\phantom{-}1.93^{(12)}_{(15)}$  & $-0.441^{(76)}_{(89)}$  & $\phantom{-}1.467^{(67)}_{(99)}$   \\
      $\Xi$ & $-1.609^{(37)}_{(40)}$ & $\phantom{-}0.288^{(91)}_{(96)}$  & $-1.469^{(17)}_{(16)}$ \\\hline 
      $D_J(0)/F_J(0)$ & $\phantom{-}1.79^{(11)}_{(~8)} $ & $-0.416^{(46)}_{(49)} $ &  $\phantom{-}1.530^{(54)}_{(56)} $
    \end{tabular}
  \end{ruledtabular}
\end{table}

In Fig.~\ref{fig:hyperons_gs_ratio} the combinations
Eq.~\eqref{eq:foverd} are shown for the isovector scalar
charges. These are compared to our SU(3) chiral limit extrapolated
results (yellow bands) and the continuum, infinite volume limit
results at the physical point (red diamonds). We find no statistically
significant symmetry breaking in this case. However, the statistical
errors are larger than for the axial case and also
$F_S>F_A$. Therefore, we cannot exclude symmetry breaking of a similar
size as for the axial charges, in particular, in the ratio of the
$\Xi$ over the $\Sigma$ baryon charge. Finally, in
Fig.~\ref{fig:hyperons_gt_ratio} we carry out the same comparison for
the tensor charges. In this case, within errors of a few per cent, no
flavour symmetry violation is seen. Moreover,
$D_T(m)/F_T(m)=D_T(0)/F_T(0)$ within errors.

\begin{figure}
  \centering
  \includegraphics[width=0.48\textwidth]{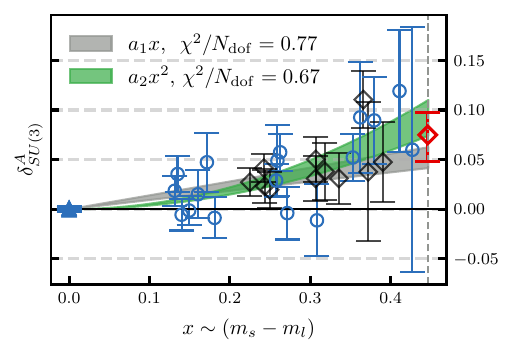}
  \caption{The SU(3) symmetry breaking ratio
    $\delta_{\text{SU(3)}}^A(x)$ (Eq.~\eqref{eq:dsu3}) for the axial
    charges as a function of $x = (M_K^2 - M_\pi^2) / (2 M_K^2 +
    M_\pi^2)$. Circles (diamonds) correspond to the $\tr M=\const$
    ($m_s=\const$) trajectories, the triangle to $m_s=m_{\ell}$.
    The grey (green) band shows the result from a linear
    (quadratic) one parameter fit including only the blue data points
    that correspond to the ensembles on the $\tr\,M =\const$ line.
    Black data points correspond to ensembles on the $m_s=\const$
    trajectory. The vertical grey dashed line indicates the physical
    point. The red diamond corresponds to the result derived from
    the values for the individual charges, see Sec.~\ref{sec:cmp}.
     \label{fig:ga_sb}}
\end{figure}

In order to quantify the symmetry breaking effect between matrix
elements involving the current $J$ as a function of the quark mass
splitting~$m_s - m_\ell$, we define
\begin{align}
  \delta_{\text{SU(3)}}^{J} = \frac{g_J^\Xi+g_J^N-g_J^\Sigma}{g_J^\Xi+g_J^N+g_J^\Sigma},\label{eq:su3}
\end{align}
where for $m_s=m_{\ell}$, $ \delta_{\text{SU(3)}}^{J} =
(2F_J-2F_J)/(2F_J+2F_J)=0$, see
Eqs.~\eqref{eq:symm1}--\eqref{eq:symm3}.
Also from these ratios some of the systematics as well as the renormalization
factors and improvement terms will cancel. We define a dimensionless
SU(3) breaking parameter $x = (M_K^2 - M_\pi^2) / (2 M_K^2 + M_\pi^2) \sim
m_s - m_{\ell}$ and assume a polynomial dependence:
\begin{align}
  \delta^J_{\text{SU(3)}} = \sum_{n>0} a^J_n \, x^n. \label{eq:dsu3}
\end{align}
The data for $\delta_{\text{SU(3)}}^{A}(x)$ depicted in
Fig.~\ref{fig:ga_sb} become more and more positive as the physical
point (vertical dashed line) is approached.  This observation agrees
with findings from earlier
studies~\cite{Lin:2007ap,Erkol:2009ev,Alexandrou:2016xok,Savanur:2018jrb,Smail:2023eyk}.
We fit to data for which the average quark mass is kept
constant (blue circles).  However, there is no significant difference
between these and the $m_s\approx\text{const}.$ points (black
squares). Both linear and quadratic fits in $x$ ($a^A_n=0$ for $n\neq 1$
and $a^A_n=0$ for $n\neq2$, respectively) give adequate descriptions of
the data and agree with our continuum, infinite volume limit extrapolated physical
point result (red diamond)
\begin{equation}
  \delta^A_{\text{SU(3)}}=0.075^{(23)}_{(27)},
\end{equation}
derived from the values for the individual charges.  Effects of this
sign and magnitude were also reported previously. ETMC~\cite{Alexandrou:2016xok} find
$g_A^N+g_A^{\Xi}-g_A^{\Sigma}=0.147(24)$, whereas Savanur and
Lin~\cite{Savanur:2018jrb} quote
$(g_A^N+g_A^\Xi-g_A^{\Sigma})/g_A^N=0.087(15)$.

\begin{figure}
  \centering
  \includegraphics[width=0.48\textwidth]{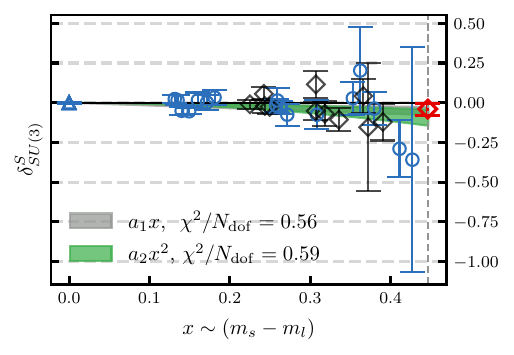}\\
  \includegraphics[width=0.48\textwidth]{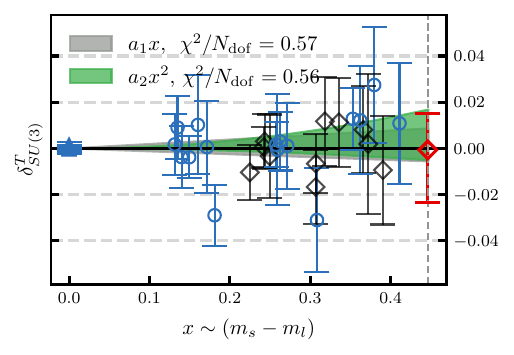}
  \caption{The same as Fig.~\ref{fig:ga_sb} for the scalar (top) and tensor (bottom) charges.\label{fig:gst_sb}}
\end{figure}

For $J\neq A$ no statistically significant effects were observed.
Nevertheless, for completeness we carry out the same analysis for $J=S$
and $J=T$, see Fig.~\ref{fig:gst_sb}. Our continuum, infinite volume limit
extrapolated physical point results
\begin{equation}
  \delta^S_{\text{SU(3)}}=-0.040^{(37)}_{(41)},\quad
  \delta^T_{\text{SU(3)}}=-0.001^{(16)}_{(23)}
\end{equation}
provide upper limits on the relative size of SU(3) flavour violation
at the physical point.

\subsection{The up and down quark mass difference\label{sec:ud_diff}}

Our results on the scalar charges, in particular, $g_S^{\Sigma}$,
enable us to determine the quark mass
splitting \mbox{$\delta_m = m_u-m_d$}. While we simulate the isosymmetric theory,
in Nature this symmetry is broken. The extent of isospin symmetry
breaking is determined by two small
parameters, $\delta_m/\Lambda_{\QCD}$ and the fine structure
constant $\alpha_{\QED}$, which are similar in size.
The vector Ward identity relates $\delta_m$ to the QCD contributions to
baryon mass splittings within an isomultiplet. In particular, to leading order
in $\delta_m/\Lambda_{\QCD}$ and $\alpha_{\QED}$,
the difference between the $\Sigma^+$ and $\Sigma^-$ baryon masses
is a pure QCD effect from which, with our knowledge of $g_S^{\Sigma}$,
we can extract $\delta_m$ without additional assumptions.

We consider isospin multiplets of baryons
$B^Q\in\{N^Q,\Sigma^Q,\Xi^Q\}$ with electric charges
$Q=I_3+\tfrac12(1+S)\in\{0,\pm 1\}$ ($N^+=p$, $N^0=n$) and define the
mass differences $\Delta m_{B^{Q+1}}=m_{B^{Q+1}}-m_{B^Q}$. Note that
for the $\Sigma$ there are two differences~\cite{Workman:2022ynf},
\begin{align}
  \Delta m_{\Sigma^+}&=m_{\Sigma^+}-m_{\Sigma^0}=-3.27(7)\,\MeV, \\
    \Delta m_{\Sigma^0}&=m_{\Sigma^0}-m_{\Sigma^-}=-4.81(4)\,\MeV.
\end{align}
The other splittings read~\cite{Workman:2022ynf}
\begin{align}
  \Delta m_{\Xi}=-6.85(21)\,\MeV, \quad \Delta m_N\approx -1.293\,\MeV.
\end{align}
The mass differences can be split into QCD ($\sim\delta_m$) and QED
($\sim\alpha_{\QED}\Lambda_{\QCD}$) contributions:
\begin{equation}\Delta m_B=\Delta m_B^{\QCD}+\Delta m_B^{\QED}.
\end{equation}
The splitting depends on the scale,
the renormalization scheme and the matching conventions between QCD
and QCD+QED. The Cottingham formula~\cite{Cottingham:1963zz} relates
the leading QED contribution to hadron masses to the total electric charge
squared times a function of the unpolarized Compton forward-amplitude,
i.e., to leading order in $\alpha_{\QED}$ the electric contribution to
charge-neutral hadron masses should vanish (as was suggested in the
massless limit by Dashen~\cite{Dashen:1969eg}).  Moreover, for
$\delta_m=0$ this implies that the leading QED contributions to the
masses of the $\Sigma^+$ and $\Sigma^-$ baryons are the
same. Therefore, up to
$\mathcal{O}(\alpha_{\QED},\delta_m/\Lambda_{\QCD})\cdot \delta_m$
terms,
\begin{align}
\Delta m_{\Sigma}^{\QCD}&=\frac12(m_{\Sigma^+}-m_{\Sigma^-})\nonumber\\
  &=-4.04(4)\,\MeV,\\
\Delta m_{\Sigma^+}^{\QED}&=\frac12(m_{\Sigma^+}+m_{\Sigma^-})-m_{\Sigma^0}\nonumber\\
&=0.77(5)\,\MeV=-\Delta m_{\Sigma^0}^{\QED}.
\end{align}
From the Ademollo-Gatto
theorem~\cite{Ademollo:1964sr} we know that the leading isospin breaking
effects on the vector charges $g_V^N=g_V^\Xi=1$
and $g_V^\Sigma=2$ are quadratic functions of $\delta_m/\Lambda_{\QCD}$
and $\alpha_{\QED}$, whereas the scalar charges
$g_S^B$ are subject to linear corrections in $\alpha_{\text{QED}}$
and $\delta_m/\Lambda_{\QCD}$.

The Lorentz decomposition of the on-shell QCD matrix element
for the isovector vector current
between baryons $B'=B^{Q+1}$ and $B=B^Q$ (that differ by $\Delta I_3=1$
in their isospin) gives (see Eq.~\eqref{eq:decomp})
\begin{align}
  i\partial^{\mu}\langle B'(p')|&\bar{d}\gamma_{\mu}u|B(p)\rangle=g_V^{B'B}i\partial^{\mu}
  \bar{u}_{B'}(p')\gamma_\mu u_{B}(p)\nonumber\\
  &=g_V^{B'B}\Delta m_B^{\QCD}[1+\mathcal{O}(\delta_m/\Lambda_{\QCD})],
\end{align}
where the leading correction is due to $q_0=p_0'-p_0=\Delta
m^{\QCD}_B=|\mathbf{q}|$. In the last step we used the equations
of motion.  Combining this with the vector Ward identity
$i\partial^{\mu}\bar{d}\gamma_{\mu}u=(m_u-m_d)\bar{d}u$ gives
\begin{equation}
  g_V^{B'B}\Delta m^{\QCD}_B=g_S^{B'B}(m_u -m_d)
  \label{eq:massssh}
\end{equation}
as the QCD contribution to the mass difference, with corrections that
are suppressed by powers of the symmetry breaking parameters. Note
that the normalization convention of the charges $g_J^B$ defined in
Eq.~\eqref{eq:decomp2}, 
\begin{equation}
  g_J^{pn}=g_J^N, \quad g_J^{\Sigma^+\Sigma^0}=-g_J^{\Sigma}/\sqrt{2}, \quad
  g_J^{\Xi^0\Xi^-}=-g_J^{\Xi},\nonumber
\end{equation}
cancels in the above equation
so that we can replace $g_J^{B'B}\mapsto g_J^B$ to obtain
\begin{equation}
  \delta_m=m_u -m_d=\frac{g_V^{B}}{g_S^B}\Delta m^{\QCD}_B,
  \label{eq:masssh}
\end{equation}
which we refer to as the CVC relation.\footnote{ Note that also the
relations between $g_S^{B'B}$ and $g_S^B$ receive
$\mathcal{O}(\alpha_{\text{QED}})$ corrections. Therefore terms
$\propto m_{\ell}\alpha_{\text{QED}}$, $\propto
\delta_m\alpha_{\text{QED}}$ and $\propto \delta_m^2/\Lambda_{\QCD}$
can be added to Eq.~\eqref{eq:masssh}. Since $m_{\ell}$ is similar in
size to $\delta_m$, we can neglect the first of these terms too, whose
appearance is related to the mixing in QCD+QED of $m_{\ell}$ and
$\delta_m$ under renormalization.  Using the $\MS$ scheme at
$\mu=2\,\GeV$ corresponds to the suggestion of
Ref.~\cite{Gasser:2003hkw}, however, for quark masses
$m_\ell\sim\delta_m$ this additional scale-dependence can be neglected
with good accuracy, as pointed out above. In addition, there are small
$\mathcal{O}(\alpha^2_{\QED}\Lambda_{\QCD})$ terms due to the QED
contributions to the $\beta$- and $\gamma$-functions, which are also
of higher order.}

Using our physical point, continuum and infinite volume limit result
$g_S^{\Sigma}=3.98^{(22)}_{(24)}$, assuming $g_V^\Sigma = 2$ and applying Eq.~\eqref{eq:masssh}
for the $\Sigma$ baryon,
we obtain in the $N_f=3$ $\MS$ scheme
at $\mu=2\,\GeV$
\begin{equation}
  m_u-m_d=-2.03^{(12)}_{(12)}\,\MeV.\label{eq:quarkdiff}
\end{equation}
We expect
$|\mathcal{O}(\delta_m/\Lambda_{\QCD},\alpha_{\QED})|\lesssim
1\%$ corrections from higher order effects to this result, which we
can neglect at the present level of accuracy.

\begin{table}
  \def\arraystretch{1.2}
  \caption{Comparison of the light quark mass difference.
    \label{tab:deltamq}}
  \begin{ruledtabular}
  \begin{tabular}{lll}
    & $N_f$  &  $\delta_m$/MeV\\ 
    \cmidrule{1-3}
    RM123~\cite{Giusti:2017dmp}                       & 2+1+1 & $-2.38(18)$ \\
    FNAL-MILC\footnotemark[1]~\cite{Bazavov:2017lyh}  & 2+1+1 & $-2.55^{(9)}_{(7)}$ \\
    MILC\footnotemark[1]~\cite{MILC:2018ddw}          & 2+1+1 & $-2.57^{(11)}_{(6)}$ \\
    BMWc~\cite{Fodor:2016bgu}                         & 2+1   & $-2.41(12)$ \\
    This work                                         & 2+1   & $-2.03(12)$
  \end{tabular}
  \end{ruledtabular}
  \footnotetext[1]{FNAL-MILC~\cite{Bazavov:2017lyh} and MILC~\cite{MILC:2018ddw}
    only quote the ratios $m_u/m_d = 0.4556^{(131)}_{(93)}$ and 
    $m_u/m_d=0.4529^{(157)}_{(82)}$, respectively (all errors added in quadrature). 
    Using the FLAG~21~\cite{Aoki:2021kgd} average $m_{\ell} = 3.410(43)$,
    we combine these results to form $\delta_m=2m_{\ell}(m_u/m_d-1)/(m_u/m_d+1)$ 
    and compute the error by error propagation.}
\end{table}

We can compare our value of $\delta_m$ with results from the
literature in Table~\ref{tab:deltamq}. This includes the $N_f=2+1$
result of BMWc~\cite{Fodor:2016bgu} and the $N_f=2+1+1$ continuum
limit results of RM123~\cite{Giusti:2017dmp},
FNAL-MILC~\cite{Bazavov:2017lyh} and
MILC~\cite{MILC:2018ddw}. In the latter two cases we convert results for
$m_u/m_d$ into $\delta_m$ as described in the table caption. We see
a tension between the previous determinations and our result on the two to three
$\sigma$ level.

We remark that all the previous results utilize the dependence of
the pion and kaon masses on the quark masses and the electromagnetic
coupling. We consider our method of determining the quark mass
splitting from the scalar coupling $g^{\Sigma}_S$ and the mass
difference between the $\Sigma^+$ and the $\Sigma^-$ baryons as more
direct.  In Ref.~\cite{Gasser:2020mzy} $\Delta
m_N^{\text{QCD}}=-1.87(16)\,\text{MeV}$ (which agrees within errors
with lattice determinations, including ours, see below) is determined
from experimental input.  A larger (negative) QCD difference would require a
larger QED contribution to the proton mass. As discussed above, the
QED contribution to the mass of the $\Sigma^+$ baryon is
$0.77(5)\,\MeV$ (similar in size to $\Delta
m_N^{\QED}=0.58(16)\,\MeV$~\cite{Gasser:2020mzy}) and it would be
surprising if this increased when replacing a strange quark by a down quark.
Assuming $\Delta m_N^{\text{QCD}}=-1.87(16)\,\text{MeV}$ and a
value $\delta_m\approx -2.50(10)\,\text{MeV}$ as suggested by
Refs.~\cite{Fodor:2016bgu,Giusti:2017dmp,Bazavov:2017lyh,MILC:2018ddw}
would require a coupling $g_S^N=0.75(7)$ to satisfy the CVC relation
Eq.~\eqref{eq:masssh}. This in turn is hard to reconcile with the
majority of lattice results compiled in Fig.~\ref{fig:cmp_nucleon_gs}.
With a lower value for $|\delta_m|$ (and/or a larger $|\Delta m_N^{\QCD}|$)
this inconsistency disappears.

\subsection{QCD and QED isospin breaking effects on the baryon masses}
\begin{figure}
  \centering
  \includegraphics[width=0.47\textwidth]{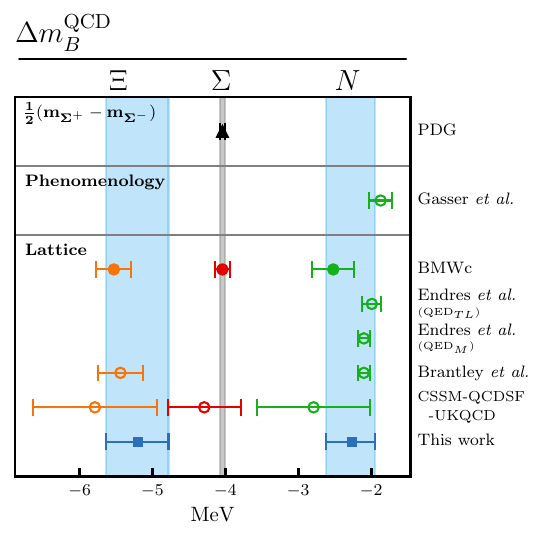}
  \caption{Comparison of the QCD contributions to isospin mass 
    splittings~\cite{Workman:2022ynf,Borsanyi:2014jba,Endres:2015gda,Brantley:2016our,CSSM:2019jmq,Gasser:2020mzy}.
    Note that in our normalization $m_{\Sigma^+}-m_{\Sigma^-}\approx 2\Delta m_{\Sigma}^{\QCD}$.
    Values with filled symbols were obtained via a quark mass, 
    continuum and finite volume extrapolation.
    Endres {\em et al.}~\cite{Endres:2015gda} only quote values for 
    $\Delta m_{N}^{\QED}$ from which we compute $\Delta m_{N}^{\QCD}$ employing 
    the experimental proton-neutron mass splitting.
    \label{fig:deltambaryon}}
\end{figure}

We proceed to compute the QED contributions to the proton and $\Xi^-$
masses, $\Delta m_N^{\QED}$ and $-\Delta m_\Xi^\QED$:
\begin{align}
  \Delta m^{\QED}_N&=\Delta m_N-g_S^N(m_u-m_d)\nonumber\\
  &=\Delta m_N-\frac{2g_S^N}{g_S^{\Sigma}}\Delta m_{\Sigma}^{\QCD},\\
  \Delta m^{\QED}_{\Xi}&=\Delta m_{\Xi}-
  \frac{2g_S^\Xi}{g_S^{\Sigma}}\Delta m_{\Sigma}^{\QCD}.
\end{align}
This gives
\begin{align}
    \Delta m_N^{\QED}&=0.97^{(31)}_{(36)}\,\MeV,&\!\!\!\!\Delta m_N^{\QCD}&=-2.26^{(31)}_{(36)}\,\MeV,\label{eq:massdiff1}\\
    \Delta m_\Sigma^{\QED}&=0.77(05)\,\MeV,&\!\!\!\!\Delta m_\Sigma^{\QCD}&=-4.04(04)\,\MeV,\\
    \Delta m_{\Xi}^{\QED}&=-1.65^{(37)}_{(39)}\,\MeV,&\!\!\!\!\Delta m_\Xi^{\QCD}&=
    -5.20^{(42)}_{(44)}\,\MeV.\label{eq:massdiff3}
\end{align}
For completeness we included the values for the $\Sigma$ baryons that
we determined from the experimental masses alone, without lattice
input. The above mass splittings agree with the
BMWc~\cite{Borsanyi:2014jba} continuum limit results from simulations
of QCD plus QED, see Fig.~\ref{fig:deltambaryon}~(errors
added in quadrature).  Nevertheless, as mentioned above, the value of
$\delta_m$, reported by BMWc~\cite{Fodor:2016bgu} from simulations of
QCD with quenched QED, differs by 2.2 standard deviations from our
result in Eq.~\eqref{eq:quarkdiff}.  Also other lattice results on the
QCD contribution to the mass-splittings~(summarized in
Fig.~\ref{fig:deltambaryon}), obtained at a single lattice spacing
from Endres {\em et al.}~\cite{Endres:2015gda} using
$\QED_{\text{TL}}$ and $\QED_{\text{M}}$, Brantley {\em et
  al.}~\cite{Brantley:2016our} and
CSSM-QCDSF-UKQCD~\cite{CSSM:2019jmq} agree within errors.

We mention the possibility of an enhancement
of the (higher order) $\delta_m^2/\Lambda_{\QCD}$ correction to
the $\Sigma^0$ mass due to the possibility of mixing
with the $\Lambda^0$, which, however, appears to be a very small
effect~\cite{Horsley:2014koa}. A positive contribution to the
$\Sigma^0$ mass would increase $\Delta m_{\Sigma}^{\QED}$ but leave
$\Delta m_{\Sigma}^{\QCD}$ (and therefore the
quark mass difference Eq.~\eqref{eq:quarkdiff}) invariant.

The electromagnetic contributions to the $p$, $\Sigma^{\pm}$ and
$\Xi^-$ masses are all similar to $1\,\MeV$, with an enhancement for
the heavier, more compact cascade baryon. Recently, combining the
Cottingham formula~\cite{Cottingham:1963zz} with experimental input
from elastic scattering and parton distribution functions, the value
$\Delta m_N^{\QED}=0.58(16)\,\MeV$ was determined in
Ref.~\cite{Gasser:2020mzy}. While within errors our result
Eq.~\eqref{eq:massdiff1} agrees with this value, the number obtained
in Ref.~\cite{Gasser:2020mzy} is more inline with the suggested
ordering $\Delta m_N^{\QED}<\Delta m_{\Sigma}^{\QED}<-\Delta
m_{\Xi}^{\QED}$. Combining their value with our determination of
$g_S^N$ gives $\delta_m=-1.69^{(28)}_{(26)}\MeV$, somewhat smaller
in modulus than our result Eq.~\eqref{eq:quarkdiff} and certainly in tension with,
e.g., $\delta_m =  -2.41(12)\,\MeV$~\cite{Fodor:2016bgu}.

We find that the effect of
$m_d>m_u$ on the $\Xi$ and $\Sigma$ mass splittings is much
bigger than for the nucleon since this is proportional to
$g_S^B/g_V^B$ and $g_S^N<g_S^{\Sigma}/2<g_S^{\Xi}$. This hierarchy
is due to $g_S^N\approx F_S+D_S$, $g_S^\Sigma/2\approx F_S$ and
$g_S^{\Xi}\approx F_S-D_S$ with $F_S>0$ and $D_S<0$.
Interestingly, the pion baryon $\sigma$ terms
$\sigma_{\pi B}=\sigma_{u B}+\sigma_{d B}$
that encode the up plus down quark mass contribution to
the baryon masses exhibit the opposite ordering~\cite{RQCD:2022xux},
$\sigma_{\pi N}>\sigma_{\pi\Sigma}>\sigma_{\pi\Xi}$.

\subsection{Isospin breaking effects on the pion baryon \texorpdfstring{$\sigma$}{sigma}~terms\label{sec:isosigma}}

Having determined the quark mass differences, we can also compute
the leading isospin violating corrections to the pion baryon $\sigma$~terms
$\sigma_{\pi B}=\sigma_{u B}+\sigma_{d B}$.
One can either work with matrix elements~\cite{Durr:2015dna},
using the identity
\begin{equation}
  m_u\bar{u}u+m_d\bar{d}d=m_{\ell}\left(\bar{u}u+\bar{d}{d}\right)+\frac{\delta_m}{2}\left(\bar{u}u-\bar{d}d\right),
\end{equation}
or one can start from the Feynman-Hellmann theorem
\begin{equation}
  \sigma_{q B}=m_q\frac{\langle B|\bar{q}q|B\rangle}{\langle B|B\rangle}=m_q\frac{\partial m_B}{\partial m_q}.
\end{equation}
Writing $m_p=m_N+\Delta m_N^{\QCD}/2+\Delta m_N^{\QED}$
and $m_n=m_N-\Delta m_N^{\QCD}/2$, where
$\Delta m_N^{\QCD}=\delta_m g_S^N/g_V^N$, and realizing that the
dependence of the QED contributions on the quark masses is of higher
order in the isospin breaking, we obtain at linear order
\begin{align}
  \sigma_{\pi p}&=m_u\frac{\partial m_p}{\partial m_u}+
  m_d\frac{\partial m_p}{\partial m_d}=\sigma_{\pi N}+\frac12\Delta m_N^{\QCD},\\  \sigma_{\pi n}&=m_u\frac{\partial m_n}{\partial m_u}+
  m_d\frac{\partial m_n}{\partial m_d}=\sigma_{\pi N}-\frac12\Delta m_N^{\QCD}.
\end{align}
The same can be carried out for the $\Sigma^{\pm}$, $\Xi^0$ and $\Xi^-$ baryons.
Using the results for the $\sigma$ terms
of the isosymmetric theory of Ref.~\cite{RQCD:2022xux},
we obtain
\begin{align}
  \sigma_{\pi p}&=42.8^{(4.7)}_{(4.7)}\,\MeV,&\sigma_{\pi n}&=45.0^{(4.7)}_{(4.7)}\,\MeV,\label{eq:sigma1}\\
  \sigma_{\pi\Sigma^+}&=21.9^{(3.8)}_{(6.1)}\,\MeV,
  &\sigma_{\pi\Sigma^-}&=29.9^{(3.8)}_{(6.1)}\,\MeV,\\
  \sigma_{\pi\Xi^0}&=8.6^{(4.5)}_{(6.4)}\,\MeV,
  &\sigma_{\pi\Xi^-}&=13.8^{(4.5)}_{(6.4)}\,\MeV, \label{eq:sigma3}
\end{align}
whereas the pion $\sigma$ term for the $\Sigma^0$ is not affected at
linear order:
$\sigma_{\pi\Sigma^0}\approx\sigma_{\pi\Sigma}=25.9^{(3.8)}_{(6.1)}\,\MeV$.
We refrain from further decomposing the pion baryon $\sigma$ terms
into the individual up and down quark contributions. However, this can
easily be accomplished~\cite{Durr:2015dna}.  It is worth noting that
$(\sigma_{\pi\Xi^-}-\sigma_{\pi\Xi^0})/\sigma_{\pi\Xi}\gg (\sigma_{\pi
  n}-\sigma_{\pi p})/\sigma_{\pi N}$, in spite of the same isospin
difference.

\section{Summary and outlook\label{sec:summary}}  

We determined the axial, scalar and tensor isovector charges of the
nucleon, sigma and cascade baryons using $N_f = 2 + 1$ lattice QCD
simulations. The analysis is based on 47 gauge ensembles, spanning a
range of pion masses from $430\,\MeV$ down to a near physical value of
$130\,\MeV$ across six different lattice spacings between $a\approx
0.039\,\fm$ and $a\approx 0.098\,\fm$ and linear spatial lattice
extents $3.0\,M_\pi^{-1} \leq L \leq 6.5\,M_\pi^{-1}$. The
availability of ensembles lying on three trajectories in the quark
mass plane enables SU(3) flavour symmetry breaking to be explored
systematically and the quark mass dependence of the charges to be
tightly constrained. Simultaneous extrapolations to the physical point
in the continuum and infinite volume limit are performed. Systematic
errors are assessed by imposing cuts on the pion mass, the lattice
spacing and the volume as well as using different sets of
renormalization factors. Our results (in the $\MS$ scheme at
$\mu = 2\, \GeV$) for the nucleon charges are
\begin{align*}
  g_A^N &=  1.284^{(28)}_{(27)} , &  g_S^N &=  1.11^{(14)}_{(16)} ,
  &g_T^N &=  0.984^{(19)}_{(29)}  .
\end{align*}
For the hyperon charges we find
\begin{align*}
  g_A^\Sigma &=  0.875^{(30)}_{(39)}  , &  g_S^\Sigma &=  3.98^{(22)}_{(24)}    , 
  &g_T^\Sigma &=  0.798^{(15)}_{(21)} ,  \\
  g_A^\Xi    &= -0.267^{(13)}_{(12)}  , &     g_S^\Xi &=  2.57^{(11)}_{(11)}    , 
  &g_T^\Xi    &=-0.1872^{(59)}_{(41)} .
\end{align*}
A comparison with previous works is presented in
Sec.~\ref{sec:cmp}. We quantify SU(3) symmetry breaking effects for
the axial charge at the physical point in terms of the combination
\[
\delta^A_{\text{SU(3)}}=\frac{g_A^\Xi+g_A^N-g_A^\Sigma}{g_A^\Xi+g_A^N+g_A^\Sigma}=0.075^{(23)}_{(27)},
\]
see Fig.~\ref{fig:ga_sb}. In particular the axial charge of the
nucleon deviates from its value in the SU(3) chiral limit, as can be
seen in Fig.~\ref{fig:hyperons_ga_ratio} and Table~\ref{tab:ratios_final}.  
No significant symmetry
breaking is observed for the other charges within current precision,
see Figs.~\ref{fig:hyperons_gs_ratio},~\ref{fig:hyperons_gt_ratio}
and~\ref{fig:gst_sb}.

To cross-check the analysis methods, the vector charges
are determined and the expected values, $g_V^N=g_V^\Xi=1$ and
$g_V^\Sigma=2$, are reproduced reasonably well:
\begin{align*}
  g_V^N &=  1.0012^{(12)}_{(11)}  ,
  &g_V^\Sigma &=  2.021^{(21)}_{(27)}   , 
  &g_V^\Xi &=  1.015^{(10)}_{(11)}   .
\end{align*}
Furthermore, we exploit the conserved vector current relation to
predict the quark mass difference
\[m_u-m_d=-2.03(12)\,\MeV\]
from the scalar charge of the $\Sigma$
baryon. We utilize this to decompose isospin mass splittings between
the baryons into QCD and QED contributions (see
Eqs.~\eqref{eq:massdiff1}--\eqref{eq:massdiff3}) and to predict the
leading isospin corrections to the pion baryon $\sigma$~terms (see
Eqs.~\eqref{eq:sigma1}--\eqref{eq:sigma3}).

A computationally efficient stochastic approach was employed in the
analysis, which allows for the simultaneous evaluation of the
three-point correlation functions of all baryons with a variety of
current insertions and momentum combinations. This work is a first
step towards determining hyperon decay form factors which are relevant
for the study of CP violation~\cite{Salone:2022lpt}. A complementary
study of the baryon octet $\sigma$~terms on the same data set as used
here is already ongoing~\cite{Petrak:2023qhx}.

\begin{acknowledgments}
  We thank all our
  \href{https://wiki-zeuthen.desy.de/CLS/}{Coordinated Lattice
    Simulations (CLS)} colleagues for discussions and the joint
  production of the gauge ensembles used. Moreover, we thank Benjamin
  Gläßle and Piotr Korcyl for their contributions regarding the
  (stochastic) three-point function code.

  S.C.\ and S.W.\ received support through the German Research
  Foundation (DFG) grant CO 758/1-1. The work of G.B.\ was funded in
  part by the German Federal Ministry of Education and Research (BMBF)
  grant no.~05P18WRFP1. Additional support from the European Union’s
  Horizon 2020 research and innovation programme under the Marie
  Sk{\l}odowska\nobreakdash-Curie grant agreement no.~813942 (ITN
  EuroPLEx) and grant agreement no.~824093 (STRONG~2020) is gratefully
  acknowledged, as well as initial stage funding through the German
  Research Foundation (DFG) collaborative research centre
  SFB/TRR\nobreakdash-55.

  The authors gratefully acknowledge the
  \href{https://www.gauss-centre.eu}{Gauss Centre for Supercomputing
    (GCS)} for providing computing time through the
  \href{http://www.john-von-neumann-institut.de}{John von Neumann
    Institute for Computing (NIC)} on the supercomputer
  JUWELS~\cite{juwels} and in particular on the Booster partition of
  the supercomputer JURECA~\cite{jureca} at
  \href{http://www.fz-juelich.de/ias/jsc/}{Jülich Supercomputing
    Centre (JSC)}. GCS is the alliance of the three national
  supercomputing centres HLRS (Universität Stuttgart), JSC
  (Forschungszentrum Jülich), and LRZ (Bayerische Akademie der
  Wissenschaften), funded by the BMBF and the German State Ministries
  for Research of Baden\nobreakdash-Württemberg (MWK), Bayern (StMWFK)
  and Nordrhein\nobreakdash-Westfalen (MIWF). Additional simulations
  were carried out on the QPACE~3 Xeon Phi cluster of
  SFB/TRR\nobreakdash-55 and the Regensburg Athene~2 Cluster. The
  authors also thank the JSC for their support and for providing
  services and computing time on the HDF Cloud cluster~\cite{hdfcloud}
  at JSC, funded via the Helmholtz Data Federation (HDF) programme.

  Most of the ensembles were generated using
  \href{https://luscher.web.cern.ch/luscher/openQCD/}{\sc
    openQCD}~\cite{Luscher:2012av} within the CLS effort. A few
  additional ensembles were generated employing the {\sc
    BQCD}\nobreakdash-code~\cite{Nakamura:2010qh} on the QPACE
  supercomputer of SFB/TRR\nobreakdash-55. For the computation of
  hadronic two- and three-point functions we used a modified version
  of the {\sc Chroma}~\cite{Edwards:2004sx} software package along
  with the {\sc Lib\-Hadron\-Analysis} library and the multigrid
  solver implementation of Refs.~\cite{Heybrock:2015kpy,Georg:2017zua}
  (see also ref.~\cite{Frommer:2013fsa}) as well as the IDFLS
  solver~\cite{Luscher:2007es} of {\sc openQCD}.  We used {\sc
    Matplotlib}~\cite{Hunter:2007} to create the figures.

\end{acknowledgments}

\appendix

\section{Further details of the three-point function measurements\label{sec:details3pt}}
\subsection{Comparison of the stochastic and sequential source methods\label{sec:stochseq}}
We computed the connected three-point functions for all the octet 
baryons utilizing the computationally efficient stochastic approach 
outlined in Sec.~\ref{sec:stoch3pts}. This approach introduces 
additional stochastic noise on top of the gauge noise. In the 
analysis presented,  for the nucleon, we make use of statistically 
more precise three-point correlation function measurements determined 
via the sequential source method as part of other projects. In the 
following, we compare the computational costs of the stochastic and 
the sequential source methods and the results for the ratios of the 
three-point over two-point functions for the nucleon.

As a typical example, we consider the measurements performed on
ensemble N200~($M_\pi=286$~MeV and $a=0.064$~fm). For our set-up,
illustrated in Figs.~\ref{fig:stoch3pt_diagram}
and~\ref{fig:src_snk_setup}, a total of $4 \times 12$ solves are
needed for the 4 source positions of the point-to-all propagators. To
form the three-point functions for the nucleon an additional
$N_{\text{sto}}=100$ light stochastic solves~(for the timeslice-to-all
propagators connecting the sink and current timeslices, the wiggly
line in Fig.~\ref{fig:stoch3pt_diagram}) are performed. This set-up
provides 8 measurements of the nucleon three-point function~(as 
shown in Fig.~\ref{fig:src_snk_setup}, with the source-sink
separations $t/a=11$, $14$, $16$, $19$) and
includes all polarizations~(and the unpolarized case) as well as a range of
sink momenta (almost) for free. In principle, decuplet baryon
three-point functions can also be constructed at the analysis stage.
This set-up is evaluated twice on each configuration leading to a total of 296
inversions. Similarly, an additional $(4 \times 12+100)\times 2$
strange solves are performed in order to form the three-point
functions for all the (octet and decuplet) hyperons, including
strangeness changing currents that we did not consider here.

In the sequential source set-up, we compute the three-point function
for the nucleon at rest, again for source-sink separations $t/a=11$, $14$,
$16$ and $19$. Ten measurements are carried out per
configuration~(corresponding to 1, 2, 3 and 4 measurements for each
$t$, respectively), where in each case the two light quark flavours ($u$ and $d$) of the
current and the four possible polarizations of the nucleon 
require $2 \times 4$ sequential sources to be constructed. This
amounts to performing $(4+10 \times 2 \times 4) \times 12 = 1008$
light solves. The additional $4\times 12$
inversions refer to the point-to-all propagators for 4 different
source positions that connect the source to the sink~(and the
current). This is three-times the cost of the stochastic approach~(for
the nucleon three-point functions), which realizes a range of sink
momenta.
 
\begin{figure*}[!hb]
  \begin{center}
    \includegraphics[width=0.7\textwidth]{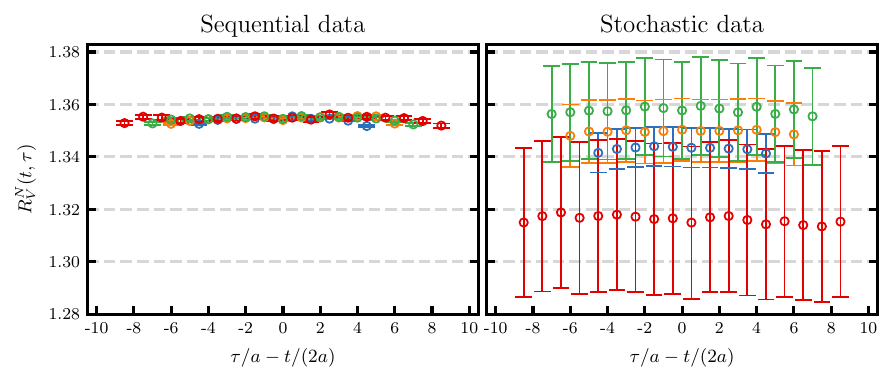}\\
    \includegraphics[width=0.7\textwidth]{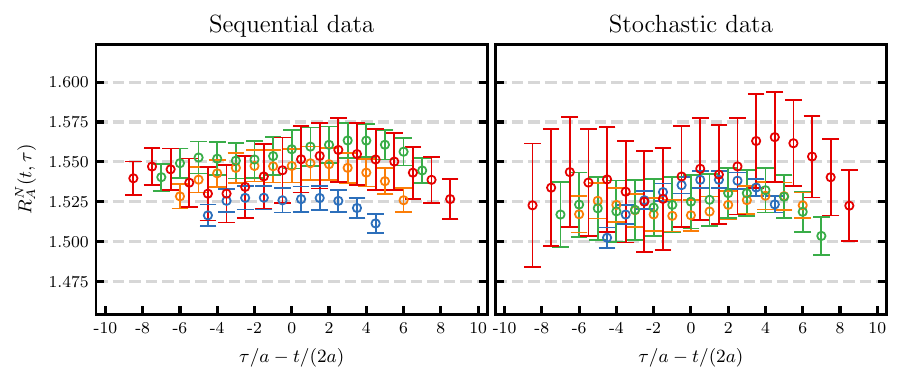}\\
    \includegraphics[width=0.7\textwidth]{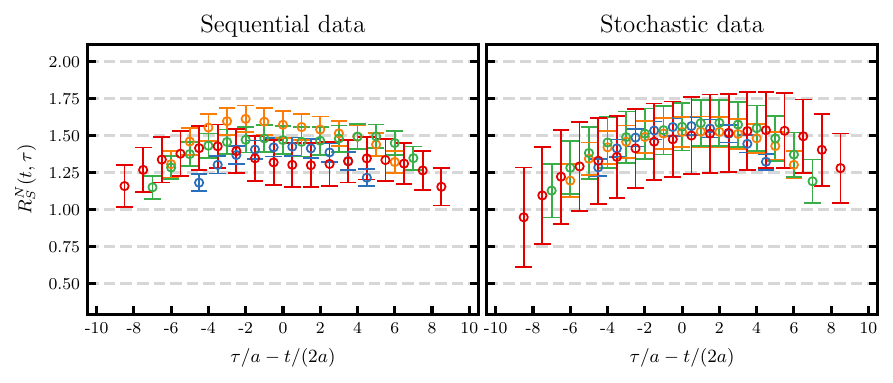}\\
    \includegraphics[width=0.7\textwidth]{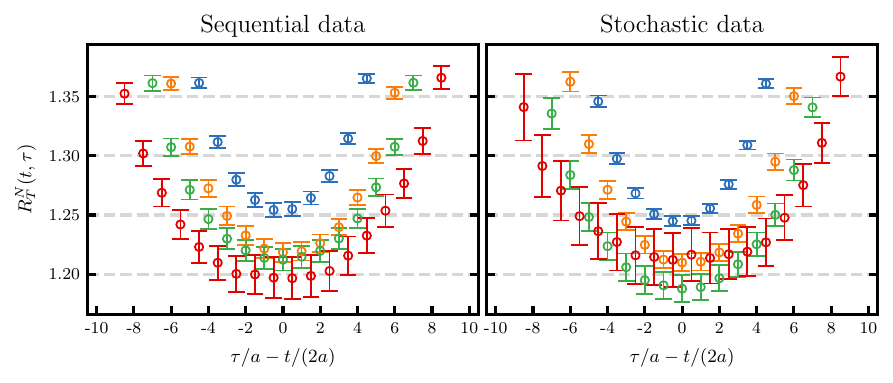}
    \caption{Unrenormalized ratio of three-point over two-point
      functions for the nucleon vector, axial, scalar and tensor
      charge (from top to bottom) on ensemble N200. The left hand side
      shows the data from the sequential source method from 1,2,3 and
      4 measurements (for increasing values of $t$) compared to the
      same measurements obtained from the stochastic approach with
      four measurements for each value of $t$ on the right hand
      side.\label{fig:gvastN200}}
  \end{center}
\end{figure*}

\begin{figure*}[!ht]
  \begin{center}
    \includegraphics[width=0.4\textwidth]{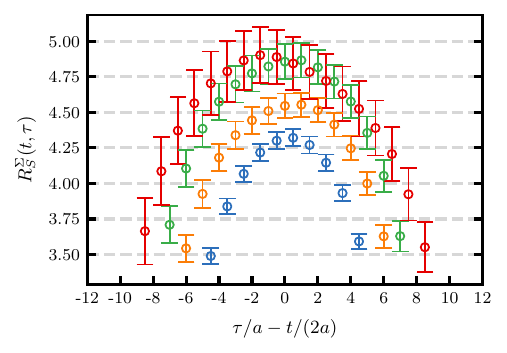}
    \includegraphics[width=0.4\textwidth]{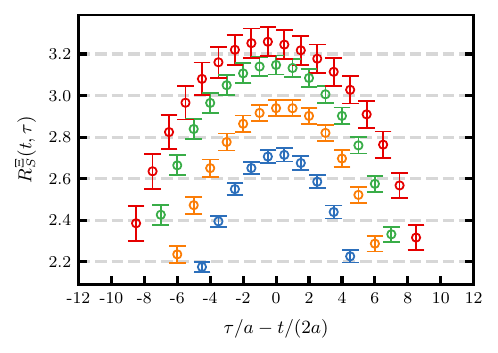}
    \caption{Unrenormalized ratio of three-point over two-point
      functions for the $\Sigma$ (left) and $\Xi$ (right) scalar
      charge on ensemble~N200.\label{fig:hypN200}}
  \end{center}
\end{figure*}

The ratios of the three-point over two-point functions for the nucleon
obtained from the two different approaches are compared in
Fig.~\ref{fig:gvastN200}. A significant part of the gauge noise
cancels in the ratio, while the (additional) stochastic noise
remains. For our set-up, this leads to larger statistical errors for
the stochastic data compared to the sequential source results.  This
difference can clearly be seen for the ratio in the vector channel,
for which the gauge noise is minimal, however, the difference is less
pronounced for the other charges. For the sigma and cascade baryons we
generally find a good statistical signal in the ratios employing the
stochastic approach, see Fig.~\ref{fig:hypN200}, although, also in
this case the ratios for the hyperon vector charges suffer from large
errors.

In the case of a large-scale analysis effort including high
statistics, as presented in this article, the disk space required to
store the stochastic three-point function data is significant. The
individual spectator~($S$) and insertion~($I$) parts as defined in
Eqs.~\eqref{eq:spe} and \eqref{eq:ins}, respectively, are stored with
all indices open. In general this amounts to $N = N[S] + N[I]$ complex
double precision floating point numbers for each gauge field
configuration where
\begin{align}
  N[S] &= N_F^{S} \cdot N_{\mathbf{p}^\prime} \cdot N_{\text{snk}} \cdot N_{\text{src}} \cdot N_{\text{sto}} \cdot N_{c} \cdot N_{s}^5 \ ,\\
  N[I] &= N_F^{I} \cdot N_{\mathbf{q}} \cdot N_{\partial_\mu} \cdot N_{\tau} \cdot N_{\text{src}} \cdot N_{\text{sto}} \cdot N_{c} \cdot N_{s}^2 \ .
\end{align}
Here $N_F^{S/I}$ denotes the number of flavour combinations for the
spectator and insertion parts (typically 4 and 2, respectively),
$N_{\mathbf{p}}$ gives the number of momentum combinations for a
maximum momentum~$|\mathbf{p}|$ (with $\mathbf{p}$ either being
the sink momentum $\mathbf{p}'$ or the momentum transfer
$\mathbf{q}$), $N_{\text{src}/\text{snk}}$ corresponds to the
number of source (4) and sink (2) positions, $N_c = 3$ and $N_s =
4$ are the dimensions of colour and spin space and $N_{\tau}$ is the
number of current insertion timeslices, usually the distance between
the two sink timeslices, see Figs.~\ref{fig:stoch3pt_diagram}
and~\ref{fig:src_snk_setup}. $N_{\partial_\mu}$
refers to the number of derivatives included in the current
insertion. We consider all currents including up to one
derivative~($N_{\partial_\mu}=1+4$), although only the currents
without derivatives are presented in this work. This adds up to a file
size of the order of GBs for a single gauge field configuration and
disk space usage of the order of TBs for a typical CLS gauge ensemble.
Storing the data with all indices open allows for a very flexible
analysis. Octet or decuplet baryon three-point functions can be
constructed from the spectator and insertion parts for different
polarizations, current insertions as well as for a large number of
momentum combinations.

\subsection{Treatment of outliers\label{sec:outlier}}
\begin{figure}[!h]
  \centering
  \includegraphics[width=0.45\textwidth]{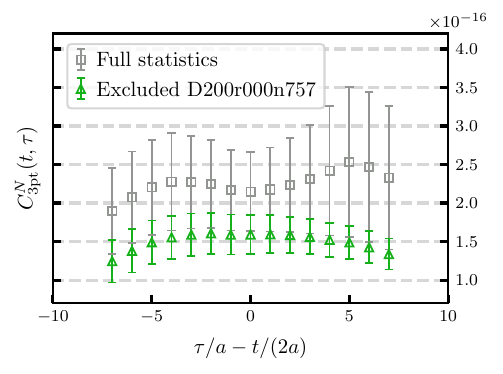}
  \caption{Nucleon three-point function for the scalar channel with a
    source-sink separation $t/a = 16$ for a single source position
    on ensemble D200. The average and standard error is computed
    including and excluding one particular gauge configuration.
   \label{fig:outlier}}
\end{figure}

When analysing the three-point functions on some of the ensembles, 
we observe a small number of three-point function results
that are by many orders of magnitude larger than the rest. These
outliers, whose origin currently remains unexplained, would have a
significant impact on the analysis, as illustrated in
Fig.~\ref{fig:outlier} for ensemble D200~($M_\pi=202$~MeV and
$a=0.064$~fm). The three-point function for the scalar charge with
source-sink separation $t/a=16$ for a single source position is
displayed. In this case one outlier is identified~(according to the
criterion given below) and one sees a substantial change in the
configuration average and a reduction in the standard deviation if
this measurement (on a particular configuration) is excluded. The
Mainz group~\cite{Agadjanov:2023jha} reported similar outliers when
determining the nucleon $\sigma$~terms on a subset of the CLS gauge
ensembles employed here. We remark that such outliers do not seem to
occur in the distributions of the two-point function measurements.
\begin{figure}[!h]
  \centering
  \includegraphics[width=0.45\textwidth]{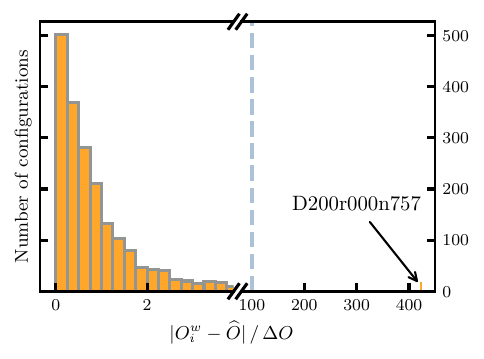}
  \caption{The numbers of configurations separated by a distance $|
    O_i^w - \widehat{O} | / \Delta O$ from the central value
    $\widehat{O}$ for the three-point function data of
    Fig.~\ref{fig:outlier} on the insertion timeslice $\tau/a =
    15$. The blue dashed line indicates the cutoff $K = 100$, which
    is exceeded by a single measurement. Note the cut and different
    scales of the ordinate. Due to this representation a few histogram
    entries are not shown.
   \label{fig:outlier_hist}}
\end{figure}
To overcome this obstacle we constrain the analysed data
and discard `outlier' configurations on which measurements give
contributions that are very far away from the expected central
value. As discussed in Sec.~\ref{sec:ensembles} we employ the
appropriate reweighting factors where $w_i = w_i^\ell w_i^s$ is the
product of the light and strange reweighting factors, determined on
each configuration $i$, see in particular the discussion in Appendix
G.2 of Ref.~\cite{RQCD:2022xux}. In order to have a robust estimate
for the variance of the reweighted data
\begin{align}
  O_i^w = \frac{w_i O_i}{w},
\end{align}
where $w =N_{\text{cnfg}}^{-1} \sum_i w_i$, we determine the lower and
upper boundary values $O^w_{\text{low}}$ and $O^w_{\text{up}}$ of the
central 68\% interval of this distribution. We then remove
configurations~$i$ with
\begin{align}\label{eq:cut}
  | O_i^w - \widehat{O} | > K \, \Delta O,
\end{align}
where
\begin{align}
  \widehat{O} = \frac{O^w_{\text{up}} + O^w_{\text{low}}}{2}  ,\quad
  \Delta O = \frac{O^w_{\text{up}} - O^w_{\text{low}}}{2}  ,
\end{align}
setting the cutoff $K$ to a large value ($K=100$).
In Fig.~\ref{fig:outlier_hist} we show the distribution of the
measurements for the nucleon three-point function for a scalar current
(at a single source position and current insertion time) on the ensemble
D200 discussed above (see Fig.~\ref{fig:outlier}). The outlier is more
than $400\,\Delta O$ away from $\widehat{O}$.  Considering all the
three-point function measurements across the different ensembles, we
uniformly set the cutoff in Eq.~\eqref{eq:cut}
to $K = 100$ and remove all configurations from the
analysis on which at least one measurement satisfies this criterion.
\newpage
\begin{widetext}
\section{Additional tables\label{sec:lattice_data}}
In Table~\ref{tab:data_lat} we include further details on the gauge
ensembles. In Tables~\ref{tab:data_nucleon}--\ref{tab:data_xi}
the results for the unrenormalized charges for the three octet baryons
are collected.
\FloatBarrier
\begin{table*}[t]
  \caption{The gauge ensembles utilized in this work: the light and
    strange hopping parameters $\kappa_\ell$ and $\kappa_s$, the
    gradient flow scale parameter $t_0/a^2$ and the pion~($M_\pi$) and
    kaon masses~($M_K$)~\cite{RQCD:2022xux}.\label{tab:data_lat}}
  \centering
  \begin{ruledtabular}
  \begin{tabular}{lllllll}
    Ensemble & $\kappa_\ell$ & $\kappa_s$ & $t_0 / a^2$ & $aM_\pi$    & $aM_K$     \\ 
    \cmidrule{1-6}
    A653       & 0.1365716          & 0.1365716          & 2.1729(50)  & 0.21235(94) & 0.21235(94) \\ 
A650       & 0.136600           & 0.136600           & 2.2878(72)  & 0.1833(13)  & 0.1833(13)  \\ 
A654       & 0.136750           & 0.136216193        & 2.1950(77)  & 0.1669(11)  & 0.22714(91) \\ 
H101       & 0.13675962         & 0.13675962         & 2.8545(81)  & 0.18283(57) & 0.18283(57) \\ 
U103       & 0.13675962         & 0.13675962         & 2.8815(57)  & 0.18133(61) & 0.18133(61) \\ 
H107       & 0.13694566590798   & 0.136203165143476  & 2.7193(76)  & 0.15913(73) & 0.23745(53) \\ 
H102r002   & 0.136865           & 0.136549339        & 2.8792(90)  & 0.15490(92) & 0.19193(77) \\ 
U102       & 0.136865           & 0.136549339        & 2.8932(63)  & 0.15444(84) & 0.19235(61) \\ 
H102r001   & 0.136865           & 0.136549339        & 2.8840(89)  & 0.15311(98) & 0.19089(78) \\ 
rqcd021    & 0.136813           & 0.136813           & 3.032(15)   & 0.14694(88) & 0.14694(88) \\ 
H105       & 0.136970           & 0.13634079         & 2.8917(65)  & 0.1213(14)  & 0.20233(64) \\ 
N101       & 0.136970           & 0.13634079         & 2.8948(39)  & 0.12132(58) & 0.20156(30) \\ 
H106       & 0.137015570024     & 0.136148704478     & 2.8227(68)  & 0.1180(21)  & 0.22471(67) \\ 
C102       & 0.13705084580022   & 0.13612906255557   & 2.8682(47)  & 0.09644(77) & 0.21783(36) \\ 
C101       & 0.137030           & 0.136222041        & 2.9176(38)  & 0.09586(64) & 0.20561(33) \\ 
D101       & 0.137030           & 0.136222041        & 2.910(10)   & 0.0958(11)  & 0.20572(45) \\ 
S100       & 0.137030           & 0.136222041        & 2.9212(91)  & 0.0924(31)  & 0.20551(57) \\ 
D150       & 0.137088           & 0.13610755         & 2.9476(30)  & 0.05497(79) & 0.20834(17) \\ 
B450       & 0.136890           & 0.136890           & 3.663(11)   & 0.16095(49) & 0.16095(49) \\ 
S400       & 0.136984           & 0.136702387        & 3.6919(74)  & 0.13535(42) & 0.17031(38) \\ 
B452       & 0.1370455          & 0.136378044        & 3.5286(66)  & 0.13471(47) & 0.20972(34) \\ 
rqcd030    & 0.1369587          & 0.1369587          & 3.914(15)   & 0.12202(68) & 0.12202(68) \\ 
N451       & 0.1370616          & 0.1365480771       & 3.6822(46)  & 0.11067(32) & 0.17828(20) \\ 
N401       & 0.1370616          & 0.1365480771       & 3.6844(52)  & 0.10984(57) & 0.17759(37) \\ 
N450       & 0.1370986          & 0.136352601        & 3.5920(42)  & 0.10965(31) & 0.20176(18) \\ 
X450       & 0.136994           & 0.136994           & 3.9935(92)  & 0.10142(62) & 0.10142(62) \\ 
D451       & 0.137140           & 0.136337761        & 3.6684(36)  & 0.08370(31) & 0.19385(15) \\ 
D450       & 0.137126           & 0.136420428639937  & 3.7076(75)  & 0.08255(41) & 0.18354(12) \\ 
D452       & 0.137163675        & 0.136345904546     & 3.7251(37)  & 0.05961(50) & 0.18647(13) \\ 
N202       & 0.137000           & 0.137000           & 5.165(14)   & 0.13388(35) & 0.13388(35) \\ 
N204       & 0.137112           & 0.136575049        & 4.9473(79)  & 0.11423(33) & 0.17734(29) \\ 
X250       & 0.137050           & 0.137050           & 5.283(28)   & 0.11319(39) & 0.11319(39) \\ 
N203       & 0.137080           & 0.136840284        & 5.1465(63)  & 0.11245(30) & 0.14399(24) \\ 
S201       & 0.137140           & 0.13672086         & 5.1638(91)  & 0.09379(47) & 0.15220(37) \\ 
N201       & 0.13715968         & 0.136561319        & 5.0427(75)  & 0.09268(31) & 0.17040(22) \\ 
N200       & 0.137140           & 0.13672086         & 5.1600(71)  & 0.09236(29) & 0.15061(24) \\ 
X251       & 0.137100           & 0.137100           & 5.483(26)   & 0.08678(40) & 0.08678(40) \\ 
D200       & 0.137200           & 0.136601748        & 5.1793(39)  & 0.06540(33) & 0.15652(15) \\ 
D201       & 0.1372067          & 0.136546844        & 5.1378(66)  & 0.06472(42) & 0.16302(18) \\ 
E250       & 0.137232867        & 0.136536633        & 5.2027(41)  & 0.04227(23) & 0.159370(61) \\ 
N300       & 0.137000           & 0.137000           & 8.576(21)   & 0.10642(38) & 0.10642(38) \\ 
N304       & 0.137079325093654  & 0.136665430105663  & 8.322(20)   & 0.08840(33) & 0.13960(31) \\ 
N302       & 0.137064           & 0.1368721791358    & 8.539(19)   & 0.08701(41) & 0.11370(36) \\ 
J304       & 0.137130           & 0.1366569203       & 8.497(12)   & 0.06538(18) & 0.13181(14) \\ 
J303       & 0.137123           & 0.1367546608       & 8.615(14)   & 0.06481(19) & 0.11975(16) \\ 
E300       & 0.137163           & 0.1366751636177327 & 8.6241(74)  & 0.04402(20) & 0.12397(15) \\ 
J500       & 0.136852           & 0.136852           & 14.013(34)  & 0.08116(34) & 0.08116(34) \\ 
J501       & 0.1369032          & 0.136749715        & 13.928(39)  & 0.06589(26) & 0.08798(23) 
  \end{tabular}
  \end{ruledtabular}
\end{table*}
\begin{table*}[b]
  \caption{Results for the unrenormalized nucleon charges $g_{J}^{N, \latt}$ for $J \in \{A, S, T, V\}$. \#ES labels the number of excited states used to determine the ground state matrix element, see the discussion in Sec.~\ref{sec:excited}.    \label{tab:data_nucleon}}
  \centering
  \begin{ruledtabular}
  \begin{tabular}{llllllllll}
    \#ES & \multicolumn{4}{c}{$1$}  &  \multicolumn{4}{c}{$2$}   \\[3pt]
    Ensemble & $g_{A}^{N,\latt}$ & $g_{S}^{N,\latt}$ & $g_{T}^{N,\latt}$ & $g_{V}^{N,\latt}$ & $g_{A}^{N,\latt}$ & $g_{S}^{N,\latt}$ & $g_{T}^{N,\latt}$ & $g_{V}^{N,\latt}$ \\  
    \cmidrule{1-9}
    A653               & 1.563(17)   & 1.449(60)   & 1.273(13)   & 1.4140(24)  & 1.589(18)   & 1.43(10)    & 1.279(13)   & 1.4136(21)  \\ 
A650               & 1.554(13)   & 1.468(68)   & 1.236(10)   & 1.4173(35)  & 1.581(18)   & 1.41(13)    & 1.243(12)   & 1.4171(32)  \\ 
A654               & 1.545(22)   & 1.62(21)    & 1.216(30)   & 1.4353(85)  & 1.538(33)   & 1.76(34)    & 1.211(19)   & 1.4365(87)  \\ 
H101               & 1.5662(88)  & 1.760(55)   & 1.221(13)   & 1.39160(46) & 1.584(16)   & 1.795(88)   & 1.224(13)   & 1.39152(46) \\ 
U103               & 1.495(17)   & 1.61(11)    & 1.219(23)   & 1.39001(77) & 1.511(26)   & 1.68(13)    & 1.227(18)   & 1.38990(66) \\ 
H107               & 1.630(37)   & 1.46(13)    & 1.248(19)   & 1.40067(98) & 1.667(29)   & 1.34(17)    & 1.260(20)   & 1.40051(77) \\ 
H102r002           & 1.596(21)   & 1.70(15)    & 1.221(22)   & 1.3986(11)  & 1.616(29)   & 1.84(21)    & 1.231(21)   & 1.39817(87) \\ 
U102\footnotemark[1] & 1.449(41)   & 1.42(17)    & 1.180(56)   & 1.3988(22)  & 1.448(41)   & 1.32(24)    & 1.194(31)   & 1.3982(11)  \\ 
H102r001           & 1.582(15)   & 1.67(10)    & 1.230(21)   & 1.39706(67) & 1.598(32)   & 1.78(19)    & 1.235(21)   & 1.39698(62) \\ 
rqcd021            & 1.546(16)   & 1.67(12)    & 1.188(22)   & 1.383(13)   & 1.567(30)   & 1.60(23)    & 1.190(22)   & 1.383(12)   \\ 
H105               & 1.533(29)   & 1.44(20)    & 1.194(37)   & 1.4077(18)  & 1.524(51)   & 1.29(44)    & 1.199(32)   & 1.4074(15)  \\ 
N101\footnotemark[1] & 1.623(19)   & 1.662(87)   & 1.229(16)   & 1.40416(63) & 1.670(32)   & 1.64(21)    & 1.236(18)   & 1.40411(53) \\ 
H106               & 1.589(34)   & 1.30(27)    & 1.223(35)   & 1.4084(20)  & 1.597(64)   & 1.20(50)    & 1.231(37)   & 1.4081(17)  \\ 
C102               & 1.699(36)   & 1.70(32)    & 1.184(25)   & 1.4083(13)  & 1.755(58)   & 1.74(54)    & 1.198(29)   & 1.4083(12)  \\ 
C101               & 1.675(37)   & 1.67(17)    & 1.214(17)   & 1.40908(85) & 1.758(44)   & 1.76(40)    & 1.232(18)   & 1.40872(69) \\ 
D101\footnotemark[1] & 1.647(51)   & 1.74(34)    & 1.222(40)   & 1.4091(18)  & 1.700(80)   & 2.09(77)    & 1.236(36)   & 1.4088(11)  \\ 
S100\footnotemark[1] & 1.86(98)    & 1.6(1.2)    & 1.27(21)    & 1.4091(49)  & 1.75(16)    & 1.6(1.5)    & 1.257(57)   & 1.4072(25)  \\ 
D150\footnotemark[1] & 1.49(21)    & 2.8(2.9)    & 0.88(23)    & 1.430(13)   & 1.39(31)    & 5.0(4.0)    & 0.92(14)    & 1.430(15)   \\ 
B450               & 1.549(12)   & 1.642(72)   & 1.228(13)   & 1.3729(31)  & 1.587(18)   & 1.61(13)    & 1.233(14)   & 1.3724(27)  \\ 
S400               & 1.525(13)   & 1.645(90)   & 1.191(22)   & 1.37682(65) & 1.523(28)   & 1.63(17)    & 1.193(20)   & 1.37683(59) \\ 
B452               & 1.555(15)   & 1.49(10)    & 1.232(17)   & 1.3781(38)  & 1.568(30)   & 1.39(24)    & 1.235(19)   & 1.3781(36)  \\ 
rqcd030            & 1.510(14)   & 1.531(92)   & 1.178(15)   & 1.3769(33)  & 1.553(27)   & 1.47(20)    & 1.182(16)   & 1.3769(30)  \\ 
N451               & 1.595(14)   & 1.603(89)   & 1.212(15)   & 1.3828(35)  & 1.636(31)   & 1.49(21)    & 1.217(18)   & 1.3828(30)  \\ 
N401               & 1.584(26)   & 2.00(27)    & 1.166(37)   & 1.3836(11)  & 1.603(51)   & 2.27(46)    & 1.180(31)   & 1.38328(91) \\ 
N450               & 1.598(13)   & 1.67(11)    & 1.208(17)   & 1.3840(29)  & 1.625(31)   & 1.58(26)    & 1.210(19)   & 1.3838(29)  \\ 
X450               & 1.558(23)   & 1.99(18)    & 1.166(23)   & 1.3884(78)  & 1.606(51)   & 2.15(37)    & 1.174(24)   & 1.3873(69)  \\ 
D451               & 1.583(25)   & 1.42(25)    & 1.175(36)   & 1.3807(78)  & 1.554(80)   & 0.88(77)    & 1.176(38)   & 1.379(10)   \\ 
D450               & 1.623(25)   & 1.46(30)    & 1.201(22)   & 1.3761(90)  & 1.672(66)   & 1.04(77)    & 1.207(27)   & 1.3769(84)  \\ 
D452\footnotemark[2] & 1.54(21)    & 0.1(4.5)    & 0.95(19)    & 1.07(42)    & 1.53(32)    & -1.0(5.1)   & 1.00(13)    & 1.13(22)    \\ 
N202               & 1.537(12)   & 1.926(84)   & 1.170(17)   & 1.34793(29) & 1.556(20)   & 2.03(11)    & 1.179(16)   & 1.34780(28) \\ 
N204               & 1.570(13)   & 1.694(91)   & 1.200(14)   & 1.35344(39) & 1.587(25)   & 1.80(19)    & 1.205(15)   & 1.35335(37) \\ 
X250               & 1.5232(98)  & 1.814(80)   & 1.163(14)   & 1.3513(25)  & 1.546(19)   & 1.88(15)    & 1.167(14)   & 1.3512(23)  \\ 
N203               & 1.532(10)   & 1.748(68)   & 1.176(16)   & 1.35169(27) & 1.545(26)   & 1.69(15)    & 1.176(16)   & 1.35169(27) \\ 
S201\footnotemark[1] & 1.436(25)   & 1.49(21)    & 1.155(41)   & 1.35576(78) & 1.52(12)    & 0.82(98)    & 1.160(41)   & 1.3555(10)  \\ 
N201               & 1.568(18)   & 1.67(16)    & 1.158(21)   & 1.35653(50) & 1.594(36)   & 1.54(35)    & 1.162(23)   & 1.35649(49) \\ 
N200               & 1.565(21)   & 1.47(15)    & 1.174(19)   & 1.35533(37) & 1.607(40)   & 1.33(32)    & 1.179(22)   & 1.35530(34) \\ 
X251               & 1.532(25)   & 1.94(16)    & 1.135(14)   & 1.3568(56)  & 1.604(25)   & 2.18(26)    & 1.149(14)   & 1.3558(41)  \\ 
D200               & 1.582(24)   & 1.94(30)    & 1.136(32)   & 1.35952(43) & 1.617(61)   & 2.37(72)    & 1.140(32)   & 1.35997(77) \\ 
D201               & 1.562(35)   & 1.60(41)    & 1.135(48)   & 1.35922(82) & 1.597(91)   & 1.82(98)    & 1.141(45)   & 1.35922(67) \\ 
E250\footnotemark[1] & 1.66(14)    & 1.8(1.2)    & 1.134(77)   & 1.353(28)   & 2.00(20)    & 3.2(3.8)    & 1.175(55)   & 1.358(13)   \\ 
N300               & 1.479(12)   & 1.813(95)   & 1.149(20)   & 1.31543(19) & 1.485(22)   & 1.83(14)    & 1.152(19)   & 1.31540(16) \\ 
N304               & 1.495(23)   & 1.64(18)    & 1.137(28)   & 1.31926(32) & 1.501(36)   & 1.52(31)    & 1.140(27)   & 1.31924(31) \\ 
N302               & 1.498(19)   & 1.82(18)    & 1.097(29)   & 1.31893(29) & 1.523(35)   & 1.83(30)    & 1.103(27)   & 1.31889(27) \\ 
J304               & 1.529(19)   & 1.77(22)    & 1.107(22)   & 1.32310(36) & 1.567(39)   & 1.67(51)    & 1.113(23)   & 1.32305(37) \\ 
J303               & 1.518(17)   & 1.51(14)    & 1.096(29)   & 1.32206(21) & 1.526(56)   & 1.16(53)    & 1.094(36)   & 1.32198(25) \\ 
E300               & 1.557(37)   & 1.67(36)    & 1.086(25)   & 1.3135(88)  & 1.646(61)   & 1.44(92)    & 1.100(33)   & 1.312(11)   \\ 
J500               & 1.451(11)   & 1.858(92)   & 1.108(18)   & 1.29090(12) & 1.453(18)   & 1.91(11)    & 1.114(13)   & 1.290855(87) \\ 
J501               & 1.484(37)   & 2.13(30)    & 1.072(59)   & 1.29352(16) & 1.499(38)   & 2.10(28)    & 1.088(29)   & 1.29349(19) 
  \end{tabular}
  \end{ruledtabular}
  \footnotetext[1]{Ensemble only enters the analysis of the nucleon charges since no data for the hyperon charges are available.}
  \footnotetext[2]{The nucleon three-point functions are computed with the ``stochastic'' approach. (For all the other ensembles the sequential source method was used.)}
\end{table*}
\begin{table*}
  \caption{Results for the unrenormalized hyperon charges
    $g_{J}^{\Sigma, \latt}$ for $J \in \{A, S, T, V\}$. \#ES labels
    the number of excited states used to determine the ground state
    matrix element, see the discussion in Sec.~\ref{sec:excited}. The
    three-point functions are computed employing the ``stochastic''
    approach. \label{tab:data_sigma}} \centering
  \begin{ruledtabular}
  \begin{tabular}{llllllllll}
    \#ES & \multicolumn{4}{c}{$1$}  &  \multicolumn{4}{c}{$2$}   \\[3pt]
    Ensemble & $g_{A}^{\Sigma,\latt}$ & $g_{S}^{\Sigma,\latt}$ & $g_{T}^{\Sigma,\latt}$ & $g_{V}^{\Sigma,\latt}$ & $g_{A}^{\Sigma,\latt}$ & $g_{S}^{\Sigma,\latt}$ & $g_{T}^{\Sigma,\latt}$ & $g_{V}^{\Sigma,\latt}$ \\ 
    \cmidrule{1-9}
    A653               & 1.180(10)   & 4.40(14)    & 1.018(14)   & 2.8354(32)  & 1.189(16)   & 4.45(14)    & 1.023(12)   & 2.8347(27)  \\ 
A650               & 1.1764(90)  & 4.46(12)    & 0.9968(95)  & 2.8381(39)  & 1.190(16)   & 4.49(15)    & 1.000(10)   & 2.8380(37)  \\ 
A654               & 1.173(12)   & 4.01(16)    & 1.028(13)   & 2.852(12)   & 1.204(37)   & 4.11(30)    & 1.028(16)   & 2.850(14)   \\ 
H101               & 1.1877(81)  & 5.03(12)    & 0.982(10)   & 2.78711(66) & 1.198(13)   & 5.12(14)    & 0.987(10)   & 2.78686(65) \\ 
U103               & 1.137(13)   & 4.69(18)    & 0.983(16)   & 2.78535(89) & 1.147(21)   & 4.75(18)    & 0.989(16)   & 2.78507(80) \\ 
H107               & 1.207(18)   & 4.25(19)    & 1.030(12)   & 2.823(15)   & 1.232(28)   & 4.35(27)    & 1.036(13)   & 2.820(11)   \\ 
H102r002           & 1.177(13)   & 4.67(17)    & 0.988(13)   & 2.807(14)   & 1.199(25)   & 4.77(22)    & 0.992(14)   & 2.805(12)   \\ 
H102r001           & 1.157(14)   & 4.73(19)    & 0.973(16)   & 2.788(11)   & 1.146(27)   & 4.79(28)    & 0.978(17)   & 2.788(11)   \\ 
rqcd021            & 1.149(15)   & 5.27(24)    & 0.958(19)   & 2.790(14)   & 1.145(27)   & 5.35(33)    & 0.962(19)   & 2.790(13)   \\ 
H105               & 1.147(22)   & 4.83(44)    & 0.958(23)   & 2.812(20)   & 1.162(40)   & 5.25(55)    & 0.969(21)   & 2.809(16)   \\ 
H106               & 1.139(15)   & 4.24(20)    & 0.968(16)   & 2.796(15)   & 1.149(38)   & 4.27(48)    & 0.970(19)   & 2.796(15)   \\ 
C102               & 1.140(19)   & 4.32(29)    & 0.950(27)   & 2.791(23)   & 1.133(45)   & 3.94(76)    & 0.952(22)   & 2.792(18)   \\ 
C101               & 1.174(21)   & 4.58(43)    & 0.957(22)   & 2.837(21)   & 1.240(51)   & 4.80(94)    & 0.963(21)   & 2.834(17)   \\ 
B450               & 1.1743(76)  & 4.80(14)    & 0.992(11)   & 2.7464(30)  & 1.195(15)   & 4.90(18)    & 0.997(11)   & 2.7460(28)  \\ 
S400               & 1.159(16)   & 5.25(29)    & 0.945(17)   & 2.751(13)   & 1.180(23)   & 5.43(27)    & 0.962(16)   & 2.752(11)   \\ 
B452               & 1.147(13)   & 4.33(18)    & 0.990(16)   & 2.764(13)   & 1.163(28)   & 4.45(33)    & 0.994(16)   & 2.763(12)   \\ 
rqcd030            & 1.140(12)   & 5.20(22)    & 0.944(16)   & 2.7576(35)  & 1.157(22)   & 5.41(33)    & 0.950(15)   & 2.7573(33)  \\ 
N451               & 1.168(15)   & 4.90(34)    & 0.976(14)   & 2.795(18)   & 1.193(30)   & 5.43(38)    & 0.988(11)   & 2.789(12)   \\ 
N401               & 1.130(52)   & 5.22(68)    & 0.933(50)   & 2.781(42)   & 1.100(67)   & 5.23(82)    & 0.949(33)   & 2.776(31)   \\ 
N450               & 1.148(18)   & 4.11(17)    & 0.994(20)   & 2.738(24)   & 1.154(43)   & 3.54(65)    & 0.992(17)   & 2.737(21)   \\ 
X450               & 1.168(23)   & 6.05(38)    & 0.924(25)   & 2.7700(84)  & 1.195(49)   & 6.38(58)    & 0.933(25)   & 2.7690(79)  \\ 
D451               & 1.143(53)   & 4.79(95)    & 0.955(38)   & 2.67(10)    & 1.18(15)    & 3.8(3.0)    & 0.948(45)   & 2.624(56)   \\ 
D450               & 1.129(63)   & 4.9(1.1)    & 0.938(52)   & 2.721(77)   & 1.09(11)    & 3.9(2.1)    & 0.944(42)   & 2.730(48)   \\ 
D452               & 1.115(52)   & 4.8(1.1)    & 0.923(52)   & 2.789(28)   & 1.07(14)    & 5.3(2.6)    & 0.935(44)   & 2.787(29)   \\ 
N202               & 1.166(10)   & 5.68(20)    & 0.939(12)   & 2.69777(45) & 1.182(16)   & 5.85(15)    & 0.955(13)   & 2.69721(41) \\ 
N204               & 1.1465(91)  & 4.30(12)    & 0.980(11)   & 2.686(11)   & 1.160(26)   & 4.22(30)    & 0.979(12)   & 2.686(12)   \\ 
X250               & 1.1549(88)  & 5.85(20)    & 0.928(12)   & 2.7051(32)  & 1.167(16)   & 6.12(21)    & 0.940(12)   & 2.7046(27)  \\ 
N203               & 1.156(12)   & 5.60(26)    & 0.941(10)   & 2.7246(96)  & 1.182(19)   & 6.05(22)    & 0.9623(87)  & 2.7184(69)  \\ 
N201               & 1.150(17)   & 5.22(43)    & 0.953(18)   & 2.737(17)   & 1.163(42)   & 5.70(53)    & 0.965(17)   & 2.733(14)   \\ 
N200               & 1.119(13)   & 5.03(21)    & 0.947(13)   & 2.721(11)   & 1.116(26)   & 5.37(36)    & 0.954(12)   & 2.720(10)   \\ 
X251               & 1.145(16)   & 6.59(33)    & 0.903(13)   & 2.7165(69)  & 1.195(24)   & 7.06(37)    & 0.916(14)   & 2.7145(55)  \\ 
D200               & 1.082(48)   & 6.4(1.4)    & 0.841(59)   & 2.774(26)   & 1.058(82)   & 7.7(1.8)    & 0.867(38)   & 2.777(29)   \\ 
D201               & 1.125(34)   & 5.7(1.3)    & 0.924(37)   & 2.730(31)   & 1.133(62)   & 6.8(1.2)    & 0.944(24)   & 2.726(19)   \\ 
N300               & 1.121(12)   & 5.52(19)    & 0.918(17)   & 2.63208(25) & 1.126(19)   & 5.59(18)    & 0.925(16)   & 2.63197(23) \\ 
N304               & 1.106(18)   & 4.84(31)    & 0.925(21)   & 2.642(17)   & 1.130(30)   & 4.87(45)    & 0.928(20)   & 2.641(15)   \\ 
N302               & 1.086(26)   & 5.41(35)    & 0.864(34)   & 2.668(25)   & 1.118(42)   & 5.38(49)    & 0.868(26)   & 2.666(21)   \\ 
J304               & 1.091(29)   & 5.95(65)    & 0.864(30)   & 2.646(25)   & 1.149(54)   & 6.61(74)    & 0.879(23)   & 2.645(18)   \\ 
J303               & 1.100(19)   & 5.23(59)    & 0.914(22)   & 2.671(31)   & 1.198(61)   & 5.88(96)    & 0.927(17)   & 2.669(26)   \\ 
E300               & 1.047(93)   & 7.9(3.8)    & 0.853(57)   & 2.728(87)   & 1.04(12)    & 9.1(3.1)    & 0.884(41)   & 2.705(55)   \\ 
J500               & 1.100(10)   & 5.78(20)    & 0.887(13)   & 2.58250(15) & 1.106(15)   & 5.82(16)    & 0.894(11)   & 2.58241(12) \\ 
J501               & 1.053(35)   & 6.01(86)    & 0.857(46)   & 2.629(22)   & 1.038(54)   & 6.26(66)    & 0.878(22)   & 2.628(19)   
  \end{tabular}
  \end{ruledtabular}
\end{table*}

\begin{table*}
  \caption{Results for the unrenormalized hyperon charges $g_{J}^{\Xi,
      \latt}$ for $J \in \{A, S, T, V\}$. \#ES labels the number of
    excited states used to determine the ground state matrix element,
    see the discussion in Sec.~\ref{sec:excited}. The three-point
    functions are computed employing the ``stochastic''
    approach. \label{tab:data_xi}} \centering
  \begin{ruledtabular}
  \begin{tabular}{llllllllll}
    \#ES & \multicolumn{4}{c}{$1$}  &  \multicolumn{4}{c}{$2$}   \\[3pt]
    Ensemble & $g_{A}^{\Xi,\latt}$ & $g_{S}^{\Xi,\latt}$ & $g_{T}^{\Xi,\latt}$ & $g_{V}^{\Xi,\latt}$ & $g_{A}^{\Xi,\latt}$ & $g_{S}^{\Xi,\latt}$ & $g_{T}^{\Xi,\latt}$ & $g_{V}^{\Xi,\latt}$ \\  
    \cmidrule{1-9}
    A653               & -0.399(16)  & 3.13(18)    & -0.2495(54) & 1.4229(30)  & -0.401(11)  & 3.019(89)   & -0.2534(38) & 1.4206(21)  \\ 
A650               & -0.3849(96) & 3.13(11)    & -0.2375(43) & 1.4213(28)  & -0.3905(95) & 3.109(92)   & -0.2423(43) & 1.4211(20)  \\ 
A654               & -0.3528(58) & 2.65(12)    & -0.2407(47) & 1.4344(67)  & -0.354(12)  & 2.63(17)    & -0.2408(59) & 1.4349(67)  \\ 
H101               & -0.3810(53) & 3.331(91)   & -0.2342(41) & 1.39556(41) & -0.3854(82) & 3.349(93)   & -0.2371(48) & 1.39528(40) \\ 
U103               & -0.3580(67) & 3.09(11)    & -0.2389(48) & 1.39547(47) & -0.362(12)  & 3.05(14)    & -0.2380(43) & 1.39563(47) \\ 
H107               & -0.3602(71) & 2.89(16)    & -0.2413(41) & 1.4165(82)  & -0.3653(77) & 2.93(12)    & -0.2453(38) & 1.4126(64)  \\ 
H102r002           & -0.3707(54) & 3.122(91)   & -0.2336(34) & 1.4045(52)  & -0.3708(92) & 3.16(12)    & -0.2357(42) & 1.4038(48)  \\ 
H102r001           & -0.3671(52) & 3.12(10)    & -0.2362(35) & 1.3987(55)  & -0.3683(99) & 3.16(14)    & -0.2383(45) & 1.3987(52)  \\ 
rqcd021            & -0.418(17)  & 3.93(22)    & -0.2211(82) & 1.415(15)   & -0.425(16)  & 3.81(21)    & -0.2275(71) & 1.409(10)   \\ 
H105               & -0.370(11)  & 3.25(26)    & -0.2214(68) & 1.4050(86)  & -0.372(11)  & 3.24(22)    & -0.2261(72) & 1.4047(73)  \\ 
H106               & -0.3448(45) & 2.605(83)   & -0.2347(27) & 1.4052(42)  & -0.3553(84) & 2.56(22)    & -0.2359(39) & 1.4049(46)  \\ 
C102               & -0.3431(42) & 2.672(96)   & -0.2296(41) & 1.4007(52)  & -0.341(13)  & 2.52(28)    & -0.2286(50) & 1.4000(57)  \\ 
C101               & -0.373(23)  & 3.15(38)    & -0.2398(58) & 1.423(13)   & -0.387(14)  & 3.19(35)    & -0.2436(33) & 1.4177(99)  \\ 
B450               & -0.3822(73) & 3.321(81)   & -0.2322(46) & 1.3737(22)  & -0.3914(98) & 3.289(96)   & -0.2352(56) & 1.3737(19)  \\ 
S400               & -0.3560(57) & 3.20(12)    & -0.2332(41) & 1.3774(56)  & -0.3577(91) & 3.28(13)    & -0.2373(50) & 1.3778(51)  \\ 
B452               & -0.3484(52) & 2.873(96)   & -0.2309(35) & 1.3669(53)  & -0.3523(85) & 2.90(13)    & -0.2333(45) & 1.3680(57)  \\ 
rqcd030            & -0.384(11)  & 3.93(22)    & -0.2249(65) & 1.3819(34)  & -0.394(14)  & 3.96(25)    & -0.2313(67) & 1.3807(28)  \\ 
N451               & -0.3693(57) & 3.55(15)    & -0.2183(42) & 1.3969(68)  & -0.3720(94) & 3.74(17)    & -0.2248(46) & 1.3948(55)  \\ 
N401               & -0.372(13)  & 3.17(18)    & -0.2296(58) & 1.389(13)   & -0.389(16)  & 3.07(32)    & -0.2325(68) & 1.388(12)   \\ 
N450               & -0.3489(63) & 2.92(14)    & -0.2294(42) & 1.3764(77)  & -0.351(14)  & 2.87(27)    & -0.2297(59) & 1.3759(78)  \\ 
X450               & -0.403(18)  & 4.29(31)    & -0.2314(88) & 1.3809(53)  & -0.414(23)  & 4.39(34)    & -0.2369(74) & 1.3810(45)  \\ 
D451               & -0.308(19)  & 2.70(25)    & -0.226(19)  & 1.361(35)   & -0.286(36)  & 1.87(98)    & -0.225(13)  & 1.351(24)   \\ 
D450               & -0.365(21)  & 3.79(88)    & -0.2334(98) & 1.391(18)   & -0.372(29)  & 4.27(62)    & -0.2377(53) & 1.391(11)   \\ 
D452               & -0.320(13)  & 3.75(66)    & -0.2213(80) & 1.399(13)   & -0.292(33)  & 4.41(95)    & -0.2268(84) & 1.397(11)   \\ 
N202               & -0.3758(48) & 3.78(10)    & -0.2202(41) & 1.34982(24) & -0.3759(89) & 3.817(99)   & -0.2244(50) & 1.34959(23) \\ 
N204               & -0.3365(44) & 2.916(92)   & -0.2262(35) & 1.3484(47)  & -0.326(11)  & 2.91(15)    & -0.2252(46) & 1.3479(51)  \\ 
X250               & -0.3753(54) & 4.18(14)    & -0.2195(43) & 1.3541(24)  & -0.3782(80) & 4.26(14)    & -0.2257(46) & 1.3536(19)  \\ 
N203               & -0.3612(42) & 3.652(95)   & -0.2178(30) & 1.3549(53)  & -0.3649(75) & 3.74(12)    & -0.2220(36) & 1.3542(47)  \\ 
N201               & -0.350(11)  & 3.44(32)    & -0.2222(49) & 1.377(12)   & -0.355(10)  & 3.56(20)    & -0.2285(47) & 1.3714(79)  \\ 
N200               & -0.3515(42) & 3.58(13)    & -0.2147(38) & 1.3665(49)  & -0.3542(78) & 3.73(16)    & -0.2184(42) & 1.3653(43)  \\ 
X251               & -0.3942(89) & 4.68(18)    & -0.2226(60) & 1.3600(46)  & -0.409(11)  & 4.84(21)    & -0.2295(43) & 1.3576(40)  \\ 
D200               & -0.347(16)  & 4.2(1.0)    & -0.2127(76) & 1.3738(89)  & -0.350(15)  & 4.69(49)    & -0.2237(51) & 1.3707(70)  \\ 
D201               & -0.3393(78) & 3.38(26)    & -0.2082(60) & 1.3647(83)  & -0.344(17)  & 3.47(51)    & -0.2100(74) & 1.3645(84)  \\ 
N300               & -0.3596(58) & 3.74(10)    & -0.2246(45) & 1.31662(16) & -0.359(10)  & 3.78(11)    & -0.2279(49) & 1.31649(12) \\ 
N304               & -0.3307(51) & 3.12(11)    & -0.2196(43) & 1.3182(66)  & -0.325(10)  & 3.09(17)    & -0.2169(42) & 1.3175(70)  \\ 
N302               & -0.361(11)  & 3.77(19)    & -0.2136(60) & 1.341(11)   & -0.373(15)  & 3.72(22)    & -0.2170(66) & 1.3383(89)  \\ 
J304               & -0.3366(73) & 3.45(22)    & -0.2090(50) & 1.3286(83)  & -0.346(11)  & 3.63(28)    & -0.2143(58) & 1.3277(67)  \\ 
J303               & -0.3416(75) & 3.66(23)    & -0.2133(48) & 1.3306(94)  & -0.358(15)  & 3.88(34)    & -0.2208(59) & 1.3298(81)  \\ 
E300               & -0.334(17)  & 3.75(49)    & -0.194(13)  & 1.342(14)   & -0.340(31)  & 3.5(1.0)    & -0.195(14)  & 1.344(17)   \\ 
J500               & -0.3508(46) & 3.89(11)    & -0.2174(36) & 1.291575(73) & -0.3454(91) & 3.93(11)    & -0.2198(38) & 1.291515(71) \\ 
J501               & -0.353(10)  & 4.14(25)    & -0.2055(62) & 1.314(11)   & -0.356(13)  & 4.17(21)    & -0.2071(55) & 1.3135(98)  
  \end{tabular}
  \end{ruledtabular}
\end{table*}
\FloatBarrier
\section{Additional figures\label{sec:plots}}
We provide additional figures (Figs.~\ref{fig:fits_sigma}--\ref{fig:extrapol_gt_hyperon}).
\begin{figure*}[!h]
  \centering
  \includegraphics[width=0.98\textwidth]{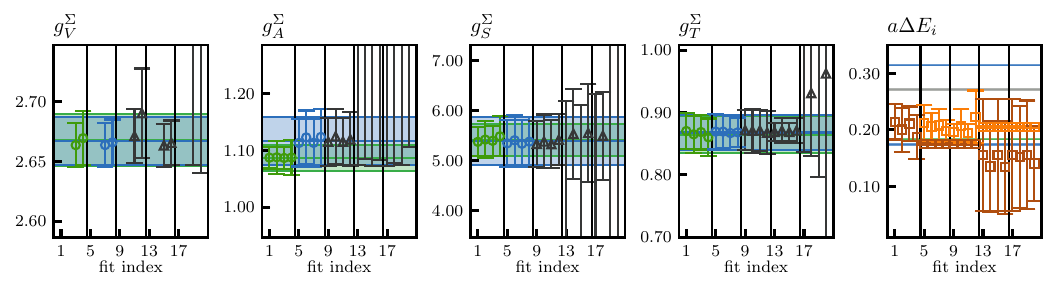}
  \caption{The same as Fig.~\ref{fig:fits_nucleon} for the sigma baryon.\label{fig:fits_sigma}}
\end{figure*}
\begin{figure*}[!h]
  \begin{center}
    \includegraphics[width=0.46\textwidth]{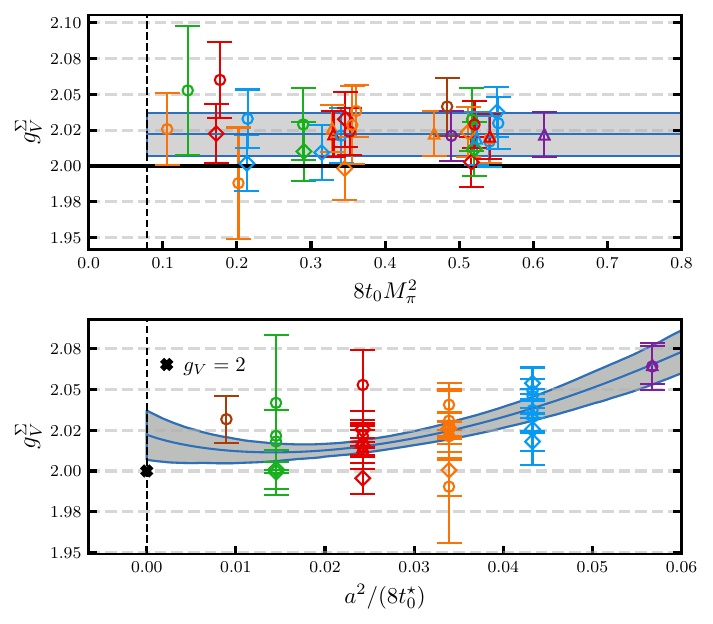}
    \includegraphics[width=0.46\textwidth]{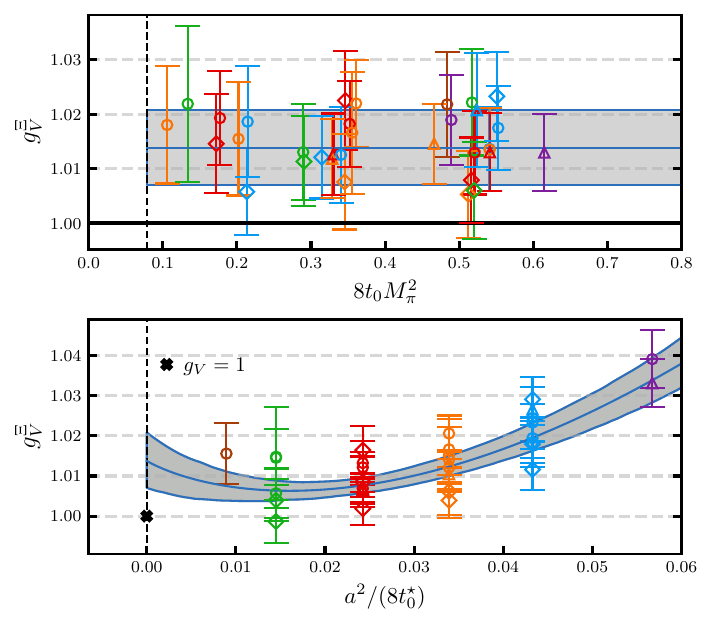}
    \caption{The same as Fig.~\ref{fig:extrapol_gv_nucleon} for the
      isovector vector charges~$g_V^B$ of the sigma baryon (left) and
      the cascade baryon (right). For better visibility, the data
      points for ensemble D451, which have large errors~(see
      Tables~\ref{tab:data_sigma} and~\ref{tab:data_xi}), are not
      displayed. See Tables~\ref{tab:data_sigma} and \ref{tab:data_xi}
      for the set of ensembles used.\label{fig:extrapol_gv_hyperon}}
  \end{center}
\end{figure*}
\end{widetext}
\begin{figure*}[!h]
  \begin{center}
    \includegraphics[width=0.45\textwidth]{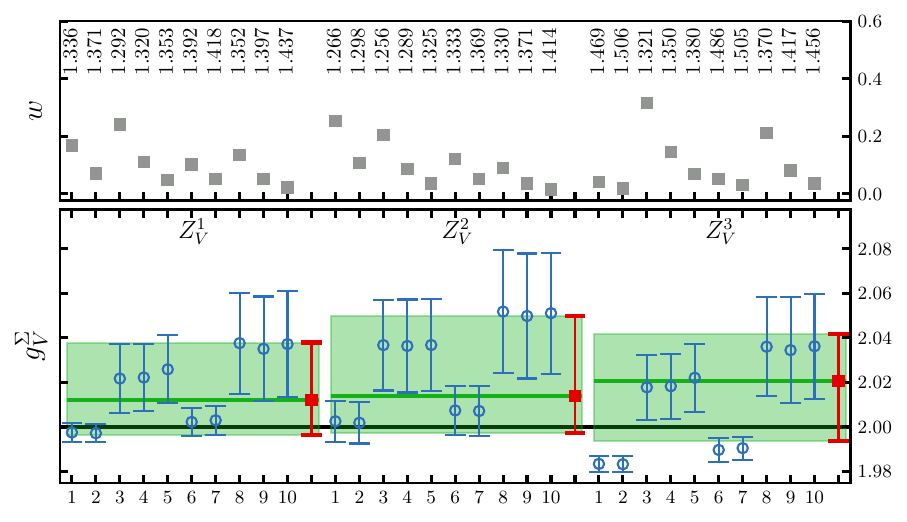}
    \includegraphics[width=0.45\textwidth]{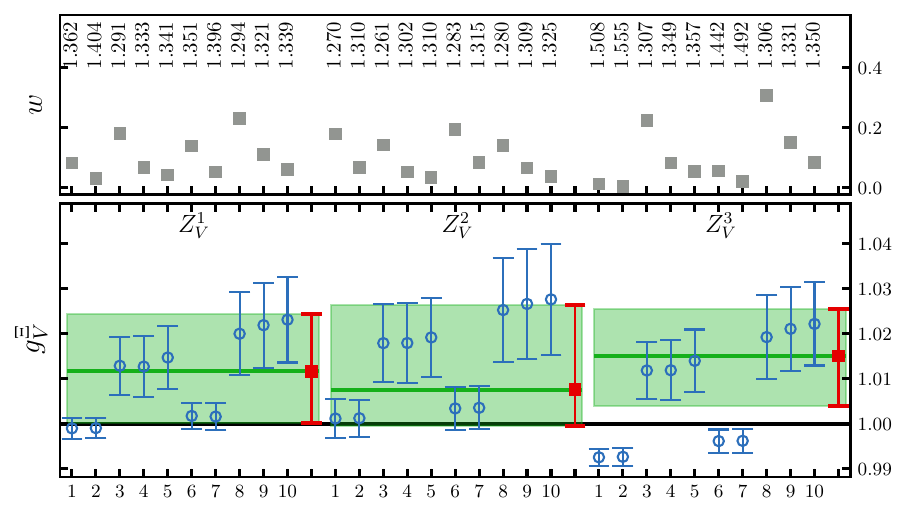}
    \caption{The same as Fig.~\ref{fig:modelavg_gv} for the vector
      charge $g_V^B$ of the sigma (right) and cascade (left)
      baryon.\label{fig:modelavg_gv_hyperon}}
  \end{center}
\end{figure*}
\begin{figure*}
  \begin{center}
    \includegraphics[width=0.45\textwidth]{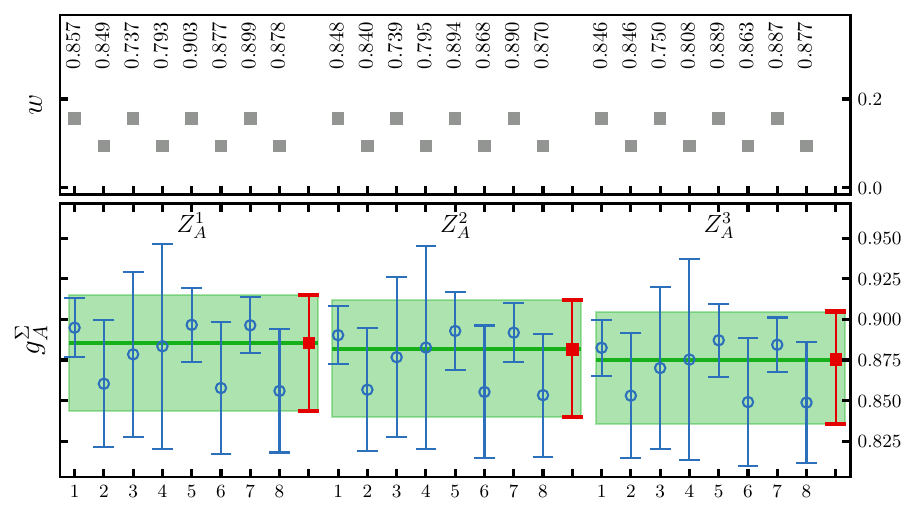}
    \includegraphics[width=0.45\textwidth]{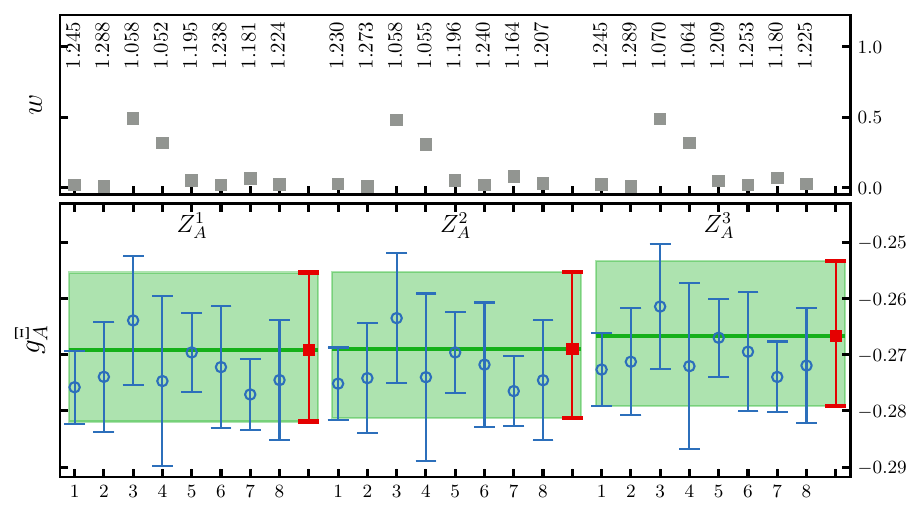}
    \caption{The same as Fig.~\ref{fig:modelavg_ga_nucleon} for the axial
    charge $g_A^B$ of the sigma (right) and cascade (left)
    baryon. \label{fig:modelavg_ga_hyperons}}
  \end{center}
\end{figure*}
\begin{figure*}
  \begin{center}
    \includegraphics[width=0.44\textwidth]{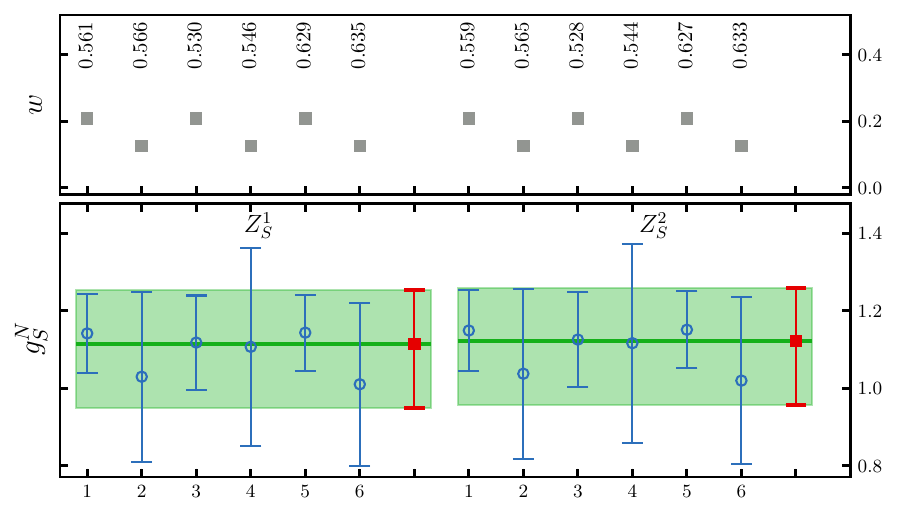}
    \includegraphics[width=0.44\textwidth]{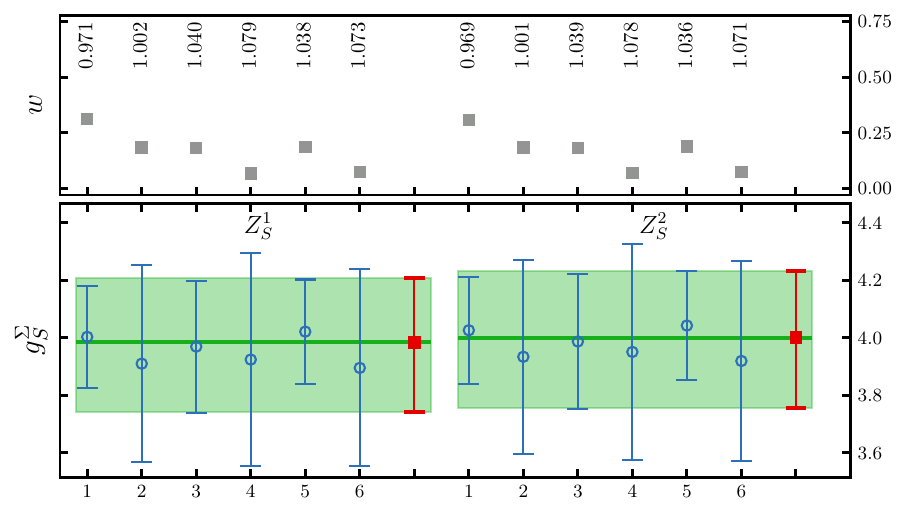}\\
    \includegraphics[width=0.44\textwidth]{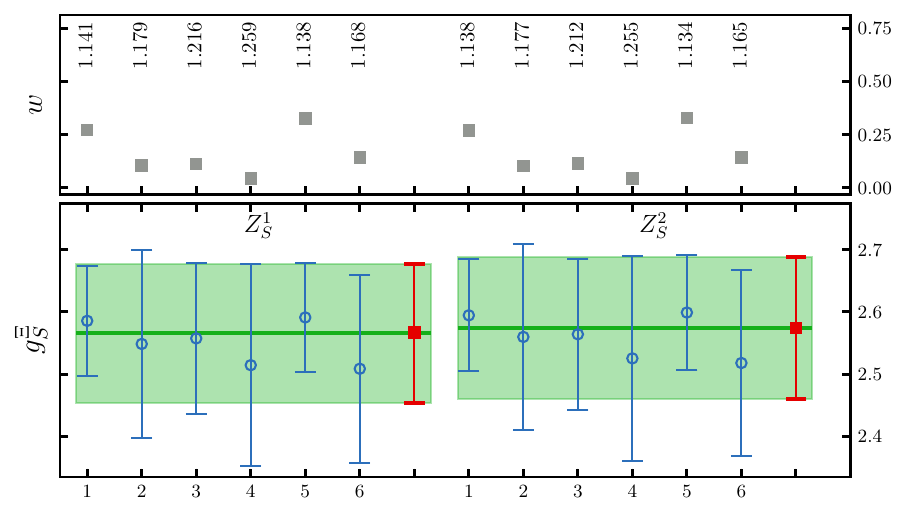}
    \caption{The same as Fig.~\ref{fig:modelavg_ga_nucleon} for the scalar
      charge $g_S^B$ of the nucleon~(top, left), sigma~(top, right)
      and cascade~(bottom) baryons. The six fits correspond to two fit
      variations, see the text, applied to three data sets,
      DS($M_\pi^{\footnotesize{<400\, \MeV}}$),
      DS($M_\pi^{\footnotesize{<400\, \MeV}}$,
      $a^{\footnotesize{<0.1\,\fm}}$) and
      DS($M_\pi^{\footnotesize{<400\, \MeV}}$,
      $LM_\pi^{\footnotesize{> 4}}$).  The data are extracted using
      two excited states in the fitting analysis, see
      Sec.~\ref{sec:excited}. Only two different sets of renormalization factors are used.\label{fig:modelavg_gs}}
  \end{center}
\end{figure*}

\begin{figure*}
  \begin{center}
    \includegraphics[width=0.45\textwidth]{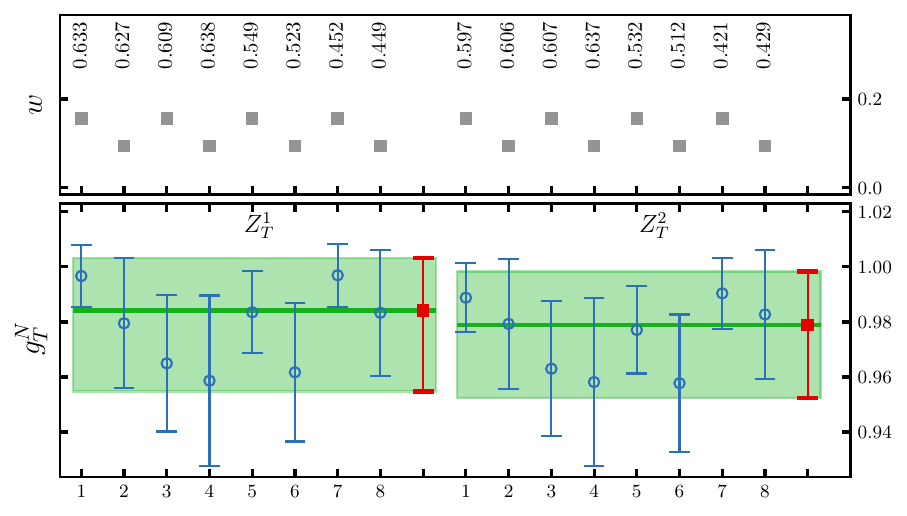}
    \includegraphics[width=0.45\textwidth]{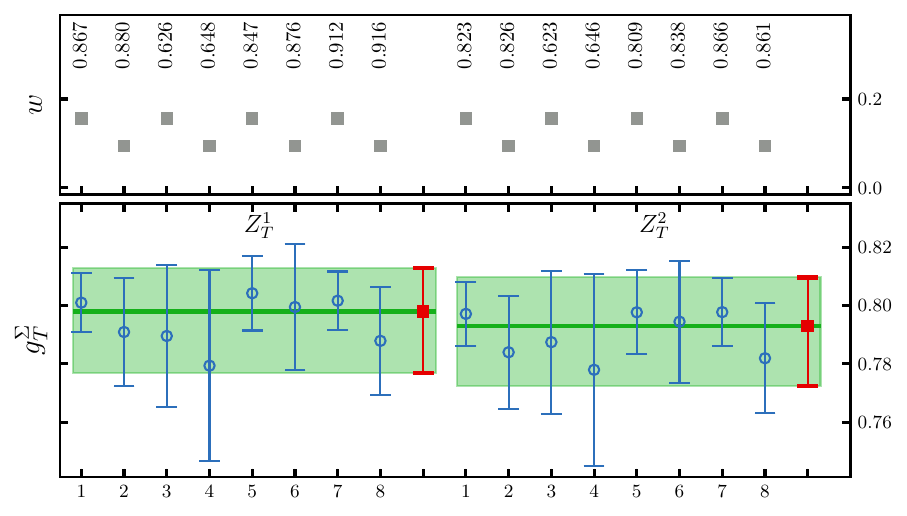}\\
    \includegraphics[width=0.45\textwidth]{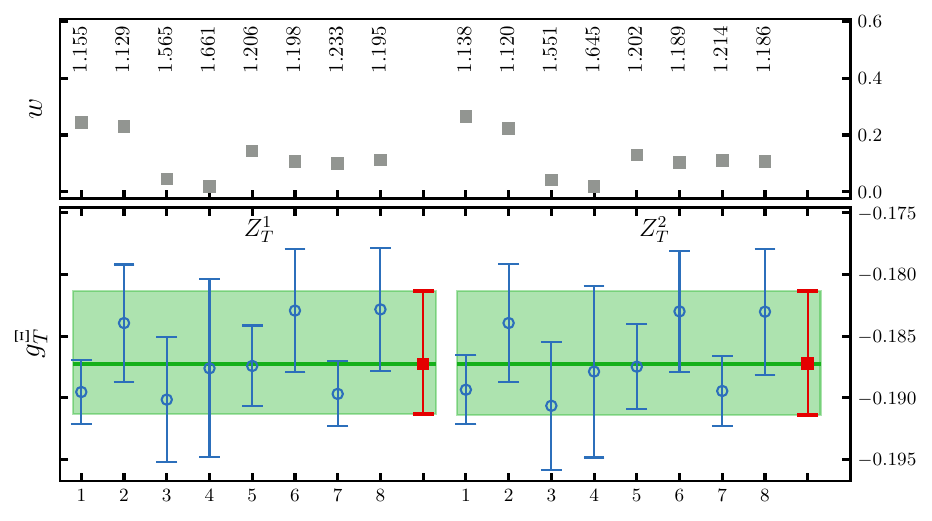}
    \caption{The same as Fig.~\ref{fig:modelavg_ga_nucleon} for the tensor charge $g_T^B$ for the nucleon~(top, left), sigma~(top, right) and cascade~(bottom) baryons.  The data are extracted using two excited states in the fitting analysis, see Sec.~\ref{sec:excited}. 
    Only two different sets of renormalization factors are used. \label{fig:modelavg_gt}}
  \end{center}
\end{figure*}
\begin{figure*}
  \begin{center}
    \includegraphics[width=0.46\textwidth]{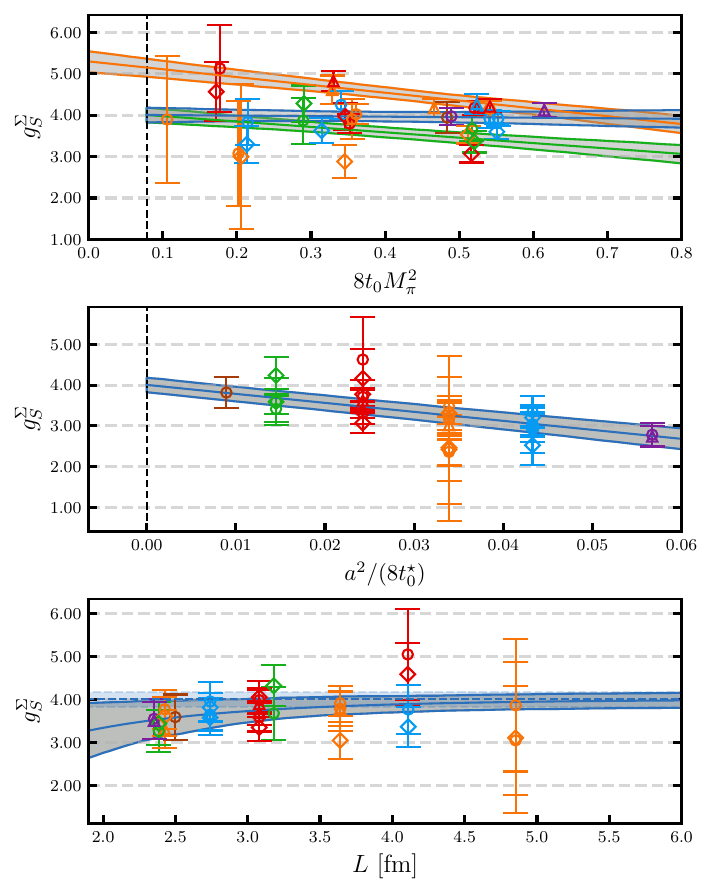}
    \includegraphics[width=0.46\textwidth]{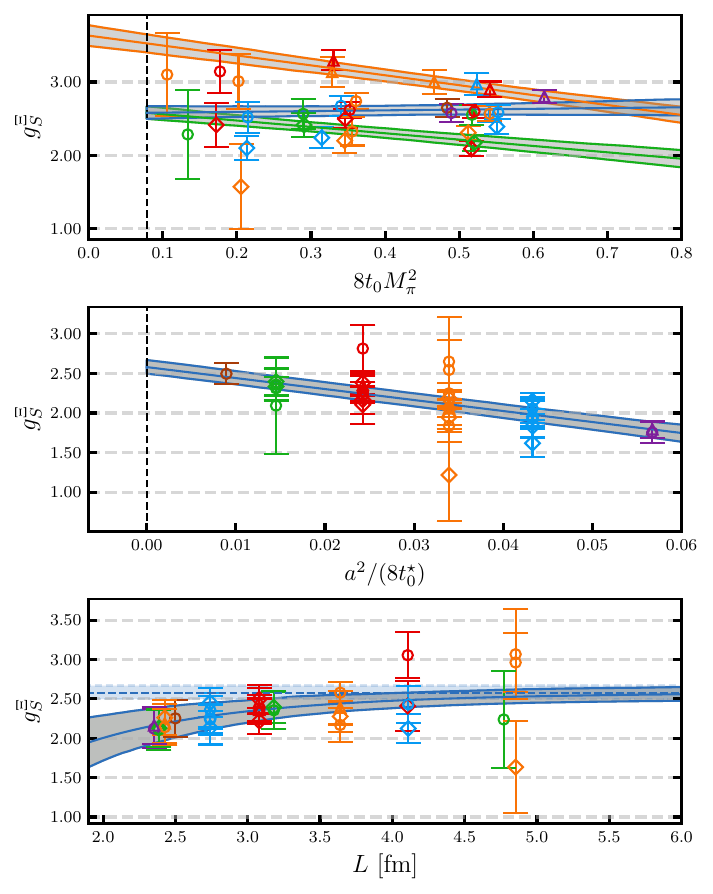}
    \caption{The same as Fig.~\ref{fig:extrapol_gs_nucleon} for the
      isovector scalar charges~$g_S^B$ of the sigma baryon (left) and
      the cascade baryon (right). For better visibility, the data
      point for ensemble E300, which has a relatively large error~(see
      Table~\ref{tab:data_sigma}), is not displayed for the sigma
      baryon. See Tables~\ref{tab:data_sigma} and \ref{tab:data_xi}
      for the set of ensembles used.\label{fig:extrapol_gs_hyperon}}
  \end{center}
\end{figure*}
\begin{figure*}
  \begin{center}
    \includegraphics[width=0.45\textwidth]{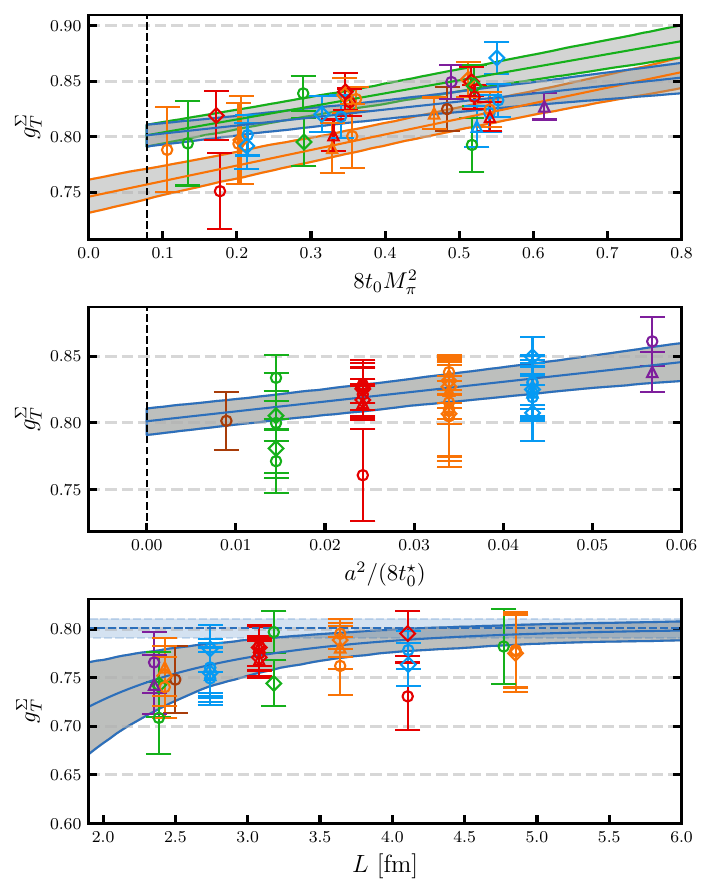}
    \includegraphics[width=0.45\textwidth]{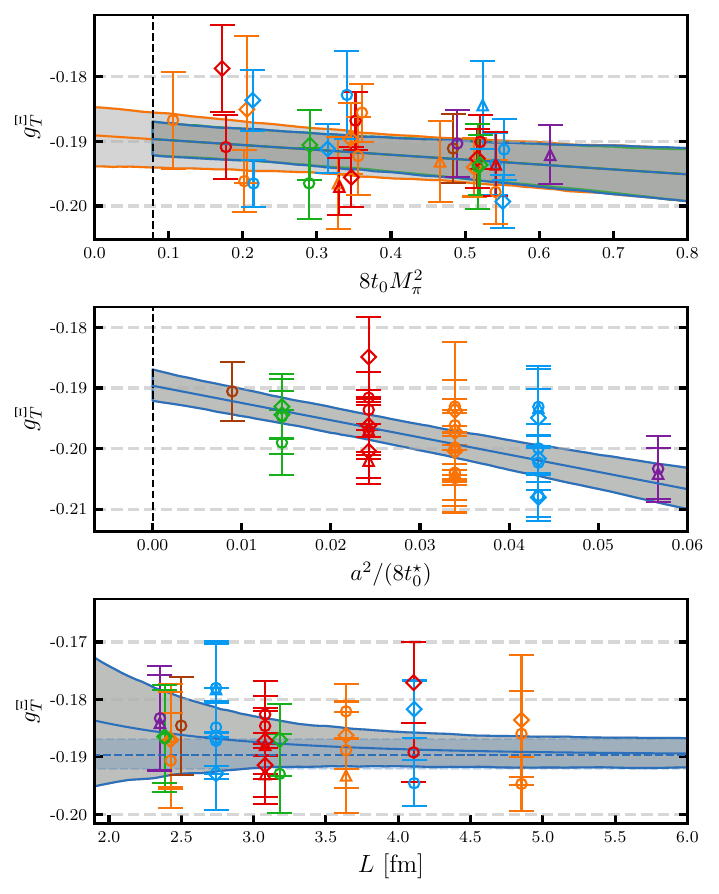}
    \caption{The same as Fig.~\ref{fig:extrapol_gt_nucleon} for the
      isovector tensor charges~$g_T^B$ of the sigma baryon (left) and
      the cascade baryon (right). For better visibility, the data
      point for ensemble E300, which has a relatively large error~(see
      Table~\ref{tab:data_xi}), is not displayed for the cascade
      baryon. See Tables~\ref{tab:data_sigma} and \ref{tab:data_xi}
      for the set of ensembles used.\label{fig:extrapol_gt_hyperon}}
  \end{center}
\end{figure*}
\FloatBarrier
\end{document}